\documentclass[useAMS,usenatbib]{mn2e}

\usepackage{graphicx}
\usepackage{times}
\usepackage{amssymb} 
\usepackage{txfonts}
\usepackage{appendix}
\usepackage{natbib}
\usepackage{longtable, portland}
\usepackage{supertabular}
\usepackage{rotating}
\usepackage{array}
\usepackage{caption}
\usepackage{gensymb}
\usepackage[T1]{fontenc}

\title[Star formation and AGN activity in the local most luminous LINERs]{Star formation and AGN activity in the most luminous LINERs in the local universe}
\author[Povi\'c et al.]{Mirjana Povi\'c$^{1}$\thanks{E-mail: mpovic@iaa.es}, Isabel M\'arquez$^{1}$, Hagai Netzer$^{2}$, Josefa Masegosa$^{1}$, Raanan Nordon$^{2}$,\newauthor Enrique P\'erez$^{1}$, and William Schoenell$^{1}$ \\
$^{1}$Instituto de Astrof\'isica de Andaluc\'ia (IAA-CSIC), Granada 18008, Spain\\
$^{2}$School of Physics and Astronomy and the Wise Observatory, The Raymond and Beverly Sackler Faculty of Exact Sciences, Tel-Aviv University, Tel-Aviv 69978, Israel\\}
\begin{document}

\date{Accepted ??. Received ??; in original form ??}

\pagerange{\pageref{firstpage}--\pageref{lastpage}} \pubyear{2016}

\maketitle

\label{firstpage}

\begin{abstract}
This work presents the properties of 42 objects in the group of the most luminous, highest star formation rate LINERs at z\,=\,0.04\,-\,0.11. We obtained long-slit spectroscopy of the nuclear regions for all sources, and FIR data (Herschel and IRAS) for 13 of them. We measured emission line intensities, extinction, stellar populations, stellar masses, ages, AGN luminosities, and star-formation rates. We find considerable differences from other low-redshift LINERs, in terms of extinction, and general similarity to star forming (SF) galaxies. We confirm the existence of such luminous LINERs in the local universe, after being previously detected at z\,$\sim$\,0.3 by \cite{tommasin12}. The median stellar mass of these LINERs corresponds to 6\,-\,7\,$\times$\,10$^{10}$\,M$_{\odot}$ which was found in previous work to correspond to the peak of relative growth rate of stellar populations and therefore for the highest SFRs. Other LINERs although showing similar AGN luminosities have lower SFR. We find that most of these sources have LAGN\,$\sim$\,LSF suggesting co-evolution of black hole and stellar mass. In general among local LINERs being on the main-sequence of SF galaxies is related to their AGN luminosity.

\end{abstract}


\section{Introduction}
\label{sec_intro}

\indent \indent Low Ionization Nuclear Emission line Regions (LINERs) are the most common active galactic nuclei (AGN), with numbers that exceed those of 'high ionization AGN' (type-I and type-II Seyfert galaxies and quasars) \citep{heckman80,ho08,heckman14}. At least in the local universe they make up 1/3 of all galaxies and 2/3 of AGN population \citep{kauffmann03a,yan06,ho08}. LINERs are normally classified by their narrow emission line ratios, e.g. [OIII]$\lambda$5007/H$\beta$, [NII]$\lambda$6584/H$\alpha$, and [OI]$\lambda$6300/H$\alpha$ \citep{baldwin81,kauffmann03a,stasinska06,kewley06}. In general, they have lower luminosities than Seyfert galaxies, but there is a big overlap between the groups in terms of properties like stellar mass, X-ray and radio luminosity, etc.  \citep{ho08,netzer09,leslie16}.\\
\indent Different mechanisms were proposed to explain the nature of LINERs. This includes shock excitation \citep[e.g.][]{dopita97,nagar05}, photoionisation by young, hot, massive stars \citep{terlevich85}, photoionisation by evolved post-asymptotic giant branch (pAGB) stars \citep[e.g.][]{stasinska08,annibali10,cid11,yan12,singh13}, and photoionisation by a central low-luminosity AGN \citep[e.g.][]{ferland83,ho08,gonzalezmartin06}. The first two proposals failed to explain the properties of large samples of LINERs. The third possibility  of pAGB stars was suggested for LINERs with the weakest emission lines, located in galaxies with predominately old stars. They can be distinguished from strong-line LINERs using the equivalent widths (EW) of their emission lines, e.g., EW([OIII]$\lambda$5007)\,$<$\,1\,\AA\,\citep{capetti11} or EW(H$\alpha$)\,$<$\,3\,\AA\,\citep{cid11}. Several works however questioned this possibility, arguing that a population that is less luminous and more numerous than pAGB stars would be needed to produce the luminosities observed in weak LINERs \citep{brown08,rosenfield12,heckman14}. However, most LINERs are powered by an AGN, especially those with stronger emission lines (e.g., EW(H$\alpha$)\,$>$\,3\AA) and unresolved hard X-ray emission \citep[e.g., ][and references therein]{gonzalezmartin06,gonzalezmartin09a,gonzalezmartin09b,heckman14}. Like other AGNs, LINERs can be divided into type-I (broad and narrow emission lines) and type-II (only narrow emission lines). Their emission lines are characterised by lower levels of ionization than in Seyferts, and their normalized accretion rates (Eddington ratio) are 1-5 orders of magnitude smaller.\\
\indent The best studied nearby LINERs \citep[e.g.][]{ho97,ho08,kauffmann03a,leslie16} are found in nuclei of galaxies with little or no evidence of active star formation (SF). They are usually characterised as being hosted by massive early-type galaxies (rarely spirals), and massive black holes in their centres, old stellar populations, small amounts of gas and dust, with low extinctions. Such LINERs show  weak and small-scale radio jets \citep{ho08,heckman14}. \\
\indent \cite{tommasin12} studied SF in LINERs from the COSMOS field at z\,$\sim$\,0.3 using Herschel/PACS observations. They showed that: a) The SF luminosities of 34 out of 97 high luminosity LINERs are on average 2 orders of magnitude higher than SF luminosities of lower AGN luminosity, nearby LINERs. b) Even if assumed that all the observed H$\alpha$ flux is due to SF (a wrong assumption since much of it must be due to AGN excitation) it is still impossible to recover the SF rate (SFR) indicated by the FIR observations. Given this result, we suspect that active SF in LINER host galaxies has escaped the attention of most earlier studies that focused on the innermost part of nearby galaxies. In this work we focus on the most luminous LINERs in the local (0.04\,$<$\,z\,$<$\,0.11) universe and study their SF and AGN activity, in order to understand the LINER phenomenon in relation to star-forming galaxies and to compare their properties with those of the LINERs at z\,$\sim$\,0.3. Many properties of these sources are known from SDSS spectroscopy and/or GALEX observations, e.g., emission line luminosities, locations on the BPT diagrams, SFRs based on Dn4000 estimations, etc. Unfortunately, the 3\,arc-sec SDSS fibre does not allow to resolve the nuclear region and hence to separate AGN excited from SF excited emission lines. The goals of the present study are to carry out a detailed, ground based spectroscopy of the central regions of the most luminous LINERs, and to measure, together with Herschel and IRAS FIR data, their SFRs in a careful way. \\
\indent The paper is organised as follows: in Section~\ref{sec_sample_selection} we describe the sample selection. Reduction procedure for our new spectroscopic data, together with our own or archival FIR data are described in Section~\ref{sec_observations}. In Section~\ref{sec_analysis} we summarise all our measurements, including spectral fittings, emission line and extinction measurements, and estimations of Dn4000 and H$\delta$ indices, AGN luminosities, and SFRs. The main results are presented in Section~\ref{sec_results_discussion} where we discuss the general properties of the most luminous LINERs in the local universe, co-evolution between the SF and AGN activity, and the location of our sample on the main sequence of SF galaxies. \\
\indent We assumed the following cosmological parameters throughout the paper: $\Omega_{\Lambda}$\,=\,0.7, $\Omega_{M}$\,=\,0.3, and H$_0$\,=\,70 km s$^{-1}$ Mpc$^{-1}$.

\section[]{Sample selection}
\label{sec_sample_selection}

\begin{figure}
\centering
\includegraphics[width=0.49\textwidth,angle=0]{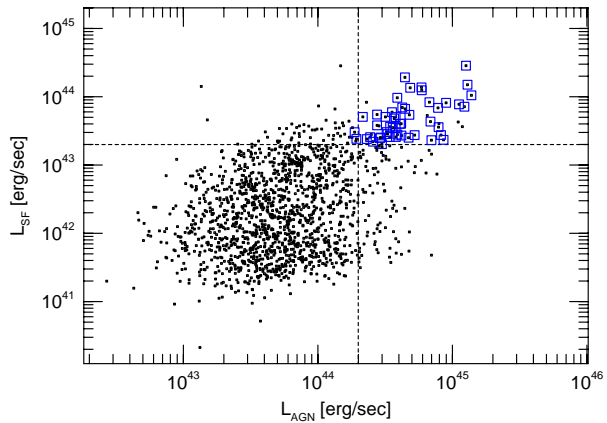}
\caption{The entire 0.04\,$<$\,z\,$<$\,0.11 SDSS/DR4 LINER sample used in this work (small black squares) and the sub-sample used for the Herschel proposal and the follow up spectroscopy (large blue squares). The dashed lines mark
the lower limits on LAGN and LSF (based on Dn4000 index) used for the selection of the targets.
\label{fig_sample_selection}}
\end{figure}

\begin{figure*}
\centering
\includegraphics[width=0.8\textwidth,angle=0]{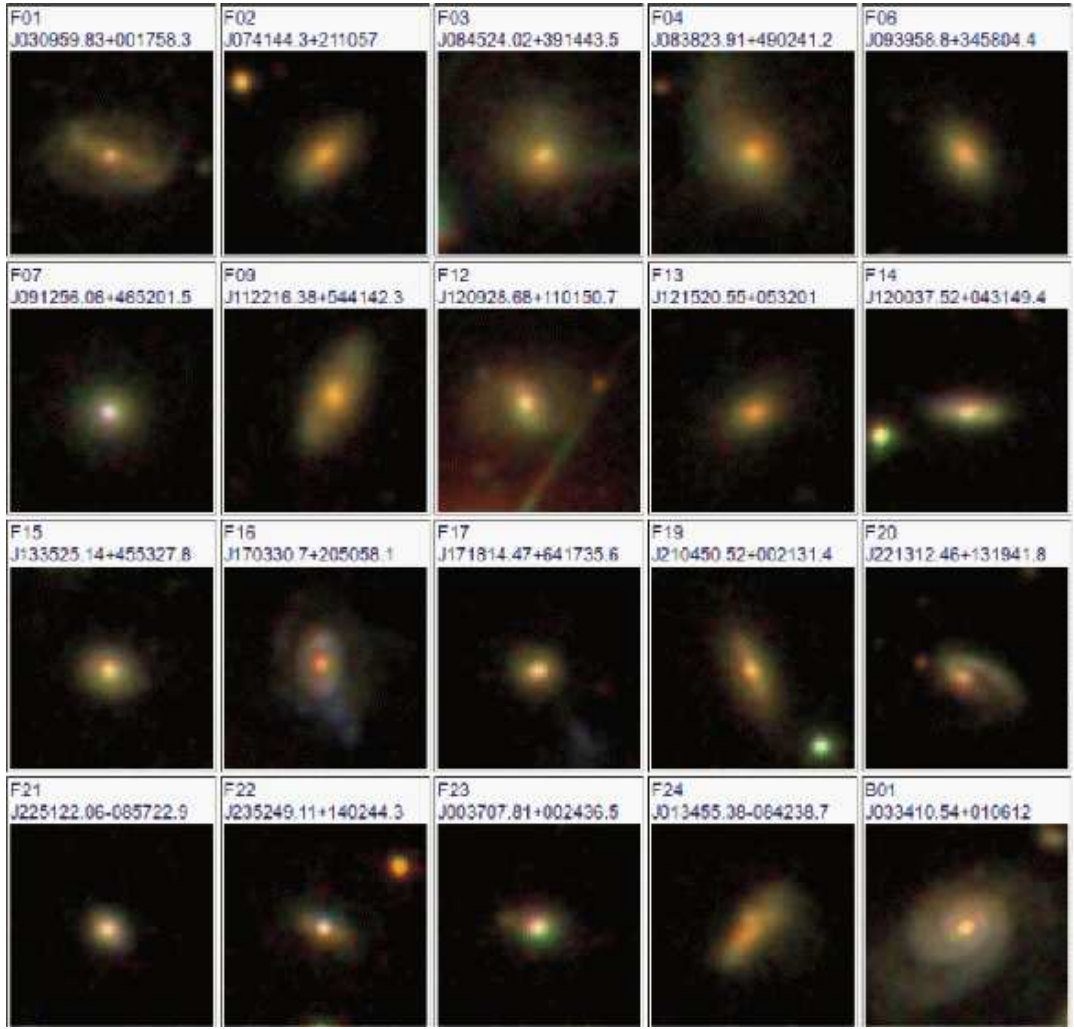}
\caption{SDSS gri colour images of our selected sample of the most luminous local LINERs. The top and bottom identifications correspond to our and SDSS ones, respectively.
\label{fig_thumbnails_part1}}
\end{figure*}

\indent \indent The sources were initially selected from the SDSS/DR4 \citep{kauffmann03,brinchmann04} catalogue in Garching MPA-JHU based on the Sloan Digital Sky Survey (SDSS\footnote{http://www.sdss.org/}) DR4 data \citep[][and references therein]{adelman06}. LINERs were first selected using both [NII]$\lambda$6584/H$\alpha$ and [OI]$\lambda$6300/H$\alpha$ criteria of \cite{kewley06}. Taking into account the completeness of the SDSS survey, only LINERs with 0.04\,$<$\,z\,$<$\,0.11 were selected \citep{netzer09}. To eliminate LINERs ionised by pAGB stars we selected only those galaxies with H$\alpha$ equivalent width EW(H$\alpha$)\,$>$\,2.5\AA ~\citep{cid11}.  \\ 
\indent The next step was the selection of the most luminous LINERs within the chosen redshift interval. We measured first their AGN luminosity (LAGN) using the [OIII]$\lambda$5007 and [OI]$\lambda$6300 method of \cite{netzer09} (see Section~\ref{sec_agn_lum}). The lines were initially corrected for reddening using the observed H$\alpha$/H$\beta$ ratio and assuming galactic extinction (see Sections~\ref{sec_emission_measure} and \ref{sec_agn_lum}). We selected a certain, statistically sufficient, fraction of 147 luminous LINERs with logLAGN\,$>$\,44.3\,ergs/sec. We call these sources 'LLINERs'. Out of these sources we selected a luminosity limited sample of 47 galaxies with SF luminosity LSF\,$>$\,43.3\,ergs/sec, where LSF is based on the Dn4000 index (see Section~\ref{sec_sfr}). Of those, we were able to obtain the optical spectra for 42 LINERs and Herschel/PACS data for 6 sources. We refer to these 42 most luminous LINERs in terms of both AGN and SF luminosity as 'MLLINERs'. All observed MLLINERs are listed in Table~\ref{tab_obs}, where we provide the basic information about their properties. \\
\indent Figure~\ref{fig_sample_selection} shows the position in the LAGN vs. LSF plane of the initially classified LINERs in the selected redshift range (black dots), and the final selected sample of MLLINERs (blue squares). Using the SDSS spectroscopy we estimated the AB continuum magnitude at 6500\,\AA\,(m6500). We used these magnitudes to divide the sample into 'faint' and 'bright' galaxies (m6500\,$>$\,17.2\,mag and m6500\,$<$\,17.2\,mag, respectively). These groups are marked with F or B in Table 1. We use this classification only for observational purposes. Figs~\ref{fig_thumbnails_part1} and \ref{fig_thumbnails_part2} show SDSS colour images of all MLLINERs.

\begin{table*}
\small
\begin{center}
\caption{Summary of observations.  
\label{tab_obs}}
\begin{tabular}{| c | c | c | c | c | c | c | c | c | c | c | c | c |}
\hline
\textbf{ID}&\textbf{RA}&\textbf{DEC}&\textbf{z}&\textbf{m6500}&\textbf{morph}&\textbf{Date}&\textbf{seeing}&\textbf{pos. ang.}&\textbf{texp\_b}&\textbf{texp\_r}& \textbf{Area$_{nuc}$}& \textbf{IR data}\\
&\textbf{[deg]}&\textbf{[deg]}&&\textbf{AB [mag]}&&&\textbf{[arc-sec]}&\textbf{[deg]}&\textbf{[sec]}&\textbf{[sec]}&\textbf{[arc-sec$^2$]}& \\
\hline
F01  &  47.499332  & 0.29955   & 0.098 & 18.53 & S & 02/11/2013 & 1.2 & PA  & 3\,$\times$\,3000.0 & 3\,$\times$\,3000.0 & 3.6 &      \\
F02  &  115.434586 & 21.18252  & 0.098 & 17.96 & E & 06/03/2014 & 1.2 & 314 & 3\,$\times$\,3000.0 & 3\,$\times$\,3000.0 & 3.6 & 1, 2  \\
F03  &  131.35008  & 39.245438 & 0.109 & 17.63 & P & 08/03/2014 & 1.3 & 201 & 3\,$\times$\,2000.0 & 3\,$\times$\,2000.0 & 3.9 &      \\
F04  &  129.59967  & 49.04478  & 0.101 & 17.58 & P & 08/03/2014 & 1.3 & 206 & 3\,$\times$\,2400.0 & 3\,$\times$\,2400.0 & 3.9 &      \\
F06  &  144.995    & 34.96791  & 0.104 & 17.63 & E & 09/03/2014 & 1.4 & 220 & 3\,$\times$\,2000.0 & 3\,$\times$\,2000.0 & 4.2 & 2     \\
F07  &  138.23363  & 46.8671   & 0.051 & 17.27 & E & 09/03/2014 & 1.4 & 338 & 3\,$\times$\,1800.0 & 3\,$\times$\,1800.0 & 4.2 &      \\
F09  &  170.5683   & 54.6951   & 0.105 & 17.50 & S & 03/05/2014 & 1.4 & 149 & 3\,$\times$\,2000.0 & 3\,$\times$\,2000.0 & 4.2 & 2    \\
F12  &  182.36954  & 11.030761 & 0.107 & 17.21 & S & 02/05/2014 & 1.6 & 410 & 3\,$\times$\,2400.0 & 3\,$\times$\,2400.0 & 6.0 & 2    \\
F13  &  183.83566  & 5.533633  & 0.082 & 18.09 & E & 05/05/2014 & 1.2 & 120 & 3\,$\times$\,3000.0 & 3\,$\times$\,3000.0 & 3.6 &       \\
F14  &  180.15637  & 4.530397  & 0.094 & 17.54 & S & 04/05/2014 & 1.2 & 265 & 3\,$\times$\,2000.0 & 3\,$\times$\,2000.0 & 3.6 & 2     \\
F15  &  203.8548   & 45.891083 & 0.092 & 17.42 & E & 03/05/2014 & 1.4 & 239 & 3\,$\times$\,1800.0 & 3\,$\times$\,1800.0 & 4.2 &      \\
F16  &  255.87796  & 20.849482 & 0.08  & 18.43 & ? & 26/07/2014 & 1.0 & 184 & 3\,$\times$\,3600.0 & 3\,$\times$\,3600.0 & 3.0 & 2     \\
F17  &  259.5603   & 64.29323  & 0.104 & 17.78 & E & 06/03/2014 & 1.2 & 213 & 3\,$\times$\,2400.0 & 3\,$\times$\,2400.0 & 3.6 & 1, 2 \\
F19  &  316.2105   & 0.358728  & 0.091 & 17.90 & ? & 25/07/2014 & 1.0 & 205 & 3\,$\times$\,2800.0 & 3\,$\times$\,2800.0 & 3.0 & 1    \\
F20  &  333.30197  & 13.3283   & 0.103 & 18.53 & P & 27/07/2014 & 1.3 & 127 & 3\,$\times$\,3600.0 & 3\,$\times$\,3600.0 & 3.9 &      \\
F21  &  342.84195  & -8.956378 & 0.08  & 17.50 & E & 28/07/2014 & 1.6 & 241 & 3\,$\times$\,3000.0 & 3\,$\times$\,3000.0 & 4.8 &       \\
F22  &  358.20468  & 14.04565  & 0.096 & 18.02 & ? & 29/07/2014 & 1.2 & 238 & 3\,$\times$\,3200.0 & 3\,$\times$\,3200.0 & 3.6 &      \\
F23  &  9.282583   & 0.410139  & 0.081 & 17.42 & ? & 30/07/2014 & 1.4 & 260 & 3\,$\times$\,2800.0 & 3\,$\times$\,2800.0 & 4.2 &      \\
F24  &  23.73075   & -8.710756 & 0.092 & 18.02 & P & 09/10/2013 & 1.2 & PA  & 3\,$\times$\,3000.0 & 3\,$\times$\,3000.0 & 3.6 & 2    \\
B01  &  53.543957  & 1.103353  & 0.048 & 17.17 & S & 31/10/2013 & 1.5 & PA  & 3\,$\times$\,1600.0 & 3\,$\times$\,1600.0 & 5.6 &      \\
B02  &  124.66104  & 23.48597  & 0.103 & 16.90 & P & 07/03/2014 & 1.2 & 315 & 3\,$\times$\,1700.0 & 3\,$\times$\,1700.0 & 3.6 & 1    \\
B03  &  129.57721  & 33.57853  & 0.062 & 16.79 & P & 06/03/2014 & 1.2 & 274 & 3\,$\times$\,1600.0 & 3\,$\times$\,1600.0 & 3.6 & 1, 2  \\
B04  &  133.79796  & 0.219117  & 0.101 & 16.90 & E & 10/03/2014 & 1.3 & 255 & 3\,$\times$\,1700.0 & 3\,$\times$\,1700.0 & 3.9 &      \\
B05  &  141.73837  & 8.630544  & 0.106 & 17.09 & S & 10/03/2014 & 1.3 & 180 & 3\,$\times$\,1800.0 & 3\,$\times$\,1800.0 & 3.9 & 2    \\
B06*  &  160.26555  & 11.096189 & 0.053 & 16.50 & ? & 01/05/2013 & 0.9 & PA  & 4\,$\times$\,900.0  & 3\,$\times$\,900.0  & 3.0 &      \\
B07  &  165.55441  & 66.1674   & 0.078 & 17.17 & P & 06/03/2014 & 1.2 & 245 & 3\,$\times$\,1800.0 & 3\,$\times$\,1800.0 & 3.6 & 1    \\
B08*  &  170.29817  & -0.293878 & 0.098 & 17.11 & E & 03/05/2013 & 0.9 & PA  & 4\,$\times$\,1200.0 & 4\,$\times$\,900.0  & 3.0 &      \\
B09  &  171.66946  & -1.6938   & 0.046 & 15.93 & E & 10/03/2014 & 1.3 & 290 & 3\,$\times$\,1200.0 & 3\,$\times$\,1200.0 & 3.9 &       \\
B10  &  183.72675  & 1.916183  & 0.099 & 16.98 & ? & 10/03/2014 & 1.3 & 315 & 3\,$\times$\,1700.0 & 3\,$\times$\,1700.0 & 3.9 &      \\
B11  &  187.959    & 58.35786  & 0.103 & 17.03 & P & 03/05/2014 & 1.4 & 446 & 3\,$\times$\,1800.0 & 3\,$\times$\,1800.0 & 4.2 & 2    \\
B12  &  190.78575  & 1.728797  & 0.092 & 17.09 & E & 05/05/2014 & 1.2 & 238 & 3\,$\times$\,1800.0 & 3\,$\times$\,1800.0 & 3.6 &      \\
B13*  &  191.979    & -3.627378 & 0.09  & 16.59 & S & 03/05/2013 & 0.7 & PA  & 2\,$\times$\,1200.0 & 4\,$\times$\,900.0  & 2.3 & 2    \\
B14*  &  192.3075   & 15.252789 & 0.083 & 16.90 & S & 01/05/2013 & 0.7 & PA  & 4\,$\times$\,900.0  & 3\,$\times$\,900.0  & 2.3 &      \\
B15*  &  205.55083  & -0.293453 & 0.086 & 17.17 & E & 02/05/2013 & 0.7 & PA  & 3\,$\times$\,900.0  & 4\,$\times$\,900.0  & 2.3 &      \\
B16  &  207.66092  & 53.73111  & 0.108 & 16.95 & E & 09/03/2014 & 1.4 & 267 & 3\,$\times$\,1700.0 & 3\,$\times$\,1700.0 & 4.2 &      \\
B17  &  211.27605  & 2.771761  & 0.077 & 17.17 & P & 04/05/2014 & 1.2 & 180 & 3\,$\times$\,1800.0 & 3\,$\times$\,1800.0 & 3.6 & 2    \\
B18  &  212.88733  & 45.28614  & 0.071 & 17.14 & E & 02/05/2014 & 1.6 & 109 & 3\,$\times$\,1800.0 & 3\,$\times$\,1800.0 & 6.0 &      \\
B19*  &  230.6967   & 59.35285  & 0.076 & 17.09 & P & 01/05/2013 & 0.8 & PA  & 4\,$\times$\,1200.0 & 4\,$\times$\,900.0  & 2.6 &      \\
B20!*  &  231.55424  & 3.884864  & 0.086 & 16.79 & E & 02/05/2013 & 0.9 & PA  & 3\,$\times$\,1200.0 & 3\,$\times$\,900.0  & 3.0 &      \\
B21  &  234.29971  & 41.0717   & 0.098 & 16.68 & E & 28/07/2014 & 1.6 & 136 & 3\,$\times$\,1800.0 & 3\,$\times$\,1800.0 & 6.0 &      \\
B22!  &  245.43016  & 29.725689 & 0.098 & 16.50 & E & 29/07/2014 & 1.2 & 264 & 3\,$\times$\,1800.0 & 3\,$\times$\,1800.0 & 3.6 &      \\
B23  &  327.73575  & -6.819708 & 0.059 & 16.68 & E & 26/07/2014 & 1.0 & 151 & 3\,$\times$\,1800.0 & 3\,$\times$\,1800.0 & 3.0 &      \\
\hline
\end{tabular}
\end{center}
\begin{flushleft}
{\textbf{Column description:} \textbf{ID} - MLLINER identification (sources observed with NOT are marked with '*'; sources marked with '!' are possibly Sy2 galaxies and not LINERs as explained in Section~\ref{sec_emission_measure});  \textbf{RA}, \textbf{DEC} - J2000 right ascension and declination in degrees; \textbf{z} - redshift, from SDSS public catalogues; \textbf{m6500} - AB continuum magnitude at 6500\,\AA; \textbf{morph} - visual morphological classification where E, S, and P stand for Elliptical/S0, spiral, and peculiar (see the text); \textbf{Date} - date of observation; \textbf{seeing} - average FWHM of the seeing in arc-sec; \textbf{pos. ang.} - slit position angle in degrees (PA means that the paralactic angle was used, otherwise the angle is orientated along the major axis); \textbf{texp\_b} and \textbf{texp\_r} - total exposure time in blue and red parts in seconds; \textbf{Area$_{nuc}$} - area covered with our 'nuclear' extraction, in arc-sec$^2$ (just for comparison, the SDSS spectra cover an area of 7.08\,arc-sec$^2$); \textbf{IR data} - availability of Herschel (1) and IRAS (2) data.}
\end{flushleft}
\end{table*}

\renewcommand{\thefigure}{\arabic{figure} (Cont.)}
\addtocounter{figure}{-1}

\begin{figure*}
\centering
\includegraphics[width=0.8\textwidth,angle=0]{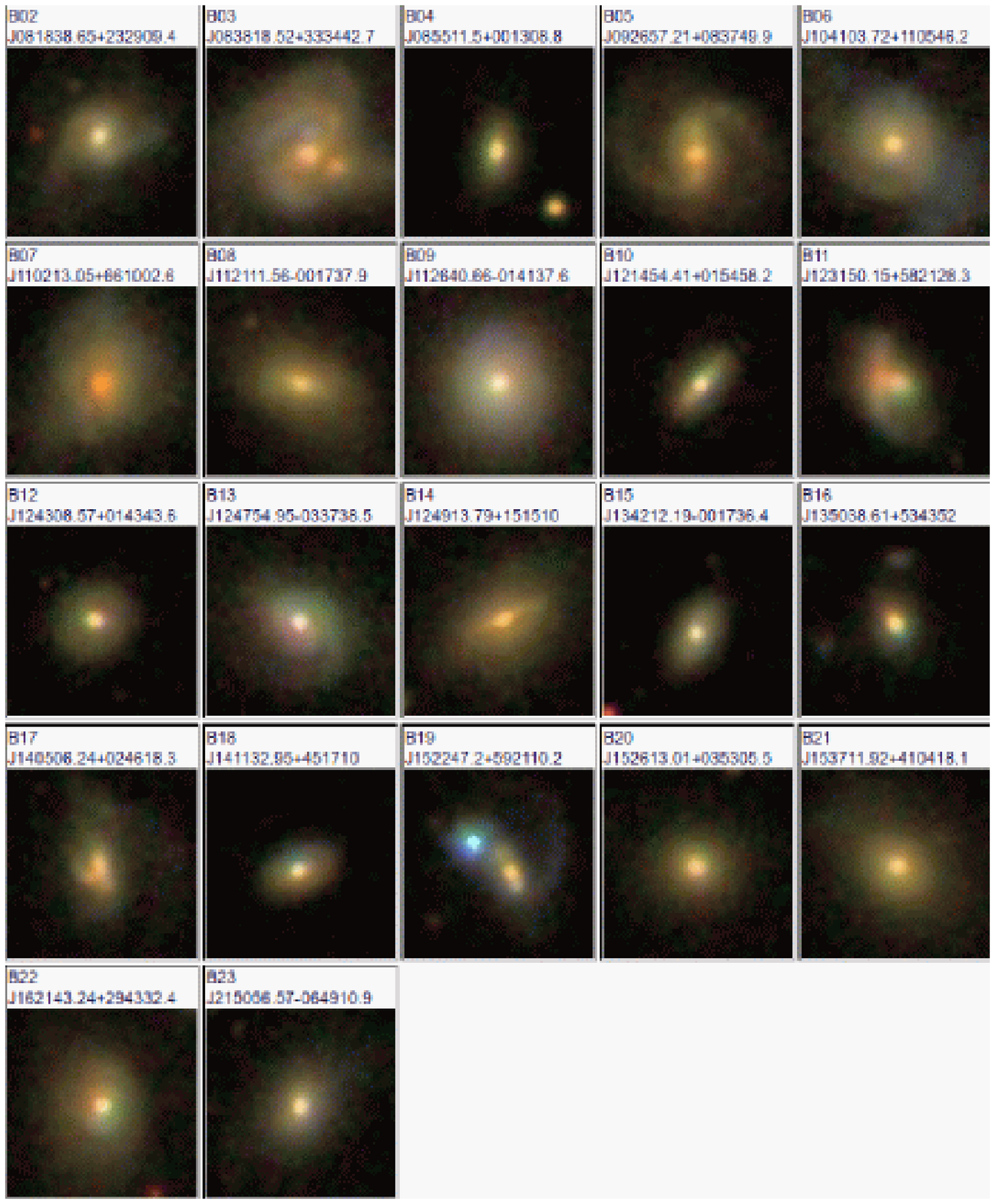}
\caption{
\label{fig_thumbnails_part2}}
\end{figure*}

\renewcommand{\thefigure}{\arabic{figure}}

\section{The Data}
\label{sec_observations}

\indent In this section we describe the optical spectroscopic observations and data reduction that we carried out for the 42 MLLINERs. We also describe the Herschel and IRAS FIR observations used in this project. To deal with catalogues we made use of TOPCAT \citep{taylor05}, while for spectral and displaying purposes we used SIPL code (Perea J.\footnote{http://www.iaa.es/$\sim$jaime/}, priv. communication). 

\subsection{Optical spectroscopy}
\label{sec_opt_spec_data}

\indent \indent The observations were carried out during six runs (PI I. M\'arquez), between October 2013 and July 2014, using the Cassegrain Twin Spectrograph (TWIN) attached to the 3.5\,m telescope at Calar Alto Observatory (CAHA\footnote{http://www.caha.es/}, Almer\'ia, Spain). Table~\ref{tab_obs} summarises the information related with observations, including the date of observation, average seeing, position angle, and exposure times. As mentioned in the previous section, we observed 42 LINERs in total. We used the T01 (red) grating during all runs, covering a spectral range of 6700\,\AA\,-\,8300\,\AA. In the blue, we used the T08 (3500\,\AA\,-\,6500\,\AA) grism during the first two runs (October and November 2013), and T13 (3700\,\AA\,-\,7000\,\AA) in the following ones. The spectral sampling for T01, T08, and T13 is 0.8, 1.1, and 2.1\,\AA/pix, respectively. The size of the slit used is 1.2\,arc-sec for seeing\,$<$\,1.5\,arc-sec, and 1.5\,arc-sec for seeing\,$\ge$\,1.5\,arc-sec. The values of seeing are listed in Table~\ref{tab_obs}.\\
\indent Additionally, ten bright MLLINERs were observed during four nights in May 2013 (PI I. M\'arquez) with the Andaluc\'ia Faint Object Spectrograph and Camera (ALFOSC) of the 2.5\,m telescope at the Nordic Optical Telescope (NOT\footnote{http://www.not.iac.es/}, Roque de los Muchachos Observatory, La Palma, Canary Islands, Spain). For six sources the S/N ratio was higher than for CAHA observations, and were consequently used throughout this work (marked with * in Table~\ref{tab_obs}). We used \#6 and \#8 gratings, covering the spectral ranges 3200\,\AA\,-\,5550\,\AA\,and 5825\,\AA\,-\,8350\,\AA, in the blue and red, with a typical spectral sampling of 1.4 and 1.3\,\AA/pixel, respectively. We used a slit of 1.3\,arc-sec in all observations. Several target exposures were taken (see Table 1) for cosmic rays and bad pixel removal. Arc lamp exposures were obtained before and after each target observation. At least two standard stars (up to four) were observed at the beginning and at the end of each night through a 10 arcsec width slit. For the final flux calibration we only considered the combination of those stars where the difference of their computed instrumental sensitivity function was lower than 10\%.  \\

\indent Spectroscopic data reduction was carried out using IRAF\footnote{http://iraf.noao.edu}. We followed the standard steps of bias subtraction, flat-field correction, wavelength calibration, atmospheric extinction correction, and flux calibration. The sky background level was determined by taking median averages over two strips on both sides of the galaxy signal, and subtracting it from the final combined galaxy spectra. As a sanity check, we compared the reduced and calibrated spectra with the SDSS ones, scaling our data to map similar areas. Good agreement was found between the two data sets, with differences lower than 20\% in both, blue and red parts of the spectra. 

\indent Morphological classification was done visually, by three independent classifiers, using the SDSS $gri$ colour images shown in Figs~\ref{fig_thumbnails_part1} and \ref{fig_thumbnails_part2}. We separated all galaxies between early-type (E: ellipticals and lenticulars), spiral (S), and peculiar (P). The type represented in Table~\ref{tab_obs} is the one assigned by the majority of the classifiers (three or two). When the classification results in three different types, we leave the source unclassified (symbol '?' in the table). P class was assigned to those sources showing a clear presence of interactions, additional structures (e.g., tails, rings), and/or irregular shapes. More discussion about galaxy morphology is given in Section~\ref{sec_discussion_general_properties}. 

\subsection{Far-infrared photometry}
\label{sec_data_ir}

\subsubsection{Herschel/PACS}
\label{sec_data_ir_herschel}

\indent We obtained FIR data for 6 objects in our sample (symbol 1 in column 13, Table~\ref{tab_obs}) using the Photo detector Array Camera and Spectrometer (PACS) on board of the Herschel Space Observatory\footnote{http://www.esa.int/herschel}. The data are part of a large LINER proposal (PI H. Netzer) out of which 6 targets were observed. We obtained 3\,$\sigma$ photometry with PACS blue and red bands, at 70 and 160\,$\mu$m, respectively. The data were processed using the standard procedure and Herschel Interactive Processing Environment (HIPE) tool \citep{ott06}. We extracted flux densities and their errors using again the standard HIPE tools. The fluxes and their errors are listed in Table~\ref{tab_obs_fir}.  

\subsubsection{IRAS}
\label{sec_data_ir_iras}

\indent We collected the available FIR flux measurements made by the Infrared Astronomical Satellite (IRAS\footnote{http://irsa.ipac.caltech.edu/IRASdocs/toc.html}). Using the catalogue of galaxies and QSOs, Point Source Catalog (PSC), and Faint Source Catalog (FSC), we found 13 sources in total with flux densities measured or estimated as upper limits in all four IRAS bands, at 12, 25, 60 and 100\,$\mu$m. All these sources are listed in the last column of table~\ref{tab_obs}, while the flux densities are provided in table~\ref{tab_obs_fir}. In the 60\,$\mu$m band, all detections have quality flag\,=\,3 (high quality), while for the 100\,$\mu$m band, 10 detections have flag\,=\,2 (moderate), and 3 sources have flag\,=\,1 (upper limit). We only used the data with flags\,=\,3 or =\,2. For sources with flag\,=\,1, we only used the information from the 60\,$\mu$m band (see Section~\ref{sec_sfr} for more information). \\
\indent Three of the IRAS observed sources (F02, F17, and B03) were also observed with Herschel/PACS. We compared the fluxes between PACS 70\,$\mu$m and IRAS 60\,$\mu$m, as well as the total SFRs measured with both surveys, and found only small differences. In the following analysis we will use the Herschel/PACS measurements for these three sources. 

\begin{table}
\small
\begin{center}
\caption{Summary of FIR observations with Herschel and IRAS.  
\label{tab_obs_fir}}
\begin{tabular}{| c | c | c | c | c | c | c | c | c |}
\hline
\textbf{ID}&\textbf{Herschel\_70}&\textbf{Herschel\_160}&\textbf{IRAS\_60}&\textbf{IRAS\_100}\\
\hline  
F02 &0.2911\,$\pm$\,0.001 &  0.3909\,$\pm$\,0.0024 & 0.2369 (3) & 1.811 (1) \\     
F06 & & & 0.2366 (3) & 0.9363 (1) \\     
F09 & & & 0.3957 (3) & 0.9608 (2) \\     
F12 & & & 0.4728 (3) & 0.9564 (2) \\     
F14 & & & 0.3074 (3) & 0.6414 (2) \\     
F16 & & & 0.5082 (3) & 0.9774 (2) \\     
F17 & 0.2894\,$\pm$\,0.0036 & 0.2431\,$\pm$\,0.0068& 0.2993 (3) & 0.4687 (1) \\     
F19 & 0.02\,$\pm$\,0.0011 & 0.0604\,$\pm$\,0.0025&  &\\     
F24 & & & 0.2772 (3) & 0.6128 (2) \\     
B02 & 0.1153\,$\pm$\,0.0037 & 0.1859\,$\pm$\,0.0069 &   &  \\     
B03 & 0.7758\,$\pm$\,0.0037& 1.2186\,$\pm$\,0.007 & 0.784 (3) & 1.356 (2) \\     
B05 & & & 0.2821 (3) & 0.8639 (2) \\     
B07 & 0.1229\,$\pm$\,0.0037& 0.3408\,$\pm$\,0.0069 &  &  \\        
B11 & & & 0.289 (3) & 0.6007 (2) \\     
B13 & & & 0.7087 (3) & 0.8789 (2) \\     
B17 & & & 0.5019 (3) & 0.9337 (2) \\     
\hline  
\end{tabular}
\end{center}
\begin{flushleft}
{\textbf{Column description:} \textbf{ID} - MLLINER identification;  \textbf{Herschel\_70} and \textbf{Herschel\_160} - FIR flux and its error in the 70\,$\mu$m and 160\,$\mu$m Herschel/PACS bands, respectively, in Jy; \textbf{IRAS\_60} and \textbf{IRAS\_100} - IRAS FIR flux and the quality flag in 60\,$\mu$m and 100\,$\mu$m bans, respectively, in Jy (quality flag is given between the brackets, where 3 means high quality, 2 moderate quality, and 1 an upper limit).}
\end{flushleft}
\end{table}

\section{Data Analysis and Measurements}
\label{sec_analysis}

\subsection[]{Dn4000 and H$\delta$ measurements}
\label{sec_dn4000_hdelta}

\indent Using the flux calibrated spectra, we measured the strength of 4000\,$\AA$ break (Dn4000) and Balmer absorption-line index H$\delta$. These two indices are known to be important for tracing the star formation histories (SFH) in galaxies \citep{kauffmann03}. Dn4000 was measured as explained in \cite{balogh99}, as the ratio between the average flux density in the continuum bands 4000\,-\,4100\,$\AA$ and 3850\,-\,3950\,$\AA$. To obtain the H$\delta$ index we used the definition of \cite{worthey97}. We first measured the average fluxes in two continuum bandpasses, blue (4041.60\,-\,4079.75\,$\AA$), and red (4128.50\,-\,4161.00\,$\AA$). The two average fluxes defined the continuum which we used to measure the H$\delta$ index, carrying out the integration within the feature in the band 4083.50\,-\,4122.25\,$\AA$ and expressing it in terms of the equivalent width. Table~\ref{tab_dn4000_hdelta} lists all these values. The main purpose of measuring Dn4000 is for using it later as a SFR indicator, while H$\delta$ was mainly used as an additional parameter of consistency of our measurements when comparing it with Dn4000. Previous works showed that the typical values for early-type galaxies are Dn4000\,$>$\,1.7 and H$\delta$\,$<$\,1 \citep{kauffmann03a}.\\
\indent We compared our Dn4000 and H$\delta$ measurements with those from the MPA-JHU DR7 database measured on SDSS spectra \citep{brinchmann04}. In general, for both parameters we found a good agreement between the two, with Spearman's rank correlation coefficients p\,=\,0.81 and 0.84, when comparing Dn4000 and H$\delta$, respectively.   

\begin{table}
\small
\begin{center}
\caption{Dn4000 and H$\delta$ measurements.  
\label{tab_dn4000_hdelta}}
\begin{tabular}{| c c c || c c c |}
\hline
\textbf{ID}&\textbf{Dn4000}&\textbf{H$\delta$}&\textbf{ID}&\textbf{Dn4000}&\textbf{H$\delta$}\\
\hline
  F01 & 1.32\,$\pm$\,0.37 & 0.90 & B03 &                   & 3.16  \\      
  F02 & 1.37\,$\pm$\,0.28 & 3.70 & B04 & 1.47\,$\pm$\,0.31 & 4.67 \\      
  F03 & 1.45\,$\pm$\,0.34 & 2.81 & B05 & 1.41\,$\pm$\,0.32 & 3.54 \\      
  F04 & 1.51\,$\pm$\,0.34 & 0.85 & B06 & 1.30\,$\pm$\,0.22 & 5.59 \\      
  F06 & 1.35\,$\pm$\,0.27 & 5.20 & B07 & 1.39\,$\pm$\,0.34 & 1.26 \\      
  F07 &                   & 7.46 & B08 & 1.43\,$\pm$\,0.24 & 2.07 \\      
  F09 & 1.44\,$\pm$\,0.35 & 2.97 & B09 &                   &       \\      
  F12 & 1.49\,$\pm$\,0.40 & 5.44 & B10 & 1.28\,$\pm$\,0.23 & 4.10 \\      
  F13 &                   & 5.59 & B11 & 1.26\,$\pm$\,0.15 & 3.54 \\      
  F14 & 1.37\,$\pm$\,0.30 & 4.16 & B12 & 1.33\,$\pm$\,0.28 & 0.83 \\      
  F15 & 1.33\,$\pm$\,0.27 & 4.93 & B13 & 1.01\,$\pm$\,0.12 & 4.52 \\      
  F16 & 1.16\,$\pm$\,0.25 & 0.66 & B14 & 1.52\,$\pm$\,0.26 & 1.28 \\      
  F17 & 1.26\,$\pm$\,0.29 & 7.80 & B15 & 1.27\,$\pm$\,0.25 & 7.04 \\      
  F19 & 1.39\,$\pm$\,0.33 & 0.24 & B16 & 1.36\,$\pm$\,0.30 & 6.23 \\      
  F20 & 1.16\,$\pm$\,0.20 & 2.25 & B17 & 1.21\,$\pm$\,0.27 & 6.90 \\      
  F21 & 1.37\,$\pm$\,0.30 & 5.98 & B18 & 1.34\,$\pm$\,0.26 & 6.00 \\      
  F22 & 1.17\,$\pm$\,0.18 & 4.94 & B19 & 1.32\,$\pm$\,0.27 & 7.21 \\      
  F23 & 1.30\,$\pm$\,0.27 & 6.48 & B20! & 1.42\,$\pm$\,0.27 & 3.20 \\      
  F24 & 1.32\,$\pm$\,0.56 & 8.26 & B21 & 1.45\,$\pm$\,0.27 & 0.24 \\      
  B01 & 1.23\,$\pm$\,0.29 & 5.49 & B22! & 1.22\,$\pm$\,0.23 & 5.79  \\      
  B02 & 1.16\,$\pm$\,0.22 & 6.35 & B23 & 1.46\,$\pm$\,0.27 & 5.96 \\      
\hline
\end{tabular}
\end{center}
\begin{flushleft}
{! possibly Sy2 galaxies (see Section~\ref{sec_emission_measure})}
\end{flushleft}
\end{table}

\begin{figure}
\centering
\includegraphics[width=7.5cm,angle=0]{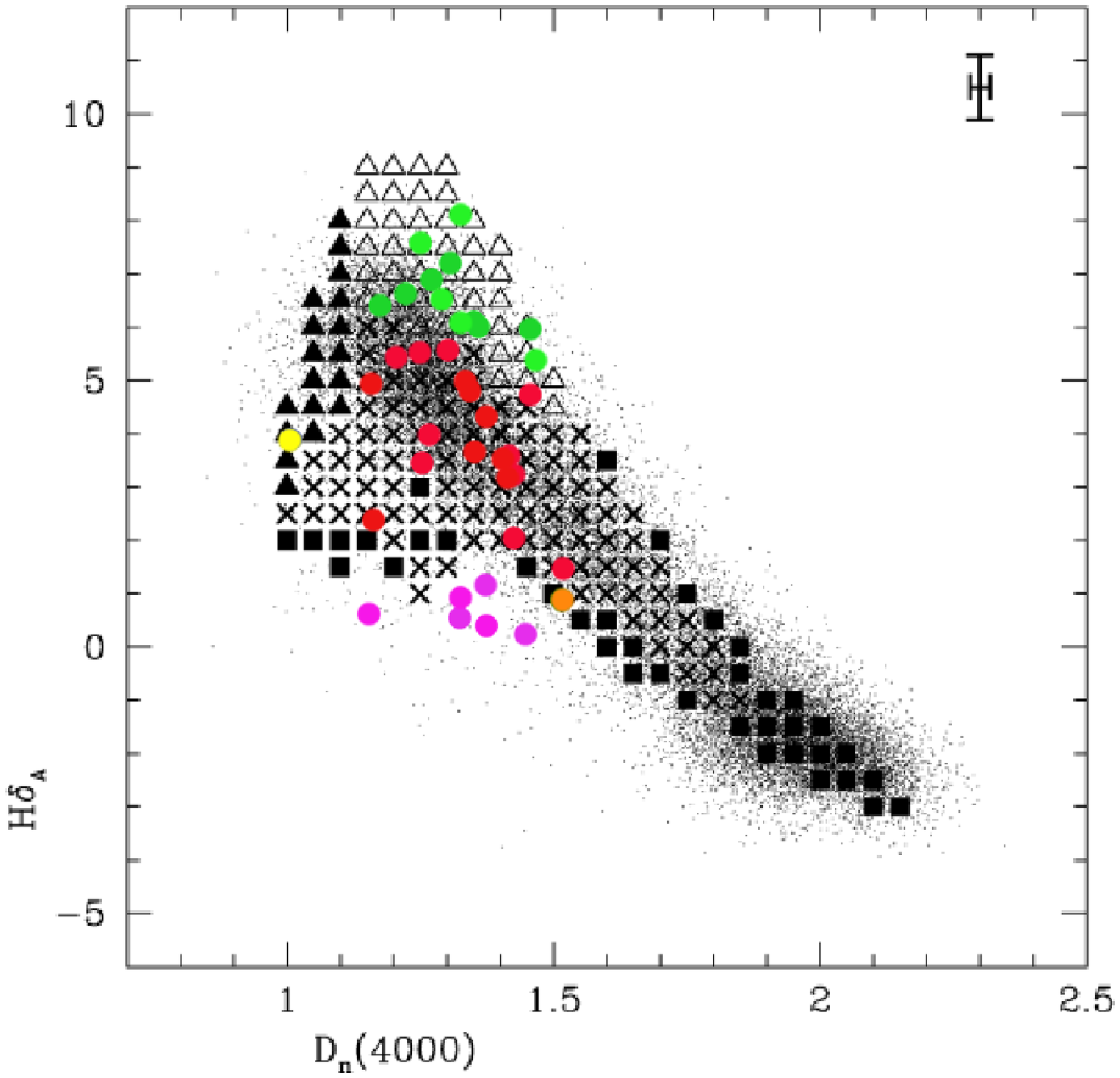}
\caption[ ]{Figure 6 of \cite{kauffmann03}, showing the relation between the Dn4000 and H$\delta$ indices for their sample of SDSS sources, with our sample over-plotted (colour filled circles). Open and solid triangles show the indices with high confidence that the galaxy has experienced a burst over the past 2\,Gyr (our green and yellow circles, respectively). Open (solid) triangles indicate regions where 95\% of the model galaxies have F$_{burst}$\,$>$\,0.05 and the burst occurred more than (less than) 0.1\,Gyr ago. Solid squares indicate regions where 95\% of the model galaxies have F$_{burst}$\,$=$\,0 (our orange circle). Regions marked with crosses contain a mix of bursty and continuous star formation models (our red circles). Violet circles lie outside the range covered by models.}
\label{fig_Dn4000_Hdelta}
\end{figure}

\indent Figure~\ref{fig_Dn4000_Hdelta} shows the relation between the Dn4000 and H$\delta$ indices obtained by \cite{kauffmann03} for the SDSS DR4 sample (see their figure 6). They used a library of 32,000 different SFH, where for each SFH they have a corresponding Dn4000 and H$\delta$ indices, as well as the fraction of the total stellar mass of the galaxy formed in the bursts over the past 2\,Gyr (F$_{burst}$). In their figure the bins are coded according to the fraction of model SFHs with F$_{burst}$ in a given range (see the caption of their Fig.~\ref{fig_Dn4000_Hdelta}). We used this figure and overplotted our Dn4000 and H$\delta$ measurements (coloured filled circles). In general our measurements are consistent with the models by \cite{kauffmann03}. More details about star formation histories are given in Section~\ref{sec_discussion_general_properties}.

\subsection[]{STARLIGHT spectral fittings of the nuclear regions}
\label{sec_starlight_fits}

\indent We extracted what we call the nuclear spectra by selecting a central region equal to 2.5 times the FWHM of the seeing. In the case of CAHA the extraction size is 5\,-8\,pixels (depending on the used slit), while in the case of NOT data the central 10 pixels were extracted. The total area covered by the nuclear extraction is given in the last column of Table~\ref{tab_obs} for all MLLINERs. \\
\indent Modelling of the nuclear stellar spectra of our sources was performed with the STARLIGHT\footnote{http://astro.ufsc.br/starlight} V.04 synthesis code \citep{cid05,cid09}. All spectra were previously corrected for galactic extinction, K-corrected, and moved to rest-frame. To correct for the galactic extinction we used the \texttt{pystarlight\footnote{https://pypi.python.org/pypi/PySTARLIGHT}} library within the astrophysics Python package\footnote{https://pythonhosted.org/Astropysics/} and \cite{schlegel98} maps of dust IR emission. Our fittings are based on the templates from \cite{bruzual03}, with solar metallicity and 25 different stellar ages, from 0.001\,$\times$\,10$^9$ to 18\,$\times$\,10$^9$. Considering that we are dealing with nuclear spectra of large galaxies, this approximation should be fine for our sources \citep{ho03, ho08}. We masked in all spectra the emission line regions, areas with atmospheric absorptions, and regions with bad pixels. To measure the signal-to-noise (S/N) we checked visually all spectra to select the continuum region free of bad pixels, using always the blue range and a width of at least 80\AA. In most cases we selected the region around $\sim$\,4600\AA\, or $\sim$\,5600\AA. S/N measurements are listed in Table~\ref{tab_starlight_fits}. \\
\indent In this work we used the \cite{cardelli89} extinction law. This law was widely used in different surveys for fitting the host-dominated sources \citep{stasinska06, cid11, gonzalezdelgado16}. We also tested the \cite{calzetti07} law, and made a comparison between the two. We found differences to be lower than 20\% therefore we
only show results that were obtained with the Cardelli et al (1989) extinction law. \\

\indent The basic information obtained from the best fit stellar population models is summarised in Table~\ref{tab_starlight_fits} for all MLLINERs. The \textit{adev} parameter gives the goodness of the fit, and presents the mean deviation over the all fitted pixels (in percentage); adev\,$<$\,6 and $<$\,10 stand for 'very good' and 'good' fits, respectively \citep{cid05,cid09}. We obtained very good fit in $\sim$\,80\% of the cases. The measured S/N ratio and extinction A$_V$ are given in columns 3 and 4, respectively. The best fit parameters (M\_cor\_tot and M\_ini\_tot) were used to measure two types of stellar masses, following \cite{cid05,cid09}: \\
the present mass in stars,
\begin{center}
\label{eq_mstar_starlight}
{M$_*$\,=\,M\_cor\_tot\,$\times$\,10$^{-17}\,$\,$\times$\,4$\pi$d$^2$\,$\times$\,(3.826\,$\times$\,10$^{33}$)$^{-1}$},
\end{center}
and the initial mass, that has been processed into stars throughout the galaxy life:
\begin{center}
\label{eq_mini_starlight}
{M$_*^{ini}$\,=\,M\_ini\_tot\,$\times$\,10$^{-17}\,$\,$\times$\,4$\pi$d$^2$\,$\times$\,(3.826\,$\times$\,10$^{33}$)$^{-1}$}.
\end{center}
The results regarding the best stellar population mixture are summarised in columns 7\,-\,9. They are represented through the light-fraction population vector (corresponds to the same wavelength selected for measuring S/N, see above) for three stellar ages: young (with age\,[yr]\,$\le$\,10$^8$), intermediate (10$^8$\,$<$\,age\,[yr]\,$\le$\,10$^9$), and old (age\,[yr]\,$>$\,10$^9$). We discuss stellar populations in more detail in Section~\ref{sec_discussion_general_properties}. Finally, we calculated the light-weighted mean ages of our MLLINERs, using as a reference \cite{cid13}:
\begin{center}
\label{eq_age_starlight}
{$<$logt$>$\,=\,$\sum\limits_{t,Z}$\,$x_{t,Z}$logt},
\end{center}
where $x_{t,Z}$ is a fraction of light at stellar age t in our best-fit model and metallicity Z (Z$_{\odot}$ in our case). The example with the best-fit models (red lines) and original spectra (blue lines) are shown in Appendix A (Figs.~\ref{fig_spectra_1}).

\begin{table*}[h]
\scriptsize
\begin{center}
\caption{The best stellar population mixture found by STARLIGHT.  
\label{tab_starlight_fits}}
\begin{tabular}{| c | c | c | c | c | c | c | c | c | c || c | c | c | c | c | c | c | c | c | c |}
\hline
\textbf{ID}&\textbf{adev}&\textbf{S/N}&\textbf{A$_V$}&\textbf{M$_*$}&\textbf{M$_*^{ini}$}&\textbf{stpop1}&\textbf{stpop2}&\textbf{stpop3}&\textbf{$<$logt$>$}&\textbf{ID}&\textbf{adev}&\textbf{S/N}&\textbf{A$_V$}&\textbf{M$_*$}&\textbf{M$_*^{ini}$}&\textbf{stpop1}&\textbf{stpop2}&\textbf{stpop3}&\textbf{$<$logt$>$}\\
\hline
F01 & 9.62& 15.16 & 0.138 & 0.76 & 1.46 & 4.83   & 22.08  & 71.42 & 9.03 &B03 & 2.64& 44.51 & 0.877 & 2.89 & 5.79  & 15.81  & 14.67  & 67.75 & 9.19\\ 
F02 & 4.70& 21.69 & 1.667 & 2.50 & 4.70  & 15.17  & 18.92  & 66.47 & 8.97&B04 & 3.54& 44.18 & 0.584 & 1.93 & 3.41  & 0.0    & 0.0    & 93.51 & 8.61\\      
F03 & 4.59& 25.82 & 0.584 & 1.67 & 3.12  & 6.38   & 0.0    & 95.8  & 9.53&B05 & 4.39& 30.96 & 0.753 & 3.77 & 7.20  & 1.72   & 3.4    & 88.68 & 8.77\\      
F04 & 4.27& 26.74 & 0.827 & 2.17 & 4.07  & 6.89   & 0.0    & 97.81 &9.82 &B06 & 2.68& 45.25 & 1.209 & 3.35 & 6.65  & 0.0    & 55.28  & 42.89 & 8.95\\       
F06 & 3.74& 36.17 & 1.306 & 1.56 & 2.64  & 5.64   & 10.51  & 79.16 & 8.51&B07 & 4.26& 37.31 & 2.209 & 6.81 & 13.63 & 8.55   & 20.57  & 70.08 & 9.46\\      
F07 & 3.37& 27.42 & 1.162 & 0.61 & 1.21  & 41.52  & 42.3   & 18.26 & 8.17&B08 & 3.43& 31.36 & 0.388 & 4.64 & 9.29  & 0.0    & 53.01  & 49.76 & 9.45\\       
F09 & 8.49& 18.36 & 1.423 & 1.93 & 3.44  & 0.0    & 10.96  & 90.84 & 9.36&B09 & 2.60& 44.41 & 0.234 & 0.96 & 1.77  & 9.18   & 0.0    & 87.64 & 8.97\\      
F12 & 5.21& 31.79 & 0.459 & 0.86 & 1.56  & 0.0    & 43.09  & 54.45 & 9.06&B10 & 3.23& 47.73 & 0.787 & 1.19 & 2.02  & 5.21   & 12.4   & 78.21 & 8.59\\      
F13 & 12.60& 23.90 & 1.515 & 0.58 & 0.99  & 0.0    & 4.24   & 90.83& 8.71&B11 & 3.43& 43.65 & 0.517 & 1.68 & 2.95  & 0.0    & 14.94  & 75.46 & 8.26\\     
F14 & 7.53& 20.31 & 0.8  & 1.10 & 2.04  & 0.0    & 89.94  & 8.7  & 8.94&B12 & 5.66& 27.73 & 0.706 & 0.57 & 0.97  &  0.0   &  24.78 &  70.88 &8.7\\     
F15 & 5.09& 28.67 & 0.991 & 1.05 & 1.87  & 0.0    & 96.77  & 3.39  & 8.99&B13 & 2.17& 50.33 & 0.673 & 1.44 & 2.47  & 7.65   & 65.0   & 26.36 & 8.18\\       
F16 & 7.51& 17.07 & 1.239 & 0.88 & 1.70  & 16.44  & 12.77  & 69.52 & 8.69&B14 & 3.39& 41.95 & 0.741 & 4.75 & 9.42  & 0.0    & 25.8   & 74.38 & 9.5\\       
F17 & 3.55& 43.21 & 1.369 & 1.02 & 1.71  & 7.1    & 32.41  & 58.34 & 8.58&B15 & 2.89& 43.62 & 0.701 & 1.03 & 1.85  & 0.0    & 92.53  & 9.72 &  9.0\\      
F19 & 7.39& 18.17 & 0.542 & 1.54 & 2.98  & 0.0    & 17.79  & 83.21 &9.68 &B16 & 3.49& 39.68 & 0.327 & 1.33 & 2.31  & 0.0    & 0.0    & 98.71 & 9.07\\      
F20 & 6.51& 22.11 & 1.154 & 0.90 & 1.69  & 6.37   & 47.45  & 46.19 & 8.68&B17 & 3.95& 45.53 & 0.997 & 1.47 & 2.83  & 0.0    & 57.78  & 42.84 & 9.07\\      
F21 & 7.49& 22.74 & 0.738 & 1.52 & 2.84  & 0.0    & 49.74  & 50.4  & 9.27&B18 & 4.38& 32.81 & 0.916 & 1.44 & 2.78  & 0.0    & 78.52  & 18.81 & 8.88\\      
F22 & 4.30& 30.01 & 0.698 & 1.45 & 2.81  & 0.0    & 49.72  & 48.15 & 8.75&B19 & 4.48& 33.10 & 1.914& 5.61 & 11.06 & 0.0    & 77.21  & 21.86 & 8.76\\       
F23 & 5.25& 34.38 & 0.702 & 1.12 & 2.13  & 0.0    & 86.75  & 9.27  & 8.58&B20! & 3.20& 31.32 & 0.312 & 5.41 & 10.82 & 0.0    & 51.55  & 53.01 & 9.9\\       
F24 & 11.55& 10.97&  1.68& 1.78 & 3.52  & 12.06  & 57.86  & 31.98 & 8.98&B21 & 4.41& 51.15 & 0.515 & 8.32 & 16.65 & 0.0    & 38.45  & 56.92 & 9.28\\      
B01 & 5.67& 23.82 &  0.681& 0.69 & 1.36  & 2.67   & 74.6   & 23.11 & 8.72&B22! & 3.71& 42.86 & 0.462 & 4.85 & 9.52  & 0.0    & 57.46  & 36.84 & 8.69\\      
B02 & 2.87& 41.10 &  1.105& 3.89 & 7.65  & 9.67   & 55.18  & 32.08 & 8.45&B23 & 4.26& 35.50 & & 1.73 & 3.31  & 0.0    & 63.79  & 34.14 & 9.11\\       
\hline
\end{tabular}
\end{center}
\begin{flushleft}
{\textbf{Column description:} \textbf{ID} - MLLINER identification ('!' - possibly Sy2 galaxies, see Section~\ref{sec_emission_measure}); \textbf{adev} - goodness of the fit (see the text);  \textbf{S/N} - measured signal-to-noise ratio (see Section~\ref{sec_starlight_fits}); \textbf{A$_V$} - extinction in V band; \textbf{M$_*$} and \textbf{M$_*^{ini}$} - current and initial mass in stars, respectively, in 10$^{10}$\,[M$_{\odot}$]; \textbf{stpop1} - fraction of young stars with age\,[yr]\,$\le$\,10$^8$ in \%; \textbf{stpop2} - fraction of intermediate stars with 10$^8$\,$<$\,age\,[yr]\,$\le$\,10$^9$ in \%; \textbf{stpop3} - fraction of old stars with age\,[yr]\,$>$\,10$^9$ in \%; \textbf{$<$logt$>$} - mean age.}
\end{flushleft}
\end{table*}

\subsection[]{Emission line measurements and classification}
\label{sec_emission_measure}

\indent We obtained the nuclear emission line spectra by subtracting from the original ones the best-fit stellar models. Fig~\ref{fig_spectra_1}\,-\,\ref{fig_spectra_5} show the final emission spectra (black solid lines) of all MLLINERs. We used these spectra to measure the properties of the emission lines. Using \texttt{splot} IRAF task, we measured the flux of the strong emission lines by fitting a single gaussian function. Table~\ref{tab_emission_lines} summarises the resulting fluxes for [OII]$\lambda$3727, H$\beta$, [OIII]$\lambda$4959, [OIII]$\lambda$5007, [OI]$\lambda$6300, [NII]$\lambda$6548, [NII]$\lambda$6584, [SII]$\lambda$6718, and [SII]$\lambda$6731 lines relative to the H$\alpha$ line. The errors were measured taking into account the rms of the continuum. We also measured the equivalent width (EW) of H$\alpha$ line by fitting again the line with a single Gaussian function and using the original spectra. \\
\indent All emission lines were corrected for extinction using the ratio of HI Balmer lines, and using H$\alpha$/H$\beta$\,=\,3.1 as the theoretical value for AGN \citep{osterbrock05}. 
Table~\ref{tab_extcorr_lines} summarises the corrected flux ratios, again relative to the H$\alpha$ line. We also summarise the measured values of extinction in the V band (A$_V$). We compared these values with those obtained from the STARLIGHT best-fit models (see Section~\ref{sec_starlight_fits} and Table~\ref{tab_starlight_fits}), finding in general important discrepancies between the two measurements, where the emission-line technique gives in general higher values of A$_V$, as has been seen previously \citep{calzetti94}.  \\

\begin{table*}
\small
\begin{center}
\caption{Properties of strong emission lines.  
\label{tab_emission_lines}}
\begin{tabular}{| c | c | c | c | c | c | c | c | c | c | c | c |}
\hline
\textbf{ID}&\textbf{[OII]$\lambda$3272}&\textbf{H$\beta$}&\textbf{[OIII]$\lambda$4959}&\textbf{[OIII]$\lambda$5007}&\textbf{[OI]$\lambda$6300}&\textbf{[NII]$\lambda$6548}&\textbf{[NII]$\lambda$6584}&\textbf{[SII]$\lambda$6716}&\textbf{[SII]$\lambda$6731}&\textbf{F$_{H\alpha}$\,$\times$\,10$^{-16}$}&\textbf{EW$_{H\alpha}$}\\
\hline
F01 & 0.61\,$\pm$\,0.19 & 0.24\,$\pm$\,0.03 & 0.17\,$\pm$\,0.04 & 0.42\,$\pm$\,0.04 &              & 0.35 \,$\pm$\,0.06 & 1.29\,$\pm$\,0.09 & 0.25\,$\pm$\,0.08 & 0.17\,$\pm$\,0.08 & 14.41\,$\pm$\,0.77 & 17.1  \\
F02 &              & 0.11\,$\pm$\,0.03 &              & 0.14\,$\pm$\,0.03 &              & 0.27 \,$\pm$\,0.06 & 0.93\,$\pm$\,0.08 & 0.24\,$\pm$\,0.05 & 0.2 \,$\pm$\,0.05 & 14.31\,$\pm$\,0.88 & 14.46 \\
F03 &              & 0.23\,$\pm$\,0.05 & 0.14\,$\pm$\,0.06 & 0.41\,$\pm$\,0.07 &              & 0.38 \,$\pm$\,0.09 & 0.84\,$\pm$\,0.11 & 0.54\,$\pm$\,0.09 & 0.37\,$\pm$\,0.08 & 14.27\,$\pm$\,1.2  & 11.44 \\
F04 &              & 0.17\,$\pm$\,0.04 & 0.07\,$\pm$\,0.02 & 0.31\,$\pm$\,0.05 &              & 0.41 \,$\pm$\,0.09 & 1.12\,$\pm$\,0.12 & 0.53\,$\pm$\,0.11 & 0.39\,$\pm$\,0.1  & 10.44\,$\pm$\,0.86 & 6.79  \\
F06 &              & 0.17\,$\pm$\,0.03 & 0.04\,$\pm$\,0.03 & 0.29\,$\pm$\,0.03 &              & 0.3  \,$\pm$\,0.05 & 0.93\,$\pm$\,0.07 & 0.24\,$\pm$\,0.04 & 0.19\,$\pm$\,0.04 & 20.82\,$\pm$\,1.0  & 10.37 \\
F07 &              & 0.12\,$\pm$\,0.06 & 0.24\,$\pm$\,0.08 & 0.66\,$\pm$\,0.13 &              & 0.42 \,$\pm$\,0.19 & 1.66\,$\pm$\,0.34 & 0.33\,$\pm$\,0.08 & 0.23\,$\pm$\,0.06 & 13.76\,$\pm$\,2.42 & 7.99  \\
F09 & 0.44\,$\pm$\,0.2  & 0.19\,$\pm$\,0.1  &              & 0.2 \,$\pm$\,0.08 &              & 0.38 \,$\pm$\,0.23 & 1.67\,$\pm$\,0.41 & 0.45\,$\pm$\,0.24 & 0.39\,$\pm$\,0.24 & 5.3  \,$\pm$\,1.12 & 1.64  \\
F12 & 0.39\,$\pm$\,0.24 & 0.12\,$\pm$\,0.07 &              & 0.13\,$\pm$\,0.09 &              & 0.23 \,$\pm$\,0.49 & 1.03\,$\pm$\,0.69 & 0.27\,$\pm$\,0.14 & 0.23\,$\pm$\,0.13 & 10.83\,$\pm$\,5.17 & 8.96  \\
F13 & 0.3 \,$\pm$\,0.1  & 0.14\,$\pm$\,0.04 & 0.1 \,$\pm$\,0.05 & 0.26\,$\pm$\,0.07 &              & 0.39 \,$\pm$\,0.08 & 1.12\,$\pm$\,0.11 & 0.39\,$\pm$\,0.1  & 0.33\,$\pm$\,0.1  & 8.74 \,$\pm$\,0.65 & 8.9   \\
F14 & 0.89\,$\pm$\,0.14 & 0.27\,$\pm$\,0.06 & 0.1 \,$\pm$\,""   & 0.35\,$\pm$\,0.07 &              & 0.68 \,$\pm$\,0.12 & 1.71\,$\pm$\,0.19 & 0.57\,$\pm$\,0.11 & 0.47\,$\pm$\,0.11 & 16.84\,$\pm$\,1.65 & 10.1  \\
F15 & 0.49\,$\pm$\,0.06 & 0.13\,$\pm$\,0.03 & 0.08\,$\pm$\,0.05 & 0.27\,$\pm$\,0.05 &              & 0.52 \,$\pm$\,0.09 & 1.1 \,$\pm$\,0.12 & 0.4 \,$\pm$\,0.07 & 0.27\,$\pm$\,0.07 & 18.4 \,$\pm$\,1.53 & 6.83  \\
F16 & 0.27\,$\pm$\,0.02 & 0.14\,$\pm$\,0.01 & 0.06\,$\pm$\,0.01 & 0.2 \,$\pm$\,0.01 &              & 0.34 \,$\pm$\,0.02 & 0.86\,$\pm$\,0.03 & 0.18\,$\pm$\,0.02 & 0.16\,$\pm$\,0.02 & 34.68\,$\pm$\,0.7  & 44.99 \\
F17 &              & 0.17\,$\pm$\,0.03 & 0.07\,$\pm$\,0.02 & 0.36\,$\pm$\,0.04 &              & 0.43 \,$\pm$\,0.06 & 1.05\,$\pm$\,0.09 & 0.22\,$\pm$\,0.04 & 0.18\,$\pm$\,0.04 & 22.58\,$\pm$\,1.33 & 16.12 \\
F19 & 0.88\,$\pm$\,0.07 & 0.21\,$\pm$\,0.02 & 0.22\,$\pm$\,0.03 & 0.54\,$\pm$\,0.04 &              & 0.26 \,$\pm$\,0.05 & 0.65\,$\pm$\,0.06 & 0.46\,$\pm$\,0.06 & 0.39\,$\pm$\,0.06 & 21.56\,$\pm$\,1.13 & 17.21 \\
F20 & 0.4 \,$\pm$\,0.03 & 0.2 \,$\pm$\,0.02 &              & 0.15\,$\pm$\,0.02 &              & 0.34 \,$\pm$\,0.03 & 0.83\,$\pm$\,0.04 & 0.27\,$\pm$\,0.02 & 0.19\,$\pm$\,0.02 & 26.29\,$\pm$\,0.81 & 36.03 \\
F21 & 0.61\,$\pm$\,0.08 & 0.13\,$\pm$\,0.05 & 0.41\,$\pm$\,0.05 & 0.41\,$\pm$\,0.05 &              & 0.56 \,$\pm$\,0.08 & 1.86\,$\pm$\,0.15 & 0.48\,$\pm$\,0.09 & 0.39\,$\pm$\,0.09 & 24.27\,$\pm$\,1.75 & 13.47 \\
F22 & 0.42\,$\pm$\,0.06 & 0.26\,$\pm$\,0.02 & 0.08\,$\pm$\,0.03 & 0.19\,$\pm$\,0.03 &              & 0.25 \,$\pm$\,0.04 & 1.13\,$\pm$\,0.05 & 0.26\,$\pm$\,0.05 & 0.2 \,$\pm$\,0.05 & 23.82\,$\pm$\,0.81 & 13.58 \\
F23 & 0.92\,$\pm$\,0.21 & 0.32\,$\pm$\,0.34 & 0.2 \,$\pm$\,0.1  & 0.62\,$\pm$\,0.13 &              & 0.47 \,$\pm$\,0.13 & 1.31\,$\pm$\,0.2  &              &              & 6.06 \,$\pm$\,0.72 & 1.43  \\
F24 &              &              &              & 0.48\,$\pm$\,0.21 &              & 0.37 \,$\pm$\,0.17 & 0.7 \,$\pm$\,0.2  & 0.66\,$\pm$\,0.25 & 0.43\,$\pm$\,0.24 & 3.59 \,$\pm$\,0.58 & 6.11  \\
B01 & 0.65\,$\pm$\,0.07 & 0.19\,$\pm$\,0.02 & 0.07\,$\pm$\,0.02 & 0.22\,$\pm$\,0.02 &              & 0.43 \,$\pm$\,0.04 & 1.07\,$\pm$\,0.06 & 0.42\,$\pm$\,0.03 & 0.35\,$\pm$\,0.02 & 61.67\,$\pm$\,2.46 & 24.33 \\
B02 &              & 0.22\,$\pm$\,0.04 & 0.07\,$\pm$\,0.03 & 0.3 \,$\pm$\,0.07 &              & 0.38 \,$\pm$\,0.06 & 0.97\,$\pm$\,0.08 & 0.29\,$\pm$\,0.08 & 0.27\,$\pm$\,0.08 & 13.63\,$\pm$\,0.82 & 4.87  \\
B03 &              & 0.11\,$\pm$\,0.01 & 0.05\,$\pm$\,0.01 & 0.15\,$\pm$\,0.01 & 0.11\,$\pm$\,0.01 & 0.3  \,$\pm$\,0.03 & 0.93\,$\pm$\,0.04 & 0.23\,$\pm$\,0.02 & 0.22\,$\pm$\,0.02 & 114.2\,$\pm$\,2.94 & 40.9  \\
B04 &              & 0.22\,$\pm$\,0.17 & 0.21\,$\pm$\,0.09 & 0.55\,$\pm$\,0.14 &              & 0.88 \,$\pm$\,0.26 & 1.75\,$\pm$\,0.4  &              &              & 11.47\,$\pm$\,2.28 & 1.48  \\
B05 &              & 0.17\,$\pm$\,0.04 & 0.05\,$\pm$\,0.03 & 0.14\,$\pm$\,0.05 &              & 0.43 \,$\pm$\,0.07 & 1.27\,$\pm$\,0.1  & 0.19\,$\pm$\,0.05 & 0.18\,$\pm$\,0.05 & 18.95\,$\pm$\,1.23 & 7.01  \\
B06 & 0.95\,$\pm$\,0.08 & 0.25\,$\pm$\,0.05 & 0.07\,$\pm$\,0.05 & 0.33\,$\pm$\,0.04 & 0.27\,$\pm$\,0.05 & 0.21 \,$\pm$\,0.05 & 1.23\,$\pm$\,0.08 & 0.54\,$\pm$\,0.07 & 0.41\,$\pm$\,0.07 & 28.11\,$\pm$\,1.49 & 4.84  \\
B07 &              & 0.1 \,$\pm$\,0.03 &              & 0.14\,$\pm$\,0.04 &              & 0.31 \,$\pm$\,0.07 & 0.93\,$\pm$\,0.09 & 0.43\,$\pm$\,0.1  & 0.33\,$\pm$\,0.09 & 12.43\,$\pm$\,0.78 & 6.94  \\
B08 & 1.27\,$\pm$\,0.1  & 0.24\,$\pm$\,0.06 & 0.13\,$\pm$\,0.04 & 0.4 \,$\pm$\,0.04 & 0.32\,$\pm$\,0.2  & 0.5  \,$\pm$\,0.08 & 1.56\,$\pm$\,0.13 & 0.75\,$\pm$\,0.11 & 0.61\,$\pm$\,0.1  & 21.48\,$\pm$\,1.53 & 9.07  \\
B09 &              & 0.37\,$\pm$\,0.05 & 0.47\,$\pm$\,0.06 & 1.57\,$\pm$\,0.13 & 0.36\,$\pm$\,0.05 & 0.32 \,$\pm$\,0.08 & 0.7 \,$\pm$\,0.09 & 0.74\,$\pm$\,0.08 & 0.59\,$\pm$\,0.07 & 55.34\,$\pm$\,4.09 & 8.37  \\
B10 &              & 0.19\,$\pm$\,0.08 &              & 0.16\,$\pm$\,0.06 &              & 0.38 \,$\pm$\,0.08 & 1.08\,$\pm$\,0.11 & 0.32\,$\pm$\,0.09 & 0.24\,$\pm$\,0.09 & 19.84\,$\pm$\,1.42 & 7.18  \\
B11 & 0.63\,$\pm$\,0.04 & 0.22\,$\pm$\,0.02 & 0.05\,$\pm$\,0.02 & 0.2 \,$\pm$\,0.02 &              & 0.44 \,$\pm$\,0.05 & 0.97\,$\pm$\,0.06 & 0.46\,$\pm$\,0.03 & 0.35\,$\pm$\,0.03 & 43.09\,$\pm$\,1.81 & 15.51 \\
B12 & 0.64\,$\pm$\,0.11 & 0.24\,$\pm$\,0.04 & 0.25\,$\pm$\,0.04 & 1.14\,$\pm$\,0.08 &              & 0.4  \,$\pm$\,0.06 & 0.92\,$\pm$\,0.07 & 0.51\,$\pm$\,0.11 & 0.39\,$\pm$\,0.11 & 17.07\,$\pm$\,0.93 & 10.37 \\
B13 & 0.4 \,$\pm$\,0.02 & 0.19\,$\pm$\,0.01 & 0.05\,$\pm$\,0.0  & 0.18\,$\pm$\,0.01 &              & 0.29 \,$\pm$\,0.02 & 0.68\,$\pm$\,0.02 & 0.19\,$\pm$\,0.02 & 0.17\,$\pm$\,0.02 & 223.1\,$\pm$\,3.95 & 47.02 \\
B14 & 0.16\,$\pm$\,0.04 & 0.26\,$\pm$\,0.06 & 0.05\,$\pm$\,0.01 & 0.28\,$\pm$\,0.04 &              & 1.07 \,$\pm$\,0.17 & 0.59\,$\pm$\,0.14 &              &              & 44.97\,$\pm$\,5.35 & 15.14 \\
B15 & 0.81\,$\pm$\,0.07 & 0.21\,$\pm$\,0.04 & 0.05\,$\pm$\,0.02 & 0.29\,$\pm$\,0.04 & 0.24\,$\pm$\,0.08 & 0.53 \,$\pm$\,0.08 & 1.63\,$\pm$\,0.13 & 0.39\,$\pm$\,0.07 & 0.32\,$\pm$\,0.07 & 29.15\,$\pm$\,1.99 & 10.58 \\
B16 &              & 0.19\,$\pm$\,0.05 & 0.22\,$\pm$\,0.06 & 0.53\,$\pm$\,0.12 &              & 0.52 \,$\pm$\,0.11 & 1.4 \,$\pm$\,0.16 & 0.61\,$\pm$\,0.13 & 0.5 \,$\pm$\,0.13 & 12.39\,$\pm$\,1.16 & 3.93  \\
B17 & 0.21\,$\pm$\,0.07 & 0.1 \,$\pm$\,0.02 & 0.1 \,$\pm$\,0.05 & 0.22\,$\pm$\,0.03 &              & 0.31 \,$\pm$\,0.07 & 1.05\,$\pm$\,0.09 & 0.41\,$\pm$\,0.12 & 0.21\,$\pm$\,0.11 & 23.88\,$\pm$\,1.49 & 8.93  \\
B18 & 0.36\,$\pm$\,0.08 & 0.25\,$\pm$\,0.07 &              & 0.19\,$\pm$\,0.07 &              & 0.5  \,$\pm$\,0.1  & 1.19\,$\pm$\,0.14 & 0.12\,$\pm$\,0.05 & 0.18\,$\pm$\,0.05 & 18.32\,$\pm$\,1.62 & 6.21  \\
B19 & 0.26\,$\pm$\,0.05 & 0.18\,$\pm$\,0.04 & 0.04\,$\pm$\,0.01 & 0.33\,$\pm$\,0.05 & 0.19\,$\pm$\,0.07 & 0.46 \,$\pm$\,0.06 & 1.36\,$\pm$\,0.09 & 0.39\,$\pm$\,0.08 & 0.32\,$\pm$\,0.08 & 15.75\,$\pm$\,0.86 & 4.32  \\
B20 & 0.99\,$\pm$\,0.06 & 0.25\,$\pm$\,0.03 & 0.57\,$\pm$\,0.04 & 1.64\,$\pm$\,0.09 & 0.35\,$\pm$\,0.06 & 0.53 \,$\pm$\,0.05 & 1.54\,$\pm$\,0.08 & 0.48\,$\pm$\,0.06 & 0.42\,$\pm$\,0.06 & 32.83\,$\pm$\,1.49 & 7.41  \\
B21 & 0.6 \,$\pm$\,0.18 & 0.14\,$\pm$\,0.07 &              & 0.42\,$\pm$\,0.13 &              &  & 1.66\,$\pm$\,0.41 & 0.29\,$\pm$\,0.16 & 0.15\,$\pm$\,0.15 & 16.94\,$\pm$\,3.59 & 3.57  \\
B22 & 0.81\,$\pm$\,0.2  & 0.19\,$\pm$\,0.11 & 0.68\,$\pm$\,0.16 & 2.04\,$\pm$\,0.41 &              & 0.52 \,$\pm$\,0.22 & 2.63\,$\pm$\,0.55 & 0.41\,$\pm$\,0.12 & 0.44\,$\pm$\,0.12 & 17.63\,$\pm$\,3.43 & 3.05  \\
B23 & 0.32\,$\pm$\,0.03 & 0.13\,$\pm$\,0.01 & 0.08\,$\pm$\,0.02 & 0.32\,$\pm$\,0.02 &              & 0.36 \,$\pm$\,0.04 & 0.84\,$\pm$\,0.04 & 0.38\,$\pm$\,0.05 & 0.31\,$\pm$\,0.05 & 63.83\,$\pm$\,2.19 & 16.05 \\
\hline
\end{tabular}
\end{center}
\begin{flushleft}
{\textbf{Column description:} \textbf{ID} - MLLINER identification ('!' - possibly Sy2 galaxies, see Sec.~\ref{sec_emission_measure}); \textbf{[OII]$\lambda$3272}, \textbf{H$\beta$}, \textbf{[OIII]$\lambda$4959}, \textbf{[OIII]$\lambda$5007}, \textbf{[OI]$\lambda$6300}, \textbf{[NII]$\lambda$6548}, \textbf{[NII]$\lambda$6584}, \textbf{[SII]$\lambda$6716}, and \textbf{[SII]$\lambda$6731} - ratio between the fluxes of the indicated emission lines and H$\alpha$ line; \textbf{F$_{H\alpha}$\,$\times$\,10$^{-16}$} - flux of the H$\alpha$ line in [erg/cm$^2$/sec]; \textbf{EW$_{H\alpha}$} - H$\alpha$ equivalent width measured in the spectrum before the subtraction of the best model for the underlying stellar population.}
\end{flushleft}
\end{table*}

\begin{table*}
\scriptsize
\begin{center}
\caption{Properties of extinction corrected strong emission lines.  
\label{tab_extcorr_lines}}
\begin{tabular}{| c | c | c | c | c | c | c | c | c | c | c | c |}
\hline
\textbf{ID}&\textbf{[OII]$\lambda$3272}&\textbf{[OIII]$\lambda$4959}&\textbf{[OIII]$\lambda$5007}&\textbf{[OI]$\lambda$6300}&\textbf{[NII]$\lambda$6548}&\textbf{[NII]$\lambda$6584}&\textbf{[SII]$\lambda$6716}&\textbf{[SII]$\lambda$6731}&\textbf{F$_{H\alpha}$\,$\times$\,10$^{-16}$}&\textbf{L$_{H\alpha}$\,$\times$\,10$^{40}$}&\textbf{A$_V$}\\
\hline
  F01 & 1.09\,$\pm$\,0.34 & 0.22\,$\pm$\,0.05 & 0.54\,$\pm$\,0.06 &              & 0.35\,$\pm$\,0.06 & 1.29\,$\pm$\,0.09 & 0.25\,$\pm$\,0.08 & 0.17\,$\pm$\,0.08 & 28.68  \,$\pm$\,6.67  & 7.03  \,$\pm$\,1.63  & 0.872  \\
  F02 &              &              & 0.36\,$\pm$\,0.11 &              & 0.27\,$\pm$\,0.08 & 0.93\,$\pm$\,0.19 & 0.24\,$\pm$\,0.07 & 0.2 \,$\pm$\,0.06 & 177.85 \,$\pm$\,49.69 & 43.74 \,$\pm$\,12.22 & 3.191  \\
  F03 &              & 0.18\,$\pm$\,0.08 & 0.55\,$\pm$\,0.1  &              & 0.38\,$\pm$\,0.09 & 0.84\,$\pm$\,0.12 & 0.54\,$\pm$\,0.09 & 0.37\,$\pm$\,0.08 & 32.09  \,$\pm$\,9.38  & 9.74  \,$\pm$\,2.85  & 1.026  \\
  F04 &              & 0.13\,$\pm$\,0.05 & 0.55\,$\pm$\,0.12 &              & 0.41\,$\pm$\,0.1  & 1.12\,$\pm$\,0.17 & 0.53\,$\pm$\,0.12 & 0.39\,$\pm$\,0.11 & 48.73  \,$\pm$\,14.38 & 12.69 \,$\pm$\,3.75  & 1.951  \\
  F06 &              & 0.07\,$\pm$\,0.06 & 0.53\,$\pm$\,0.08 &              & 0.3 \,$\pm$\,0.05 & 0.93\,$\pm$\,0.09 & 0.24\,$\pm$\,0.05 & 0.19\,$\pm$\,0.04 & 99.64  \,$\pm$\,22.31 & 27.69 \,$\pm$\,6.2   & 1.983  \\
  F07 &              & 0.57\,$\pm$\,0.3  & 1.54\,$\pm$\,0.73 &              & 0.42\,$\pm$\,0.23 & 1.66\,$\pm$\,0.63 & 0.33\,$\pm$\,0.13 & 0.23\,$\pm$\,0.1  & 134.56 \,$\pm$\,63.88 & 8.36  \,$\pm$\,3.97  & 2.888  \\
  F09 & 1.22\,$\pm$\,0.7  &              & 0.32\,$\pm$\,0.14 &              & 0.38\,$\pm$\,0.24 & 1.67\,$\pm$\,0.52 & 0.45\,$\pm$\,0.26 & 0.39\,$\pm$\,0.25 & 17.69  \,$\pm$\,8.5   & 4.96  \,$\pm$\,2.38  & 1.528  \\
  F12 & 2.64\,$\pm$\,3.66 &              & 0.29\,$\pm$\,0.34 &              & 0.23\,$\pm$\,0.51 & 1.03\,$\pm$\,0.98 & 0.27\,$\pm$\,0.23 & 0.23\,$\pm$\,0.2  & 103.78 \,$\pm$\,87.17 & 30.5  \,$\pm$\,25.61 & 2.862  \\
  F13 & 1.5 \,$\pm$\,0.61 & 0.2 \,$\pm$\,0.11 & 0.53\,$\pm$\,0.18 &              & 0.39\,$\pm$\,0.1  & 1.12\,$\pm$\,0.19 & 0.39\,$\pm$\,0.12 & 0.33\,$\pm$\,0.11 & 58.59  \,$\pm$\,16.99 & 9.72  \,$\pm$\,2.82  & 2.41   \\
  F14 & 1.23\,$\pm$\,0.21 &    & 0.4 \,$\pm$\,0.09 &              & 0.68\,$\pm$\,0.12 & 1.71\,$\pm$\,0.2  & 0.57\,$\pm$\,0.11 & 0.47\,$\pm$\,0.11 & 24.86  \,$\pm$\,7.8   & 5.51  \,$\pm$\,1.73  & 0.493  \\
  F15 & 3.16\,$\pm$\,0.95 & 0.18\,$\pm$\,0.11 & 0.61\,$\pm$\,0.17 &              & 0.52\,$\pm$\,0.12 & 1.1 \,$\pm$\,0.2  & 0.4 \,$\pm$\,0.09 & 0.27\,$\pm$\,0.08 & 167.08 \,$\pm$\,51.2  & 35.22 \,$\pm$\,10.79 & 2.794  \\
  F16 & 1.47\,$\pm$\,0.16 & 0.12\,$\pm$\,0.03 & 0.43\,$\pm$\,0.03 &              & 0.34\,$\pm$\,0.02 & 0.86\,$\pm$\,0.04 & 0.18\,$\pm$\,0.02 & 0.16\,$\pm$\,0.02 & 255.32 \,$\pm$\,36.87 & 40.48 \,$\pm$\,5.84  & 2.528  \\
  F17 &              & 0.11\,$\pm$\,0.03 & 0.64\,$\pm$\,0.09 &              & 0.43\,$\pm$\,0.07 & 1.05\,$\pm$\,0.11 & 0.22\,$\pm$\,0.05 & 0.18\,$\pm$\,0.05 & 101.99 \,$\pm$\,25.28 & 27.86 \,$\pm$\,6.91  & 1.91   \\
  F19 & 1.99\,$\pm$\,0.2  & 0.31\,$\pm$\,0.04 & 0.78\,$\pm$\,0.07 &              & 0.26\,$\pm$\,0.06 & 0.65\,$\pm$\,0.07 & 0.46\,$\pm$\,0.06 & 0.39\,$\pm$\,0.06 & 57.04  \,$\pm$\,13.16 & 11.85 \,$\pm$\,2.73  & 1.232  \\
  F20 & 1.03\,$\pm$\,0.11 &              & 0.22\,$\pm$\,0.03 &              & 0.34\,$\pm$\,0.03 & 0.83\,$\pm$\,0.05 & 0.27\,$\pm$\,0.03 & 0.19\,$\pm$\,0.02 & 79.76  \,$\pm$\,14.12 & 21.46 \,$\pm$\,3.8   & 1.406  \\
  F21 & 3.56\,$\pm$\,1.23 & 0.9 \,$\pm$\,0.24 & 0.9 \,$\pm$\,0.24 &              & 0.56\,$\pm$\,0.13 & 1.86\,$\pm$\,0.36 & 0.48\,$\pm$\,0.12 & 0.39\,$\pm$\,0.11 & 195.11 \,$\pm$\,57.69 & 30.95 \,$\pm$\,9.15  & 2.64   \\
  F22 & 0.65\,$\pm$\,0.09 & 0.09\,$\pm$\,0.04 & 0.22\,$\pm$\,0.04 &              & 0.25\,$\pm$\,0.04 & 1.13\,$\pm$\,0.05 & 0.26\,$\pm$\,0.05 & 0.2 \,$\pm$\,0.05 & 39.57  \,$\pm$\,7.31  & 9.14  \,$\pm$\,1.69  & 0.643  \\
  F23 & 0.91\,$\pm$\,0.2  & 0.19\,$\pm$\,0.1  & 0.62\,$\pm$\,0.13 &              & 0.47\,$\pm$\,0.13 & 1.31\,$\pm$\,0.2  &              &              & 6.01   \,$\pm$\,2.06  & 0.96  \,$\pm$\,0.33  & -0.011 \\
  B01 & 1.95\,$\pm$\,0.26 & 0.12\,$\pm$\,0.03 & 0.36\,$\pm$\,0.04 &              & 0.43\,$\pm$\,0.05 & 1.07\,$\pm$\,0.07 & 0.42\,$\pm$\,0.03 & 0.35\,$\pm$\,0.03 & 227.37 \,$\pm$\,45.83 & 12.22 \,$\pm$\,2.46  & 1.652  \\
  B02 &              & 0.09\,$\pm$\,0.05 & 0.43\,$\pm$\,0.1  &              & 0.38\,$\pm$\,0.07 & 0.97\,$\pm$\,0.1  & 0.29\,$\pm$\,0.08 & 0.27\,$\pm$\,0.08 & 35.35  \,$\pm$\,8.75  & 9.55  \,$\pm$\,2.36  & 1.207  \\
  B03 &              & 0.11\,$\pm$\,0.03 & 0.37\,$\pm$\,0.04 & 0.13\,$\pm$\,0.01 & 0.3 \,$\pm$\,0.03 & 0.93\,$\pm$\,0.06 & 0.23\,$\pm$\,0.03 & 0.22\,$\pm$\,0.03 & 1359.55\,$\pm$\,223.14& 126.89\,$\pm$\,20.83 & 3.137  \\
  B04 &              & 0.28\,$\pm$\,0.14 & 0.75\,$\pm$\,0.26 &              & 0.88\,$\pm$\,0.3  & 1.75\,$\pm$\,0.49 &              &              & 27.04  \,$\pm$\,12.43 & 7.03  \,$\pm$\,3.23  & 1.086  \\
  B05 &              & 0.09\,$\pm$\,0.05 & 0.25\,$\pm$\,0.09 &              & 0.43\,$\pm$\,0.08 & 1.27\,$\pm$\,0.15 & 0.19\,$\pm$\,0.05 & 0.18\,$\pm$\,0.05 & 84.22  \,$\pm$\,22.05 & 24.24 \,$\pm$\,6.35  & 1.889  \\
  B06 & 1.57\,$\pm$\,0.16 & 0.09\,$\pm$\,0.06 & 0.41\,$\pm$\,0.05 & 0.28\,$\pm$\,0.05 & 0.21\,$\pm$\,0.05 & 1.23\,$\pm$\,0.09 & 0.54\,$\pm$\,0.07 & 0.41\,$\pm$\,0.07 & 50.85  \,$\pm$\,11.73 & 3.39  \,$\pm$\,0.78  & 0.751  \\
  B07 &              &              & 0.4 \,$\pm$\,0.15 &              & 0.31\,$\pm$\,0.09 & 0.93\,$\pm$\,0.2  & 0.43\,$\pm$\,0.13 & 0.33\,$\pm$\,0.11 & 190.8  \,$\pm$\,54.47 & 28.31 \,$\pm$\,8.08  & 3.459  \\
  B08 & 2.24\,$\pm$\,0.24 & 0.17\,$\pm$\,0.05 & 0.52\,$\pm$\,0.06 & 0.34\,$\pm$\,0.21 & 0.5 \,$\pm$\,0.08 & 1.56\,$\pm$\,0.15 & 0.75\,$\pm$\,0.11 & 0.61\,$\pm$\,0.11 & 42.03  \,$\pm$\,11.27 & 10.3  \,$\pm$\,2.76  & 0.85   \\
  B09 &              & 0.42\,$\pm$\,0.05 & 1.39\,$\pm$\,0.12 & 0.35\,$\pm$\,0.05 & 0.32\,$\pm$\,0.08 & 0.7 \,$\pm$\,0.09 & 0.74\,$\pm$\,0.08 & 0.59\,$\pm$\,0.07 & 39.99  \,$\pm$\,10.88 & 2.03  \,$\pm$\,0.55  & -0.412 \\
  B10 &              &              & 0.26\,$\pm$\,0.11 &              & 0.38\,$\pm$\,0.09 & 1.08\,$\pm$\,0.16 & 0.32\,$\pm$\,0.1  & 0.24\,$\pm$\,0.09 & 72.72  \,$\pm$\,20.36 & 17.98 \,$\pm$\,5.03  & 1.645  \\
  B11 & 1.42\,$\pm$\,0.12 & 0.07\,$\pm$\,0.03 & 0.28\,$\pm$\,0.03 &              & 0.44\,$\pm$\,0.05 & 0.97\,$\pm$\,0.06 & 0.46\,$\pm$\,0.03 & 0.35\,$\pm$\,0.03 & 112.47 \,$\pm$\,23.18 & 30.65 \,$\pm$\,6.32  & 1.215  \\
  B13 & 1.21\,$\pm$\,0.08 & 0.08\,$\pm$\,0.01 & 0.28\,$\pm$\,0.01 &              & 0.29\,$\pm$\,0.02 & 0.68\,$\pm$\,0.02 & 0.19\,$\pm$\,0.02 & 0.17\,$\pm$\,0.02 & 819.2  \,$\pm$\,109.52& 167.61\,$\pm$\,22.41 & 1.647  \\
  B14 & 0.25\,$\pm$\,0.06 & 0.06\,$\pm$\,0.02 & 0.34\,$\pm$\,0.05 &              & 1.07\,$\pm$\,0.18 & 0.59\,$\pm$\,0.14 &              &              & 74.49  \,$\pm$\,25.78 & 12.79 \,$\pm$\,4.43  & 0.639  \\
  B15 & 1.93\,$\pm$\,0.26 & 0.07\,$\pm$\,0.03 & 0.42\,$\pm$\,0.07 & 0.25\,$\pm$\,0.09 & 0.53\,$\pm$\,0.08 & 1.63\,$\pm$\,0.16 & 0.39\,$\pm$\,0.07 & 0.32\,$\pm$\,0.07 & 81.78  \,$\pm$\,21.61 & 15.11 \,$\pm$\,3.99  & 1.306  \\
  B16 &              & 0.35\,$\pm$\,0.1  & 0.82\,$\pm$\,0.21 &              & 0.52\,$\pm$\,0.11 & 1.4 \,$\pm$\,0.2  & 0.61\,$\pm$\,0.14 & 0.5 \,$\pm$\,0.14 & 40.88  \,$\pm$\,12.76 & 12.26 \,$\pm$\,3.82  & 1.512  \\
  B17 & 2.07\,$\pm$\,0.9  & 0.28\,$\pm$\,0.16 & 0.59\,$\pm$\,0.16 &              & 0.31\,$\pm$\,0.08 & 1.05\,$\pm$\,0.19 & 0.41\,$\pm$\,0.13 & 0.21\,$\pm$\,0.12 & 354.13 \,$\pm$\,96.75 & 51.23 \,$\pm$\,14.0  & 3.415  \\
  B18 & 0.59\,$\pm$\,0.13 &              & 0.23\,$\pm$\,0.09 &              & 0.5 \,$\pm$\,0.1  & 1.19\,$\pm$\,0.15 & 0.12\,$\pm$\,0.05 & 0.18\,$\pm$\,0.05 & 32.35  \,$\pm$\,9.65  & 3.99  \,$\pm$\,1.19  & 0.72   \\
  B19 & 0.82\,$\pm$\,0.19 & 0.06\,$\pm$\,0.02 & 0.55\,$\pm$\,0.09 & 0.21\,$\pm$\,0.07 & 0.46\,$\pm$\,0.07 & 1.36\,$\pm$\,0.14 & 0.39\,$\pm$\,0.09 & 0.32\,$\pm$\,0.08 & 62.04  \,$\pm$\,14.86 & 8.85  \,$\pm$\,2.12  & 1.736  \\
  B20! & 1.68\,$\pm$\,0.13 & 0.72\,$\pm$\,0.06 & 2.07\,$\pm$\,0.13 & 0.37\,$\pm$\,0.06 & 0.53\,$\pm$\,0.05 & 1.54\,$\pm$\,0.09 & 0.48\,$\pm$\,0.06 & 0.42\,$\pm$\,0.06 & 61.76  \,$\pm$\,13.21 & 11.34 \,$\pm$\,2.43  & 0.8    \\
  B21 & 3.07\,$\pm$\,1.93 &              & 0.86\,$\pm$\,0.44 &              &              & 1.66\,$\pm$\,0.65 & 0.29\,$\pm$\,0.18 & 0.15\,$\pm$\,0.16 & 117.96 \,$\pm$\,59.8  & 28.62 \,$\pm$\,14.51 & 2.458  \\
  B22! & 2.44\,$\pm$\,1.12 & 1.11\,$\pm$\,0.41 & 3.32\,$\pm$\,1.16 &              & 0.52\,$\pm$\,0.24 & 2.63\,$\pm$\,0.77 & 0.41\,$\pm$\,0.15 & 0.44\,$\pm$\,0.15 & 65.01  \,$\pm$\,30.24 & 15.84 \,$\pm$\,7.37  & 1.653  \\
  B23 & 1.95\,$\pm$\,0.28 & 0.18\,$\pm$\,0.04 & 0.72\,$\pm$\,0.08 &              & 0.36\,$\pm$\,0.04 & 0.84\,$\pm$\,0.07 & 0.38\,$\pm$\,0.05 & 0.31\,$\pm$\,0.05 & 552.23 \,$\pm$\,104.81& 45.77 \,$\pm$\,8.69  & 2.733  \\

\hline
\end{tabular}
\end{center}
\begin{flushleft}
{\textbf{Column description:} \textbf{ID} - MLLINER identification ('!' - possibly Sy2 galaxies, see Section~\ref{sec_emission_measure}); \textbf{[OII]$\lambda$3272}, \textbf{[OIII]$\lambda$4959}, \textbf{[OIII]$\lambda$5007}, \textbf{[OI]$\lambda$6300}, \textbf{[NII]$\lambda$6548}, \textbf{[NII]$\lambda$6584}, \textbf{[SII]$\lambda$6716}, and \textbf{[SII]$\lambda$6731} - ratio between the extinction corrected fluxes of the indicated emission lines and H$\alpha$ line; \textbf{F$_{H\alpha}$\,$\times$\,10$^{-16}$} - flux of the extinction corrected H$\alpha$ line in [erg/cm$^2$/sec]; \textbf{L$_{H\alpha}$\,$\times$\,10$^{40}$} - luminosity of the extinction corrected H$\alpha$ line in [erg/sec]; \textbf{A$_V$} - interstellar extinction parameter in the V band in [mag].}
\end{flushleft}
\end{table*}


\begin{figure*}
\centering
\begin{minipage}[c]{0.33\textwidth}
\includegraphics[width=6.5cm,angle=0]{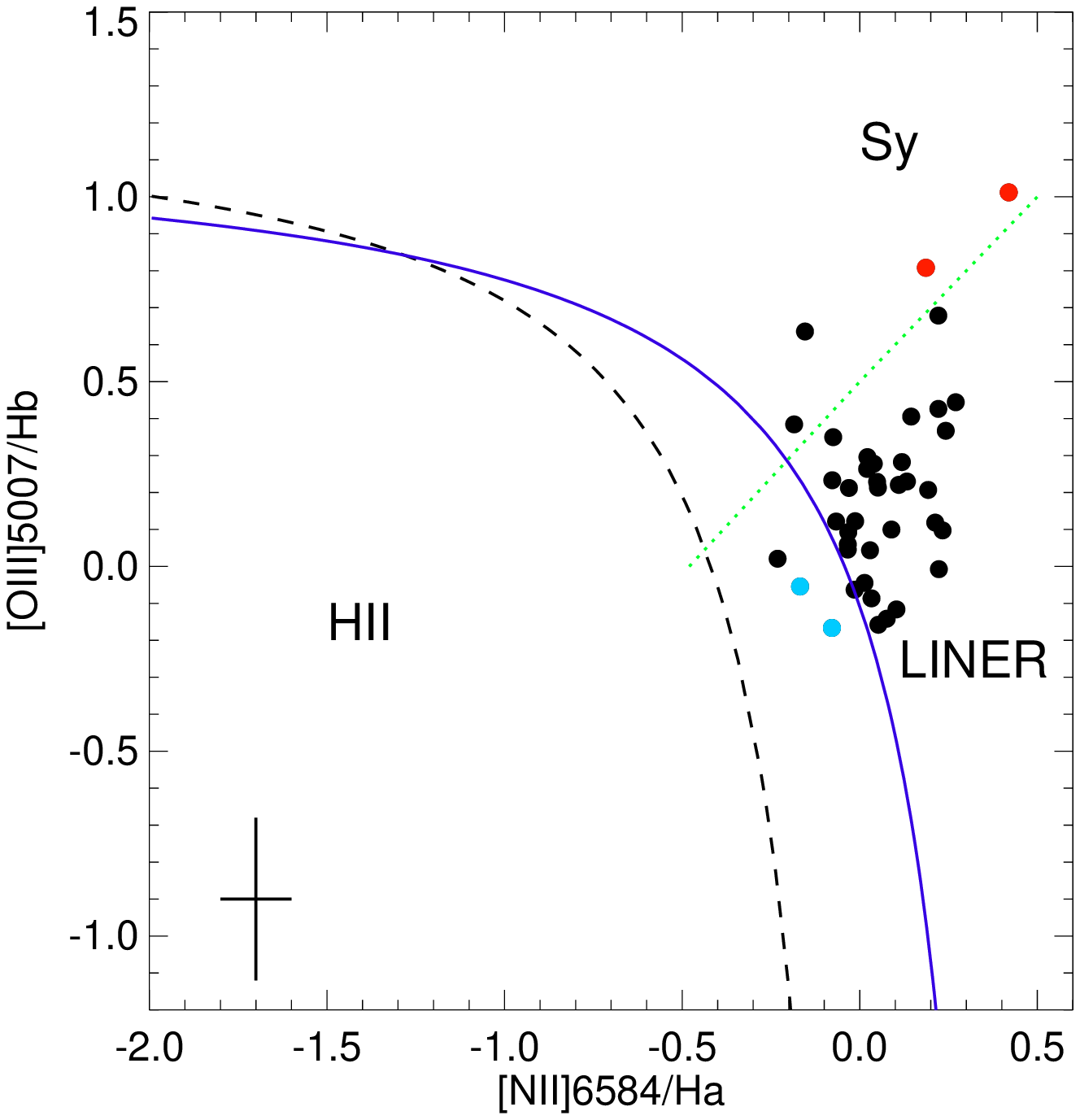}
\end{minipage}
\begin{minipage}[c]{0.33\textwidth}
\includegraphics[width=6.5cm,angle=0]{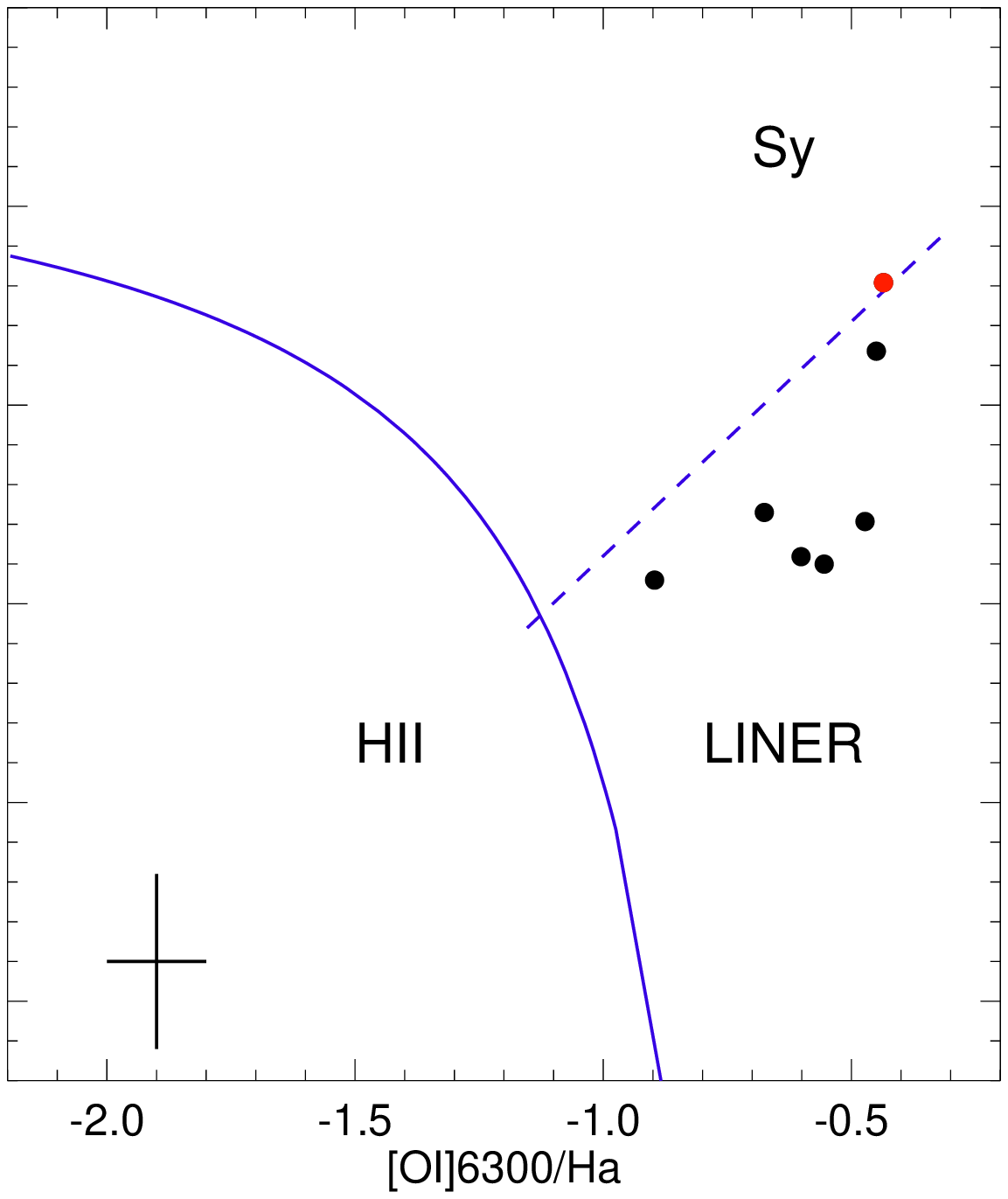}
\end{minipage}
\begin{minipage}[c]{0.33\textwidth}
\includegraphics[width=6.5cm,angle=0]{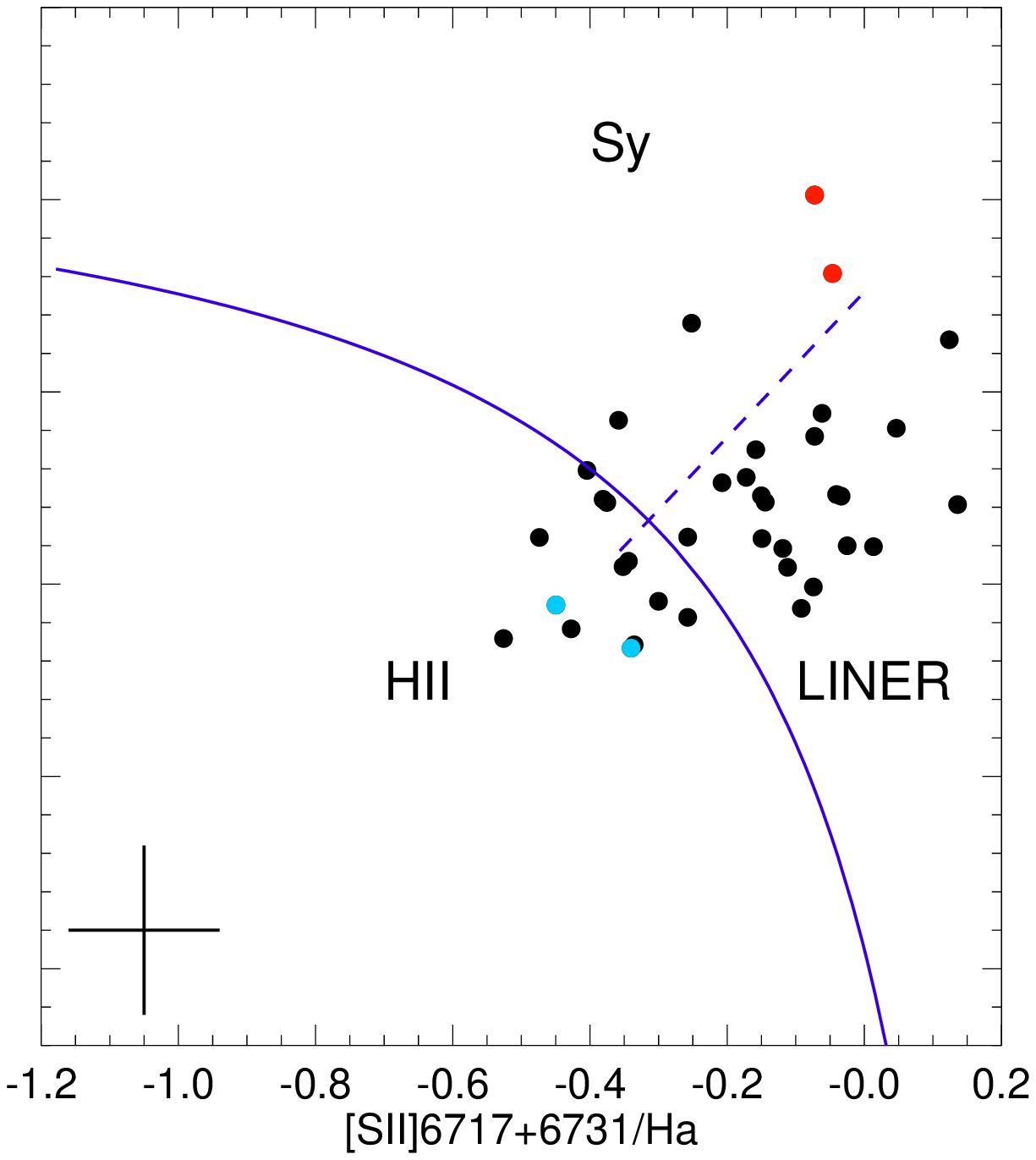}
\end{minipage}
\begin{minipage}[c]{0.33\textwidth}
\includegraphics[width=6.5cm,angle=0]{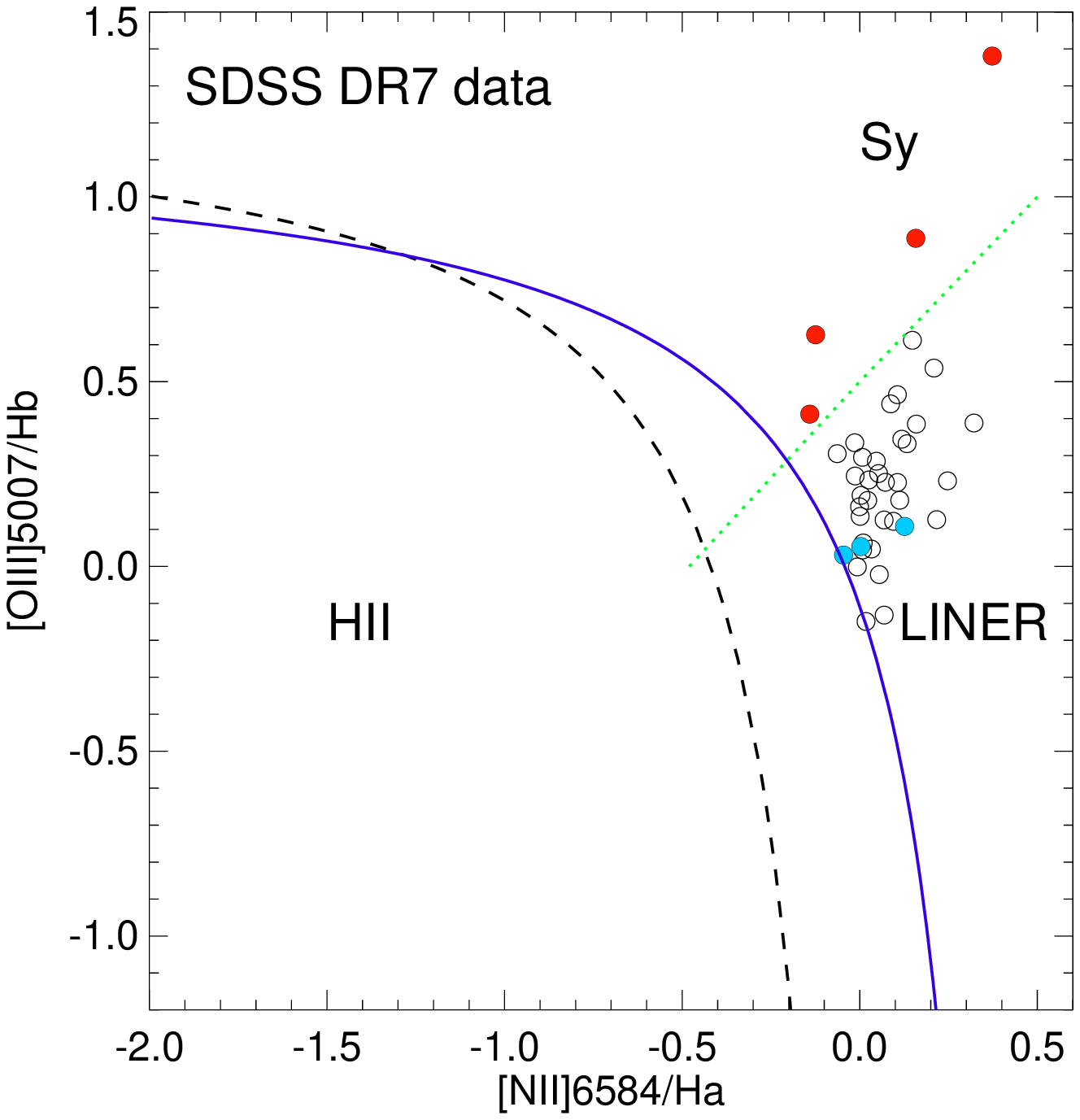}
\end{minipage}
\begin{minipage}[c]{0.33\textwidth}
\includegraphics[width=6.5cm,angle=0]{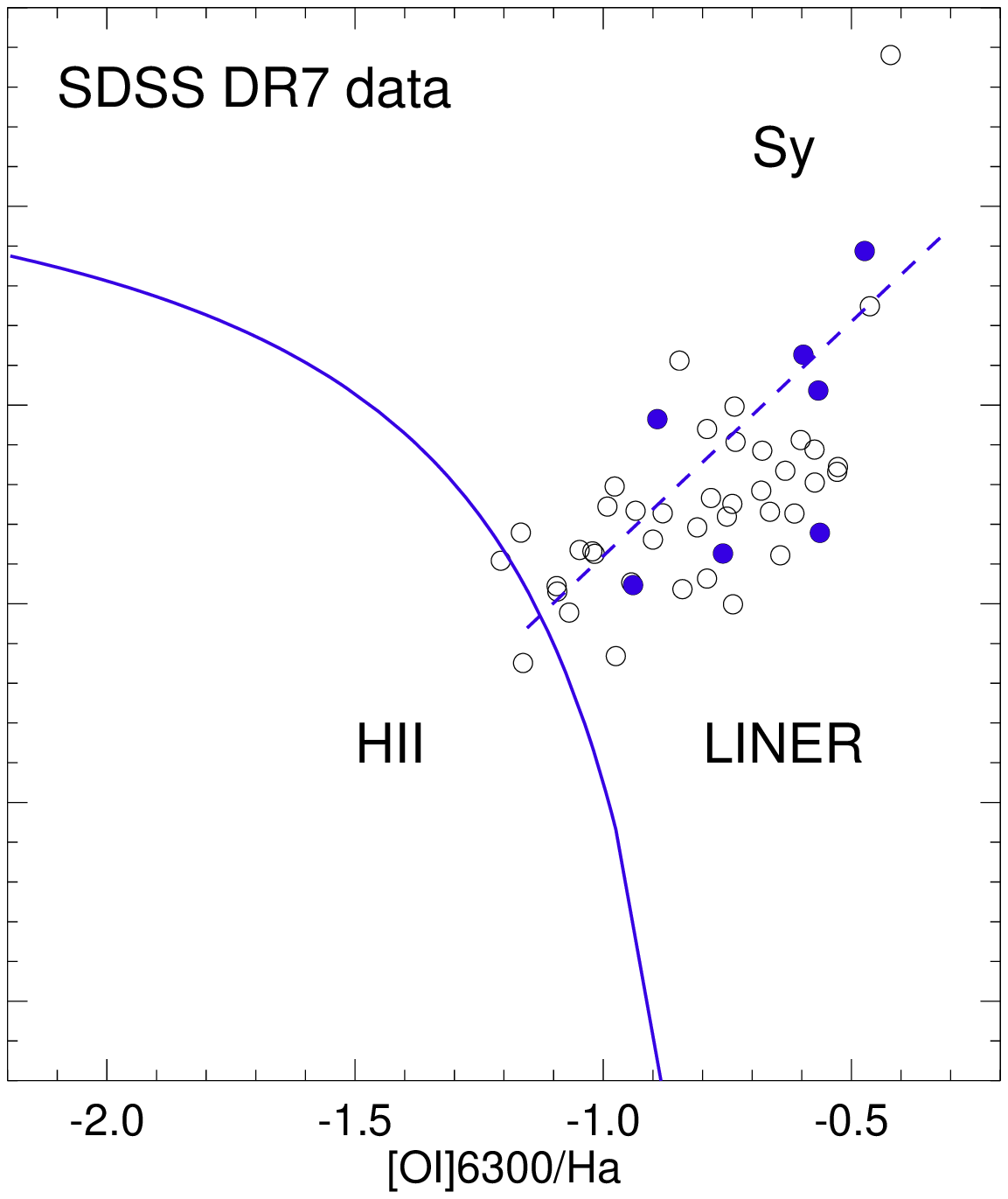}
\end{minipage}
\begin{minipage}[c]{0.33\textwidth}
\includegraphics[width=6.5cm,angle=0]{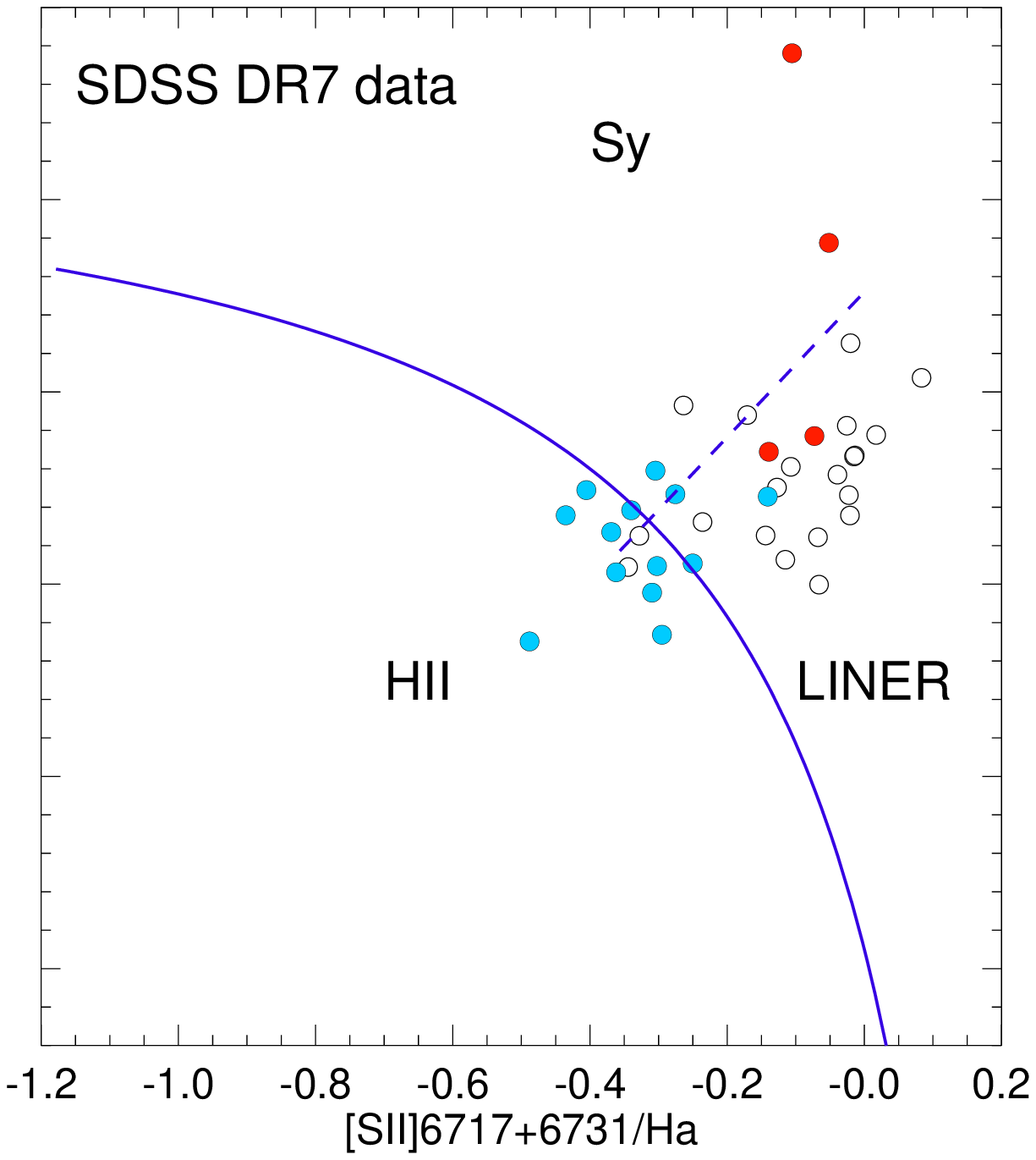}
\end{minipage}
\caption[ ]{\textit{(Top)} The BPT-NII \textit{(left)}, BPT-OI \textit{(centre)}, and BPT-SII \textit{(right)} diagrams. In the BPT-NII plot black dashed line \citep{kauffmann03} and blue solid line \citep{kewley01} separate HII regions and AGNs, while green dotted line \citep{cid10} separates Seyfert (above) and LINERs (below). The BPT-OI and BPT-SII diagrams use \cite{kewley06} limits to distinguish between different sources. In all plots red and blue filled circles show possible outliers (see the text) classified as Seyfert and transit, respectively. The median error bars are given in all plots in the bottom left corner. \textit{(Bottom)} Same as above, but using SDSS MPA-JHU DR7 data. In the BPT-NII and BPT-SII diagrams, red and blue filled circles show the position of sources being in the Seyfert and transition areas, respectively, in our plots. In the BPT-OI diagram there are fewer sources due to the availability of [OI]$\lambda$6300 line. We marked the position of all sources for we have data with dark blue filled circles.}
\label{fig_bpt}
\end{figure*} 

\indent Fig.~\ref{fig_bpt} (top plots) shows three standard BPT diagrams based on [NII]$\lambda$6584, [OI]$\lambda$6300, and [SII]$\lambda\lambda$6716+6731 emission line ratios, used to separate between the HII regions and AGN, and between Seyfert 2 galaxies and LINERs \citep{baldwin81}. The lines correspond to \cite{kewley01}, \cite{kauffmann03}, \cite{kewley06}, and \cite{cid10} (see the caption of Fig.~\ref{fig_bpt}). Following the BPT-NII diagram, 4 MLLINERs (B22, B20, B09, and F19) enter in the region of Seyfert galaxies (although they are located close to the limiting line with LINERs), while another three sources (B13, B14, and F20) stay inside the transition region. As for [OI]$\lambda$6300 we could not detect this line in most of our CAHA observations, and we only have 7 sources plotted in the BPT-OI diagram (see Table~\ref{tab_emission_lines}). All these sources enter in the region typical of LINERs. In the BPT-SII diagram again four sources are located inside the area typical of Seyfert (B20, B21, B22, and F07), while 13 lie in the transition region (in particular B03, B05, B10, B13, B18, F01, F02, F06, F12, F16, F17, F20, and F22). We considered as possible outliers those sources that at least in two of the BPT diagrams lie outside of the standard LINER region. We found two possible Seyfert galaxies (B20 and B22, marked in red), and two possible transition galaxies (F20 and B13, marked in blue). \\
\indent The discrepant classification of some of the sources is not surprising given that the original sample selection was based on the SDSS MPA-JHU DR4 data. Since then both SDSS data calibration and the analysis by the MPA-JHU have improved. We compared the positions of our MLLINERs on the BPT diagrams using the new version of MPA-JHU catalogues based on the SDSS DR7 data. These plots are presented in Fig.~\ref{fig_bpt} (bottom diagrams). B20 and B22 enter in the Seyfert region in this case as well. Therefore, we will consider these two galaxies as outliers, and although we provide their measurements in all tables, we exclude them from all diagrams showed in Section~\ref{sec_results_discussion}. In all tables these two galaxies are marked with '!'. F20 and B13 stay inside the LINER region in the SDSS DR7, so we do not consider them as outliers. 

\begin{figure}
\centering
\includegraphics[width=7.0cm,angle=0]{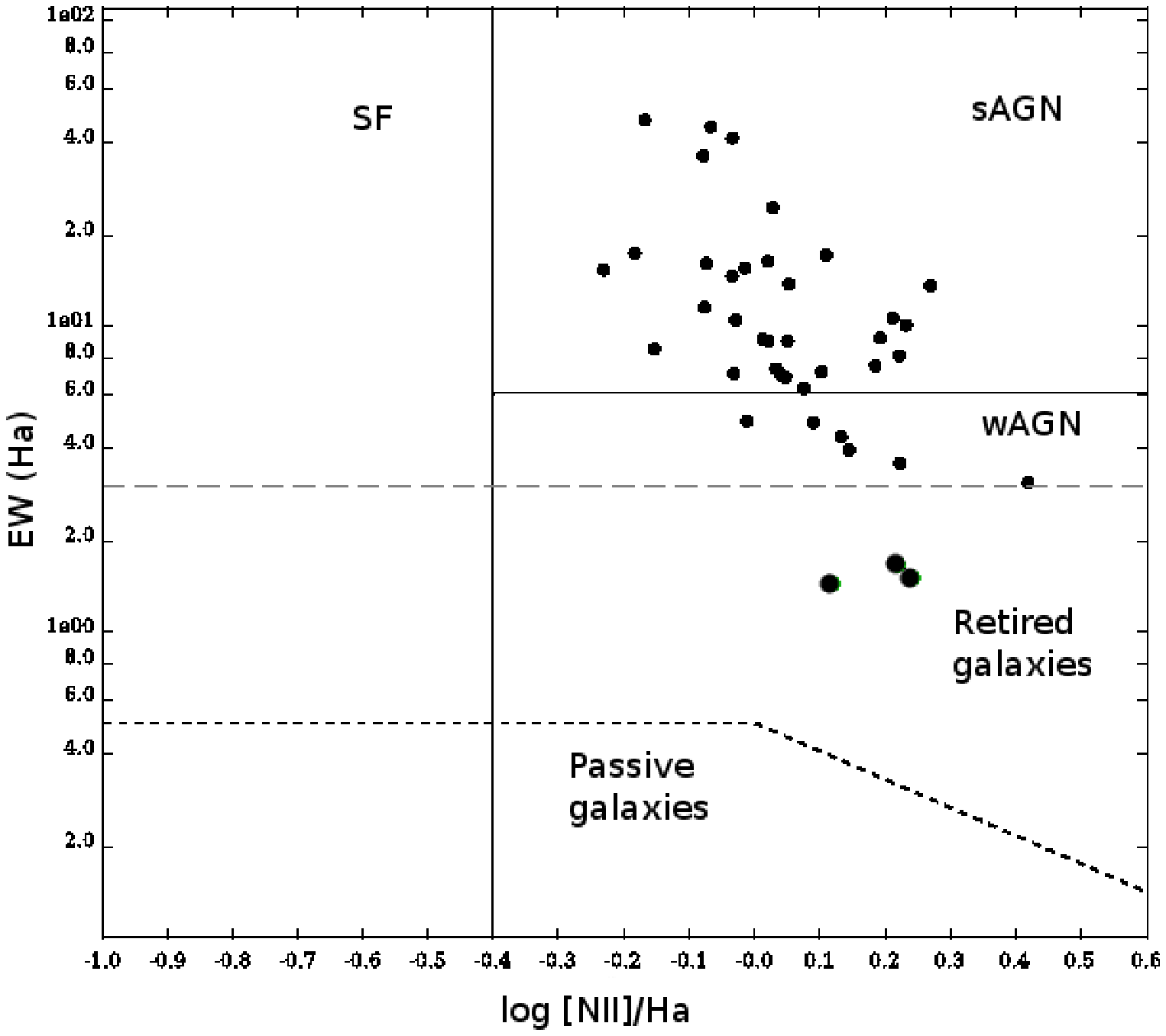}
\caption[ ]{The revised WHAN classification diagram showing the relation between EW(H$\alpha$) and [NII]/H$\alpha$. The limits are those suggested by \cite{cid11}. They are used to separate between SF galaxies, strong AGN (sAGN), weak AGN (wAGN), retired, and passive galaxies, as marked on the diagram.}
\label{fig_WHAN}
\end{figure}

\indent To test in more detail the nuclear classification of our MLLINERs, we also used the WHAN diagram by \cite{cid11}. This diagram shows the relation between EW(H$\alpha$) and the [NII]$\lambda$6584/H$\alpha$ ratio, and separates all galaxies in passive (lineless), retired, pure star-forming, and strong and weak AGN. The purpose is to distinguish 'true' from 'fake' AGN, and to separate between the two classes that overlap in the LINER region of the traditional diagnostic diagrams: galaxies hosting weak AGN, and retired galaxies that have stopped forming stars and are ionised by hot low-mass evolved (pAGB) stars. Fig.~\ref{fig_WHAN} represents the WHAN diagram for all our MLLINERs. Thirty three sources occupy the region of strong AGN having EW(H$\alpha$)\,$>$\,6\AA ~and [NII]$\lambda$6584/H$\alpha$\,$>$\,-0.4, while six (B02, B06, B16, B19, B21, and B22) are located in the area of weak AGN with 3\AA\,$<$\,EW(H$\alpha$)\,$<$\,6\AA ~and [NII]$\lambda$6584/H$\alpha$\,$>$\,-0.4. \\
\indent Three MLLINERs (B04, F09, and F23) show EW(H$\alpha$) between 1.4 and 3A and are therefore located in the area of retired galaxies. The lower limit of EW(H$\alpha$)\,=\,3\AA ~was determined by \cite{cid11} using the 3\,arc-sec SDSS fibre spectra. In our case, however, we used the nuclear spectra covering a smaller area for all MLLINERs. We compared our EW(H$\alpha$) measurements with those of SDSS MPA-JHU DR7\footnote{http://wwwmpa.mpa-garching.mpg.de/SDSS/DR7/} finding a linear correlation, but higher MPA-JHU values in all cases due to aperture differences. Since in the MPA-JHU DR7 catalogue all our sources have EW(H$\alpha$)\,$>$\,3\AA ~(as in the initially used DR4 version), we will continue to consider the entire selected sample as AGN, including the three sources that enter in the area of retired galaxies with the values from our nuclear spectra.  

\subsection[]{AGN luminosity}
\label{sec_agn_lum}

\indent Our measurements of LAGN are based on the reddening-corrected luminosity of H$\beta$ and [OIII]$\lambda$5007. We used eq. 4 from \cite{tommasin12} which is based on \cite{netzer09}:
\begin{center}
{logLAGN\,=\,logL(H$\beta$)\,+\,3.75\,+\,max[0,0.31\,$\times$\,(log([OIII]$\lambda$5007/H$\beta$)-0.6)]}. 
\end{center} 
Table~\ref{tab_SFR_AGNluminosity} summarises the obtained values for all MLLINERs. \cite{netzer09} showed that [OIII]$\lambda$5007 and [OI]$\lambda$6300 lines provide more accurate measurements of the LAGN, measured as:
\begin{center}
{logLAGN\,=\,3.8\,+\,0.25logL([OIII]$\lambda$5007)\,+\,0.75logL([OI]$\lambda$6300)}.
\end{center}
However, since [OI]$\lambda$6300 is missing in most of our spectra (see table~\ref{tab_extcorr_lines}), we are able to measure the LAGN based on this line only in the case of seven MLLINERs, and therefore for consistency we won't use these measurements in our analyses.

\subsection[]{AGN and SF contributions to the emission lines}
\label{sec_agn_contribution}

\indent Both H$\alpha$ and [OII]$\lambda$3727 lines can be used to estimate SFRs in non-active galaxies \citep{kennicutt92, kewley04, mouhcine05, moustakas06}. However, all our sources are classified as LINERs and hence much of the flux in these two lines can be due to ionization and excitation by the central non-stellar source. To assess the various contributions to L(H$\alpha$), we made an estimate of the expected H$\alpha$ luminosity (using \cite{netzerbook} expression) based on the SFRs, obtained from STARLIGHT by using stellar absorption spectra and young stellar populations (age\,$\le$\,10$^8$\,yr). We compared these values with the measured H$\alpha$ luminosities (see Tab.~\ref{tab_extcorr_lines}). Tab.~\ref{tab_agn_contribution} gives all measurements and estimated AGN contributions. For those MLLINERs without young stellar populations detected (see table~\ref{tab_starlight_fits}) we assume that all H$\alpha$ emission comes from the AGN. In almost all MLLINERs, most of the nuclear Ha is due to the AGN (all sources except 4 have AGN contribution of $>$\,60\%). Therefore we do not consider the estimators based on H$\alpha$ and [OII]$\lambda$3727 lines as reliable tracers of SF in our case.  

\begin{table}
\small
\begin{center}
\caption{AGN contribution measured through L(H$\alpha$) and STARLIGHT SFRs (obtained from young stellar populations).  
\label{tab_agn_contribution}}
\begin{tabular}{| c c c || c c c |}
\hline
\textbf{ID}&\textbf{L(H$\alpha$)\_test}&\textbf{AGN$_{cont}$}&\textbf{ID}&\textbf{L(H$\alpha$)\_test}&\textbf{AGN$_{cont}$}\\
&\textbf{$\times$\,10$^{40}$\,[erg/s]}&\textbf{[\%]}&&\textbf{$\times$\,10$^{40}$\,[erg/s]}&\textbf{[\%]}\\
\hline
F01 & 0.23   &  96.71 &   B03 & 6.49 & 94.87  \\
F02 & 3.33   &  92.37 &   B04 & 0.0  & 100.0  \\
F03 & 1.65   &  82.97 &   B05 & 1.69 & 92.99  \\
F04 & 1.65   &  86.97 &   B06 & 0.0  & 100.0  \\
F06 & 5.99   &  78.35 &   B07 & 9.83 & 65.25  \\
F07 & 3.76   &  55.01 &   B08 & 0.0  & 100.0  \\
F09 & 0.0    &  100.0 &   B09 & 1.08 & 46.37  \\
F12 & 0.0    &  100.0 &   B10 & 2.48 & 86.18  \\
F13 & 0.0    &  100.0 &   B11 & 0.0  & 100.0  \\ 
F14 & 5.19   &  5.702 &   B13 & 4.55 & 97.28  \\
F15 & 0.0    &  100.0 &   B14 & 0.0  & 100.0  \\
F16 & 1.25   &  96.89 &   B15 & 0.0  & 100.0  \\   
F17 & 6.35   &  77.18 &   B16 & 0.0  & 100.0  \\ 
F19 & 0.0    &  100.0 &   B17 & 0.0  & 100.0  \\  
F20 & 0.95   &  95.57 &   B18 & 0.0  & 100.0  \\
F21 & 0.0    &  100.0 &   B19 & 0.0  & 100.0  \\
F22 & 0.0    &  100.0 &   B20! & 0.0  & 100.0  \\
F23 & 0.0    &  100.0 &   B21 & 0.0  & 100.0  \\
B01 & 0.62   &  94.93 &   B22! & 0.0  & 100.0  \\
B02 & 11.1   &  -16.9 &   B23 & 0.0  & 100.0  \\
\hline
\end{tabular}
\end{center}
\begin{flushleft}
{\textbf{Column description:} \textbf{ID} - MLLINER identification; \textbf{L(H$\alpha$)\_test} - H$\alpha$ luminosity obtained from the STARLIGHT SFRs correspondent only to young stellar populations; \textbf{AGN$_{cont}$} - approximation of AGN contribution to H$\alpha$ luminosity in \%.}
\end{flushleft}
\end{table}

\subsection[]{Star formation rates}
\label{sec_sfr}

We measured the SFRs using different methods and data, both optical and FIR. In the following, we provide a full description for each measurement, and list the results in Table~\ref{tab_SFR_AGNluminosity}. To convert SFR to LSF, we assume a slightly rounded value of LSF\,$=$\,SFR\,$\times$\,10$^{10}$\,L$_{\bigodot}$ based on the Kroupa initial mass function (IMF). When scaling the nuclear measurements of SFRs of our MLLINERs to those of the entire galaxy, we assume that the specific star-formation rate (sSFR) is constant throughout the galaxy and therefore: SFR$_{scaled}$\,=\,SFR$_{nuclear}$/M$_{nuclear}$\,$\times$\,M$_{tot}$, where the total stellar mass was taken from the MPA-JHU DR7 catalogue and is listed in Table~\ref{tab_SFR_AGNluminosity} (last column), while M$_{nuclear}$ is the mass measured from our nuclear spectra (see Table~\ref{tab_starlight_fits}). This assumption is further tested by comparing optical and FIR measurements.\\

\indent \textbf{\textit{SFR using STARLIGHT best fits.}} We followed the equation from \cite{cid13} and obtained the mean SFR surface density, by accumulating all the stellar mass formed since a look-back time of t$_{SF}$. The mass-over-time average, is:
\begin{center}
{SFR(t$_{SF}$)\,=\,1/t$_{SF}$\,$\sum$M$_t$}, 
\end{center}
where M$_t$ is the mass of stars formed at look-back time t (corresponding to M$_*^{ini}$ in Section~\ref{sec_starlight_fits}). We measured three SFRs, for stellar populations younger than 10$^8$ years, for those younger than 10$^9$ years, and the total one corresponding to the entire initial mass processed into stars throughout the galaxy life ($<$logt$>$, column 10 in Table~\ref{tab_starlight_fits}). The SFRs are listed in columns 2, 3, and 4 in Table~\ref{tab_SFR_AGNluminosity}. We also estimated what would be the values of STARLIGHT total SFRs when scaled to map the entire galaxy (column 5 in Table~\ref{tab_SFR_AGNluminosity}), as explained above. \\

\indent \textbf{\textit{SFR using Dn4000.}} We compared our results with the models obtained by \cite{brinchmann04}, showing the relation between the sSFR and the Dn4000 index (their Figure 11). Using the nuclear M$_*$ masses from the STARLIGHT fits (see Section~\ref{sec_starlight_fits}) and our Dn4000 measurements (see Section~\ref{sec_dn4000_hdelta}) we obtained the mode SFRs. These values are again provided in Table~\ref{tab_SFR_AGNluminosity} together with the scaled SFR if mapping the entire galaxy (columns 6 and 7).  \\

\indent \textbf{\textit{SFR using FIR luminosity.}} Finally, we measured SFRs using \textit{Herschel}/PACS and IRAS FIR data (see Table~\ref{tab_obs}). We assumed that all the FIR luminosity is due to star formation, and that the total IR SF luminosity (TIR, the SF luminosity integrated over the range 8\,-\,100\,$\mu$m) is dominated by the FIR luminosity. Thus, LSF\,=\,L(TIR). In the case of six sources observed with Herschel, we performed the spectral energy distribution (SED) fitting to obtain L(FIR) through $\chi^2$ minimisation and using the templates of \cite{chary01}. To measure the SFR with IRAS data, we followed the same procedure applied in \cite{tommasin12}. L$_{FIR}$ is measured through F$_{FIR}$, using two IRAS bands, and following the expression provided in \cite{sanders96}:  
\begin{center}
{F$_{FIR}$\,=\,1.26$\times$10$^{-14}$(2.58\,$\times$\,F(60$\mu$m)\,+\,F(100$\mu$m))\,[W m$^{-2}$]}, 
\end{center}
where F(60$\mu$m) and F(100$\mu$m) are the fluxes in 60$\mu$m and 100$\mu$m IRAS bands, respectively. As in \cite{tommasin12}, we do not include the fluxes at 12$\mu$m and 25$\mu$m bands since they may be influenced by warm AGN heated dust. In the case of three MLLINERs with poor flux measurements in the 100$\mu$m band, having flag quality of 1 (see table~\ref{tab_obs_fir}), we measured the total FIR flux as F$_{FIR}$\,=\,2\,$\times$\,F(60$\mu$m) \citep[see e.g.,][]{rosario12}.

\begin{table*}
\small
\begin{center}
\caption{SFRs and AGN luminosities.  
\label{tab_SFR_AGNluminosity}}
\begin{tabular}{| c c c c c c c c c c c c |}
\hline
\textbf{ID}&\textbf{SFR1}&\textbf{SFR2}&\textbf{SFR$_{tot}$}&\textbf{SFR$_{tot\_sc}$}&\textbf{SFR$_{Dn4000}$}&\textbf{SFR$_{Dn4000\_sc}$}&\textbf{SFR$_{PACS}$}&\textbf{SFR$_{IRAS}$}&\textbf{logL$_{AGN}$}&\textbf{log(M$_{tot}$/M$_{\odot}$)}&\textbf{log(M$_{BH}$/M$_{\odot}$)}\\
\hline
F01 & 0.01 & 0.09  & 0.81\,$\pm$\,0.13 & 8.13 \,$\pm$\,2.68 & 0.95 \,$\pm$\,0.28 & 9.52 \,$\pm$\,3.94  &                     &                    & 44.11\,$\pm$\,0.15 & 10.86& \\
F02 & 0.18 & 1.24  & 2.61\,$\pm$\,0.41 & 8.12 \,$\pm$\,2.67 & 2.5  \,$\pm$\,0.21 & 7.77 \,$\pm$\,2.34  &  15.78\,$\pm$\,0.05 & 15.15\,$\pm$\,0.45 & 44.9 \,$\pm$\,0.24 & 10.88&7.45 \\
F03 & 0.09 & 0.01  & 1.73\,$\pm$\,0.26 & 13.99\,$\pm$\,4.51 & 1.18 \,$\pm$\,0.23 & 9.55 \,$\pm$\,3.33  &                     &                    & 44.25\,$\pm$\,0.20 & 11.12& \\
F04 & 0.09 & 0.01  & 2.26\,$\pm$\,0.33 & 17.43\,$\pm$\,5.57 & 1.22 \,$\pm$\,0.22 & 9.39 \,$\pm$\,3.19  &                     &                    & 44.36\,$\pm$\,0.20 & 11.21& \\
F06 & 0.33 & 1.36  & 1.47\,$\pm$\,0.24 & 7.99 \,$\pm$\,2.66 & 1.56 \,$\pm$\,0.2  & 8.49 \,$\pm$\,2.68  &                     & 17.1 \,$\pm$\,0.51 & 44.7 \,$\pm$\,0.17 & 10.91&7.87 \\
F07 & 0.21 & 0.48  & 0.67\,$\pm$\,0.12 & 0.82 \,$\pm$\,0.28 &                    &                     &                     &                    & 44.22\,$\pm$\,0.30 & 9.86&7.14  \\
F09 & 0.0  & 0.48  & 1.92\,$\pm$\,0.29 & 15.91\,$\pm$\,5.14 & 1.37 \,$\pm$\,0.24 & 11.37\,$\pm$\,3.83  &                     & 18.2 \,$\pm$\,1.0  & 43.95\,$\pm$\,0.30 & 11.19& \\
F12 & 0.0  & 2.36  & 0.87\,$\pm$\,0.14 & 9.78 \,$\pm$\,3.20 & 0.61 \,$\pm$\,0.27 & 6.88 \,$\pm$\,3.62  &                     & 20.95\,$\pm$\,1.09 & 44.74\,$\pm$\,0.31 & 10.98& \\
F13 & 0.0  & 0.27  & 0.56\,$\pm$\,0.09 & 4.51 \,$\pm$\,1.51 &                    &                     &                     &                    & 44.25\,$\pm$\,0.24 & 10.66&7.31 \\
F14 & 0.29 & 0.37  & 1.13\,$\pm$\,0.18 & 2.71 \,$\pm$\,0.91 & 1.1  \,$\pm$\,0.22 & 2.64 \,$\pm$\,0.94  &                     & 10.42\,$\pm$\,0.55 & 44.0 \,$\pm$\,0.20 & 10.41& \\
F15 & 0.0  & 10.57 & 1.04\,$\pm$\,0.16 & 4.38 \,$\pm$\,1.45 & 1.33 \,$\pm$\,0.2  & 5.59 \,$\pm$\,1.84  &                     &                    & 44.81\,$\pm$\,0.21 & 10.63&7.03 \\
F16 & 0.07 & 0.09  & 0.94\,$\pm$\,0.15 & 4.56 \,$\pm$\,1.52 & 2.2  \,$\pm$\,0.21 & 10.64\,$\pm$\,3.27  &                     & 11.89\,$\pm$\,0.61 & 44.87\,$\pm$\,0.11 & 10.61& \\
F17 & 0.35 & 3.05  & 0.95\,$\pm$\,0.16 & 3.39 \,$\pm$\,1.14 & 1.45 \,$\pm$\,0.23 & 5.15 \,$\pm$\,1.72  & 13.52\,$\pm$\,0.16  & 21.26\,$\pm$\,0.64 & 44.7 \,$\pm$\,0.16 & 10.55&7.83 \\
F19 & 0.0  & 0.49  & 1.66\,$\pm$\,0.24 & 7.25 \,$\pm$\,2.35 & 1.55 \,$\pm$\,0.24 & 6.77 \,$\pm$\,2.22  & 1.25 \,$\pm$\,0.04  &                    & 44.33\,$\pm$\,0.12 & 10.82& \\
F20 & 0.05 & 1.53  & 0.94\,$\pm$\,0.15 & 5.21 \,$\pm$\,1.74 & 2.27 \,$\pm$\,0.18 & 12.63\,$\pm$\,3.81  &                     &                    & 44.59\,$\pm$\,0.15 & 10.68& \\
F21 & 0.0  & 1.93  & 1.58\,$\pm$\,0.24 & 1.45 \,$\pm$\,0.48 & 1.52 \,$\pm$\,0.22 & 1.4  \,$\pm$\,0.46  &                     &                    & 44.75\,$\pm$\,0.26 & 10.13&7.15 \\
F22 & 0.0  & 1.84  & 1.56\,$\pm$\,0.25 & 2.35 \,$\pm$\,0.79 & 3.65 \,$\pm$\,0.15 & 5.5  \,$\pm$\,1.64  &                     &                    & 44.22\,$\pm$\,0.11 & 10.32& \\
F23 & 0.0  & 5.18  & 1.19\,$\pm$\,0.2  & 1.93 \,$\pm$\,0.65 & 1.42 \,$\pm$\,0.21 & 2.31 \,$\pm$\,0.77  &                     &                    & 43.24\,$\pm$\,0.44  & 10.25& \\
F24 & 0.21 & 1.16  & 1.95\,$\pm$\,0.31 & 8.82 \,$\pm$\,2.91 & 2.24 \,$\pm$\,0.42 & 10.11\,$\pm$\,3.49  &                     & 9.25 \,$\pm$\,0.49 &                    & 10.88& \\
B01 & 0.03 & 0.87  & 0.75\,$\pm$\,0.12 & 3.29 \,$\pm$\,1.10 & 1.37 \,$\pm$\,0.23 & 5.98 \,$\pm$\,2.04  &                     &                    & 44.35\,$\pm$\,0.18 & 10.46& \\
B02 & 0.61 & 4.45  & 4.25\,$\pm$\,0.71 & 11.47\,$\pm$\,3.82 & 9.79 \,$\pm$\,0.19 & 26.44\,$\pm$\,7,64  & 7.19 \,$\pm$\,0.17  &                    & 44.24\,$\pm$\,0.19 & 11.01& \\
B03 & 0.36 & 1.06  & 3.22\,$\pm$\,0.49 & 18.94\,$\pm$\,6.14 &                    &                     & 16.47\,$\pm$\,0.06  & 10.33\,$\pm$\,0.52 & 45.36\,$\pm$\,0.10 & 11.22& \\
B04 & 0.0  & 0.0   & 1.89\,$\pm$\,0.31 & 6.76 \,$\pm$\,2.25 & 1.22 \,$\pm$\,0.21 & 4.35 \,$\pm$\,1.47  &                     &                    & 44.11\,$\pm$\,0.37 & 10.82&7.98 \\
B05 & 0.09 & 0.25  & 4.0 \,$\pm$\,0.64 & 29.76\,$\pm$\,9.71 & 3.35 \,$\pm$\,0.22 & 24.95\,$\pm$\,7.27  &                     & 15.01\,$\pm$\,0.88 & 44.64\,$\pm$\,0.20 & 11.43& \\
B06 & 0.0  & 1.93  & 3.7 \,$\pm$\,0.58 & 7.47 \,$\pm$\,2.47 & 4.21 \,$\pm$\,0.17 & 8.52 \,$\pm$\,2.50  &                     &                    & 43.79\,$\pm$\,0.18 & 10.82& \\
B07 & 0.54 & 2.17  & 7.57\,$\pm$\,1.13 & 20.91\,$\pm$\,6.74 & 6.81 \,$\pm$\,0.24 & 18.81\,$\pm$\,5.41  & 5.31 \,$\pm$\,0.09  &                    & 44.71\,$\pm$\,0.25 & 11.26& \\
B08 & 0.0  & 0.69  & 5.16\,$\pm$\,0.77 & 20.88\,$\pm$\,6.73 & 3.29 \,$\pm$\,0.17 & 13.29\,$\pm$\,3.85  &                     &                    & 44.27\,$\pm$\,0.20 & 11.26&8.16 \\
B09 & 0.06 & 0.01  & 0.98\,$\pm$\,0.15 & 5.45 \,$\pm$\,1.81 &   		   &                     &                     &                    & 43.57\,$\pm$\,0.14 & 10.71&7.67 \\
B10 & 0.14 & 1.23  & 1.12\,$\pm$\,0.18 & 4.59 \,$\pm$\,1.54 & 1.89 \,$\pm$\,0.18 & 7.73 \,$\pm$\,2.37  &                     &                    & 44.51\,$\pm$\,0.28 & 10.68& \\
B11 & 0.0  & 0.68  & 1.64\,$\pm$\,0.28 & 11.66\,$\pm$\,3.89 & 2.66 \,$\pm$\,0.12 & 18.93\,$\pm$\,5.51  &                     & 12.02\,$\pm$\,0.63 & 44.75\,$\pm$\,0.12 & 11.06& \\
B12 & 0.0  & 1.7   & 0.54\,$\pm$\,0.09 & 6.37 \,$\pm$\,2.11 & 0.69 \,$\pm$\,0.21 & 8.15 \,$\pm$\,3.47  &                     &                    &                    & 10.82&7.43 \\
B13 & 0.25 & 3.24  & 1.37\,$\pm$\,0.24 & 7.86 \,$\pm$\,2.65 & 14.44\,$\pm$\,0.12 & 82.6 \,$\pm$\,23.91 &                     & 18.15\,$\pm$\,0.85 & 45.48\,$\pm$\,0.09 & 10.89&8.32* \\
B14 & 0.0  & 1.97  & 5.24\,$\pm$\,0.78 & 17.25\,$\pm$\,5.57 & 2.99 \,$\pm$\,0.17 & 9.87 \,$\pm$\,2.88  &                     &                    & 44.37\,$\pm$\,0.19 & 11.18& \\
B15 & 0.0  & 7.81  & 1.03\,$\pm$\,0.16 & 3.57 \,$\pm$\,1.18 & 1.63 \,$\pm$\,0.19 & 5.67 \,$\pm$\,1.79  &                     &                    & 44.44\,$\pm$\,0.19 & 10.54&7.36 \\
B16 & 0.0  & 0.0   & 1.28\,$\pm$\,0.2  & 5.14 \,$\pm$\,1.69 & 1.33 \,$\pm$\,0.22 & 5.35 \,$\pm$\,1.79  &                     &                    & 44.35\,$\pm$\,0.21 & 10.71&7.93 \\
B17 & 0.0  & 2.64  & 1.57\,$\pm$\,0.25 & 7.15 \,$\pm$\,2.36 & 2.94 \,$\pm$\,0.22 & 13.35\,$\pm$\,4.00  &                     &  10.56\,$\pm$\,0.54& 44.97\,$\pm$\,0.21 & 10.81& \\
B18 & 0.0  & 5.37  & 1.54\,$\pm$\,0.25 & 2.73 \,$\pm$\,0.91 & 1.74 \,$\pm$\,0.19 & 3.07 \,$\pm$\,0.97  &                     &                    & 43.86\,$\pm$\,0.22 & 10.39&7.21 \\
B19 & 0.0  & 6.42  & 6.15\,$\pm$\,0.99 & 4.61 \,$\pm$\,1.54 & 6.75 \,$\pm$\,0.21 & 5.06 \,$\pm$\,1.49  &                     &                    & 44.21\,$\pm$\,0.20 & 10.61& \\
B20! & 0.0  & 3.44  & 6.01\,$\pm$\,0.86 & 7.51 \,$\pm$\,2.42 & 4.82 \,$\pm$\,0.19 & 6.02 \,$\pm$\,1.76  &                     &                    & 44.38\,$\pm$\,0.15 & 10.82&8.20 \\
B21 & 0.0  & 6.95  & 9.25\,$\pm$\,1.41 & 16.77\,$\pm$\,5.44 & 5.89 \,$\pm$\,0.19 & 10.68\,$\pm$\,3.08  &                     &                    & 44.72\,$\pm$\,0.30 & 11.17&8.03\\
B22! & 0.0  & 7.84  & 5.29\,$\pm$\,0.86 & 14.64\,$\pm$\,4.83 & 9.68 \,$\pm$\,0.19 & 26.78\,$\pm$\,7.71  &                     &                    & 44.6 \,$\pm$\,0.32 & 11.11&8.16 \\
B23 & 0.0  & 4.62  & 1.84\,$\pm$\,0.29 & 3.07 \,$\pm$\,1.02 & 1.22 \,$\pm$\,0.18 & 2.04 \,$\pm$\,0.68  &                     &                    & 44.92\,$\pm$\,0.13 & 10.45&7.56 \\
\hline
\end{tabular}
\end{center}
\begin{flushleft}
{\textbf{Column description:} \textbf{ID} - MLLINER identification; \textbf{SFR1} and \textbf{SFR2} - SFR measured through the STARLIGHT best-fit model using only young (age\,$<$\,10$^8$) and young and intermediate (age\,$<$\,10$^9$) stars, respectively; \textbf{SFR$_{tot}$} - total SFR obtained from the best-fit model; \textbf{SFR$_{tot\_sc}$} - as previous, but scaled to the entire galaxy; \textbf{SFR$_{Dn4000}$} - SFR measured using the Dn4000 index; \textbf{SFR$_{Dn4000\_sc}$} - as previous, but scaled to the entire galaxy; \textbf{SFR$_{PACS}$} and \textbf{SFR$_{IRAS}$} - SFRs measured through Herschel/PACS and IRAS FIR data, respectively; \textbf{logL$_{AGN}$} - AGN luminosity in [erg/sec], measured through H$\beta$ and [OIII]$\lambda$5007 extinction corrected emission lines; \textbf{log(M$_{tot}$/M$_{\odot}$)} - total stellar mass, measured by MPA-JHU team using SDSS DR7 data; \textbf{log(M$_{BH}$/M$_{\odot}$)} - black hole mass measured for galaxies classified as E.   
}
\end{flushleft}
\end{table*}

\section{Results and Discussion}
\label{sec_results_discussion}

\subsection{General properties of the MLLINERs}
\label{sec_discussion_general_properties}

\indent In this section we describe the general properties of our MLLINERs: their masses, extinction, morphology, SFRs, and stellar populations. We compare them with the properties of other LLINERs (see Section~\ref{sec_sample_selection}), with the sample of the most-luminous LINERs at z\,$\sim$\,0.3 \citep{tommasin12}, and with the nearby and local LINER population analysed in previous studies \citep[e.g,][]{ho97,leslie16}. 

\subsubsection{Stellar and black hole mass}
\label{sec_discussion_masses}

\begin{figure}
\centering
\begin{minipage}[c]{0.33\textwidth}
\includegraphics[width=6.0cm,angle=0]{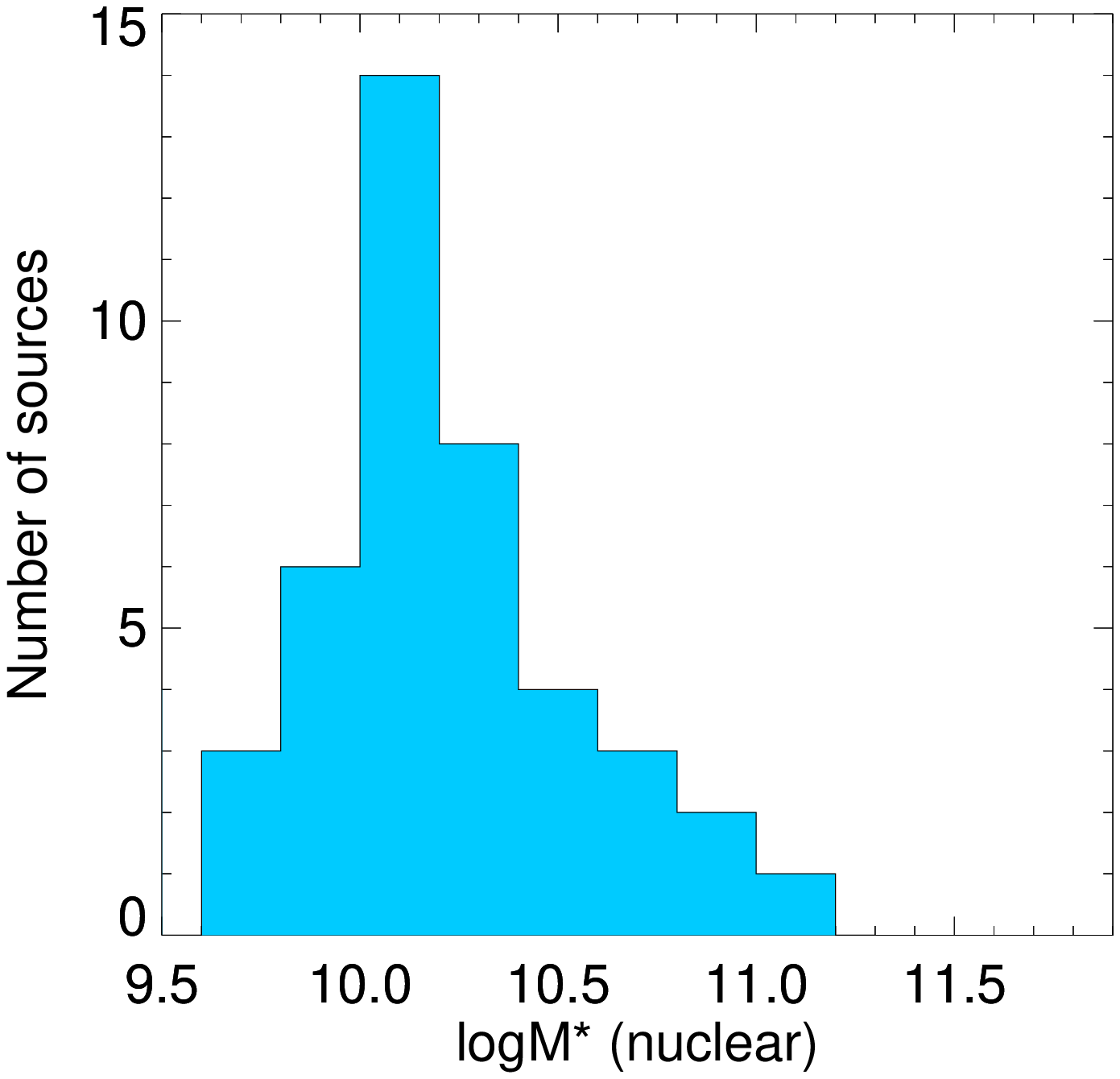}
\end{minipage}
\begin{minipage}[c]{0.33\textwidth}
\includegraphics[width=6.0cm,angle=0]{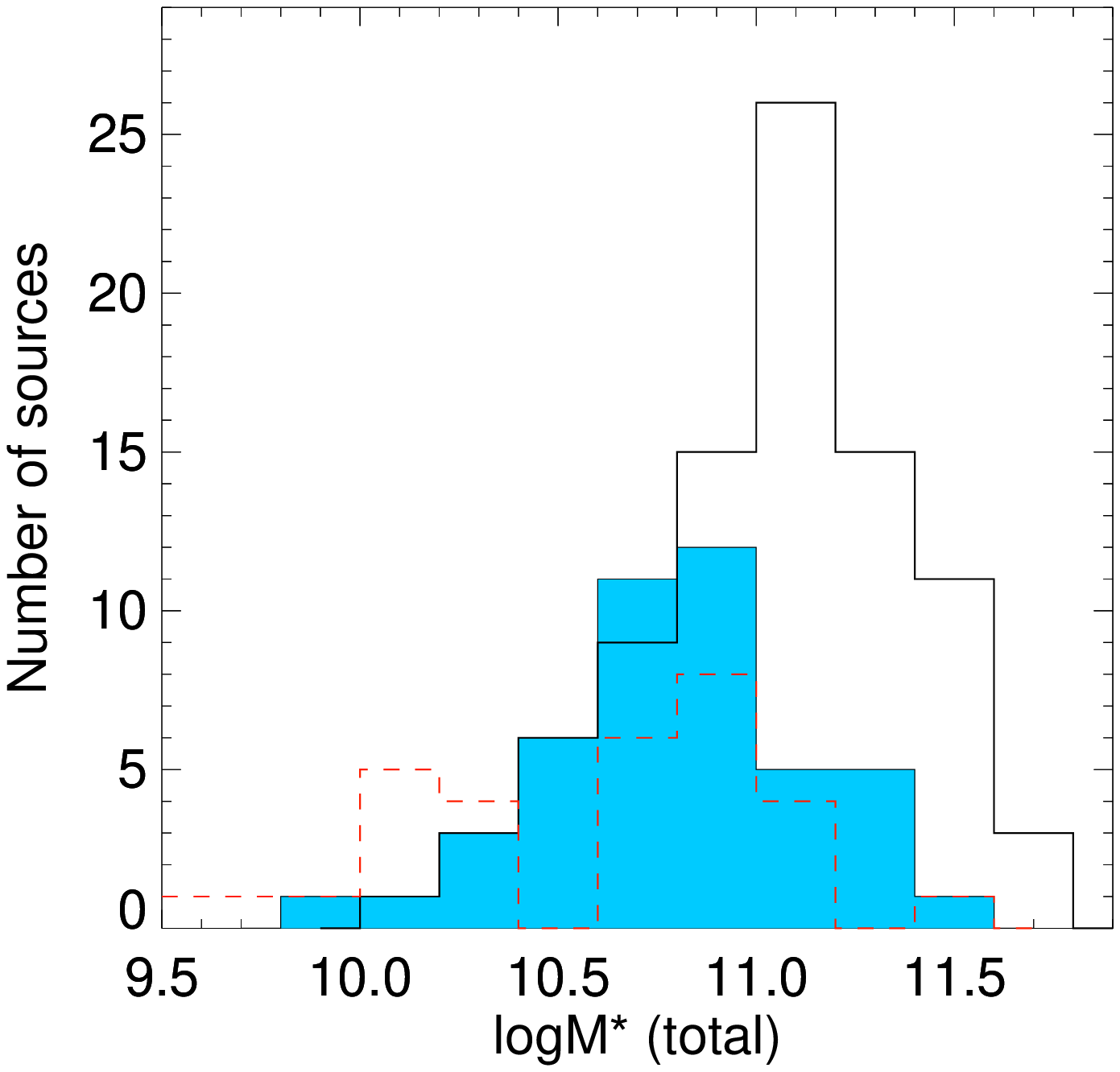}
\end{minipage}
\caption[ ]{\textit{Top:} Distribution of \textbf{nuclear} stellar masses of MLLINERs. \textit{Bottom:} Distributions of the SDSS/DR7 \textbf{total} stellar masses of MLLINERs (filled blue histogram), of the entire population of LLINERs (solid black lines), and of the most-luminous LINERs at z\,$\sim$\,0.3 (red dashed lines) from \cite{tommasin12}.}
\label{fig_masses}
\end{figure}

\indent The nuclear stellar masses of our MLLINERs cover the range between 5.7\,$\times$\,10$^9$\,M$_{\odot}$ and 8.32\,$\times$\,10$^{10}$\,M$_{\odot}$. The median stellar mass is 1.52\,$\times$\,10$^{10}$\,M$_{\odot}$ and the average mass is 2.11\,$\times$\,10$^{10}$\,M$_{\odot}$. Fig.~\ref{fig_masses} (top plot) shows the distribution of our nuclear measurements. Using the SDSS MPA-JHU DR7 measurements of total stellar masses, we found that our MLLINERs cover the range 7.21\,$\times$\,10$^9$\,M$_{\odot}$\,-\,2.71\,$\times$\,10$^{11}$\,M$_{\odot}$, with median masses of 6.58\,$\times$\,10$^{10}$\,M$_{\odot}$. In Fig.~\ref{fig_masses} (bottom plot) we compared this distribution with those of LLINERs (see Fig.~\ref{fig_sample_selection}), and with the sample of the most luminous LINERs at z\,$\sim$\,0.3 from \cite{tommasin12}. Interestingly, MLLINERs at z\,$\sim$\,0.07 and z\,$\sim$\,0.3, although hosted by massive galaxies, do not cover the region of the most massive galaxies. When comparing our sample with the sample at z\,$\sim$\,0.3 the distributions are not completely consistent (Kolmogorov-Smirnov (KS) probability factor of 0.02). A significant part (35\%) of \cite{tommasin12} LINERs have lower stellar masses, however the peak of the two distributions at log\,M$_*$\,$\sim$\,10.9\,M$_{\odot}$ is the same for both samples.\\
\indent We compared the distributions of black hole masses (MBH) between MLLINERs and LLINERs. To derive MBH, we used its correlation with stellar velocity dispersion found by \cite{tremaine02} in the nearby universe, shown to be reliable for elliptical and bulge-dominated galaxies. We recovered stellar velocity dispersions from the MPA-JHU DR7 catalogue, and we obtained MBH only for galaxies classified as ellipticals (see Section~\ref{sec_observations}). The values are given in Table~\ref{tab_SFR_AGNluminosity}. MLLINERs cover the range between log\,(MBH/M$_{\odot}$)\,=\,7.03\,-\,8.57 with a median value of log\,(MBH/M$_{\odot}$)\,=\,7.45, while LLINERs show MBH in the range log\,(MBH/M$_{\odot}$)\,=\,6.24\,-\,8.54 and median value of 
log\,(MBH/M$_{\odot}$)\,=\,8.04. Interestingly, our MLLINERs do not contain the most massive BHs in their centres. \\ 

\indent It could be surprising that MLLINERs, having on average higher SFRs than LLINERs, show in general lower stellar masses. Different works, both observational \citep{kauffmann03c,mateus06,leauthaud12,perez13} and numerical \citep{shankar06,behroozi12}, revealed a stellar mass of $\sim$\,6\,$\times$\,10$^{10}$\,M$_{\odot}$ as critical for the growth rate of stellar populations. In particular, in \cite{perez13} by studying a 3D spectroscopic sample of 105 local galaxies, the authors found that in galaxies more massive than 5\,$\times$\,10$^{10}$\,M$_{\odot}$ the inner regions ($<$\,0.5R$_{50}$) grew as much as 50\%\,-\,100\% faster than in the lower-mass galaxies. They found that the peak of relative growth rates of inner and outer galaxy regions correspond to the stellar mass of 6\,-\,7\,$\times$\,10$^{10}$\,M$_{\odot}$ (see their Fig. 5), while for lower and higher masses the growth rate decreases and therefore SFRs (LSF) as well. The median stellar mass of our MLLINERs (6.58\,$\times$\,10$^{10}$\,M$_{\odot}$) corresponds perfectly to this region, while for most LLINERs their stellar masses are already higher and correspond to lower values of the relative growth rate (lower LSF). This explains why MLLINERs having in average lower stellar masses in comparison to LLINERs, have higher LSF. \\
\indent In addition, we studied the stellar mass distributions of MLLINERs and LLINERs for the three morphological groups. Table~\ref{tab_masses_morphology} shows the median stellar masses for different morphological types. As can be seen, of three morphological types the highest difference was obtained for early-type galaxies. These galaxies represent a significant fraction of MLLINERs (40\%, see Fig.~\ref{fig_comparison_morphology}) and their median mass corresponds exactly to the highest relative growth rate of stellar populations (according to \cite{perez13}), which then could explain their high SFR values. This is not the case for early-type LLINERs that are characterised by higher stellar masses (lower growth rates) and lower SFRs.

\begin{table}
\small
\begin{center}
\caption{Median total stellar masses of MLLINERs and LLINERs in relation to morphology (given as logarithm and in M$_{\odot}$) 
\label{tab_masses_morphology}}
\begin{tabular}{| c | c | c |c | c | c | c |}
\hline
&&\textbf{All}&\textbf{E}&\textbf{S}&\textbf{P}&\textbf{unclass}\\
\hline  
\hline  
&\textbf{logM$_{tot}$}&10.83&10.73&11.0&11.08&10.69\\
\hline  
\textbf{MLLINERs}&\textbf{st.dev.}&0.34&0.34&0.33&0.22&0.22\\
\hline  
&\textbf{num.}&40&16&8&10&6\\
\hline  
\hline  
&\textbf{logM$_{tot}$}&11.04&11.14&11.05&11.14&10.94\\
\hline  
\textbf{LLINERs}&\textbf{st.dev.}&0.33&0.36&0.29&0.35&0.31\\
\hline  
&\textbf{num.}&88&33&27&10&18\\
\hline  
\end{tabular}
\end{center}
\end{table}

\subsubsection{Extinction}

\begin{figure}
\centering
\begin{minipage}[c]{0.33\textwidth}
\includegraphics[width=6.0cm,angle=0]{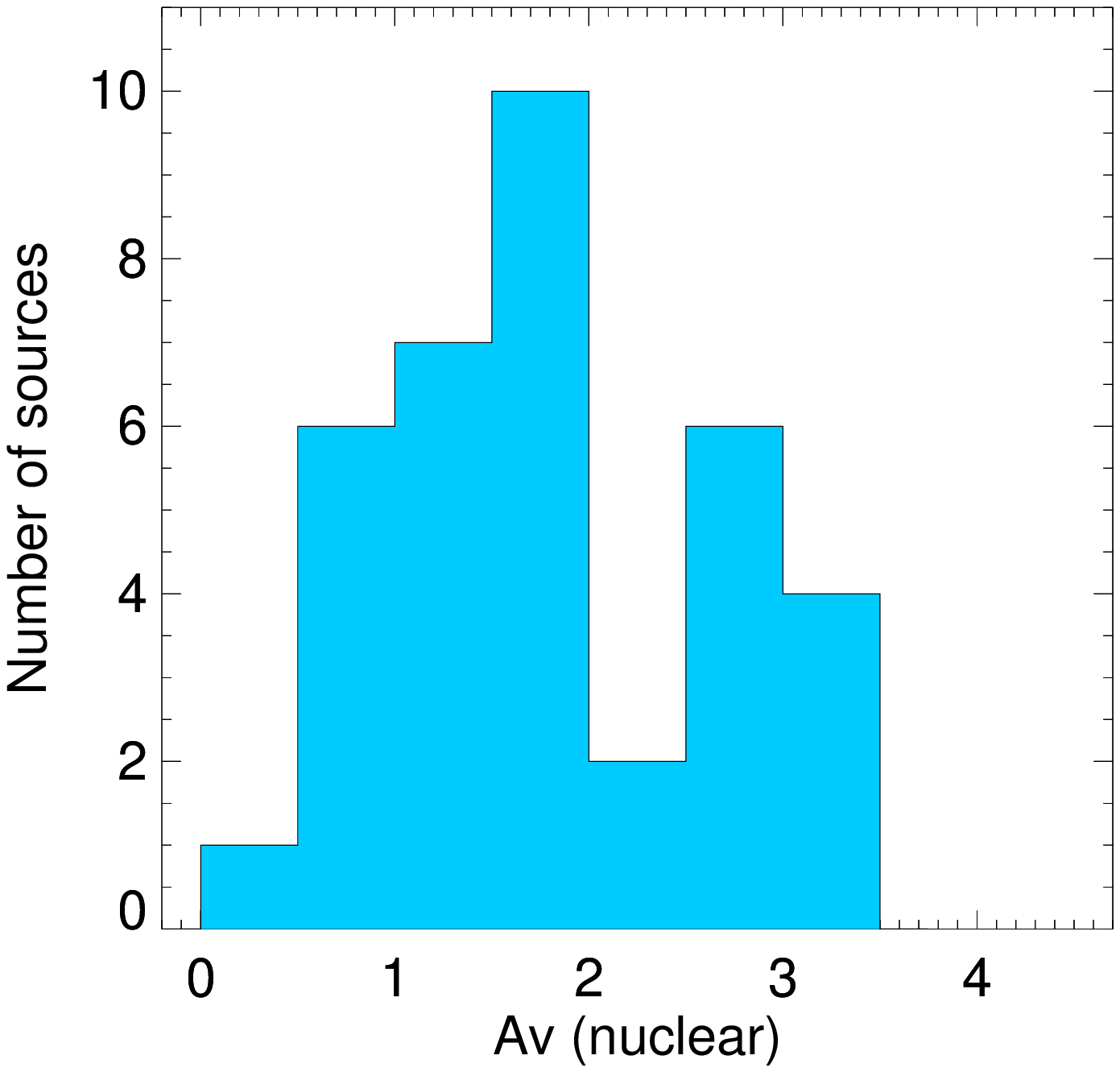}
\end{minipage}
\begin{minipage}[c]{0.33\textwidth}
\includegraphics[width=6.0cm,angle=0]{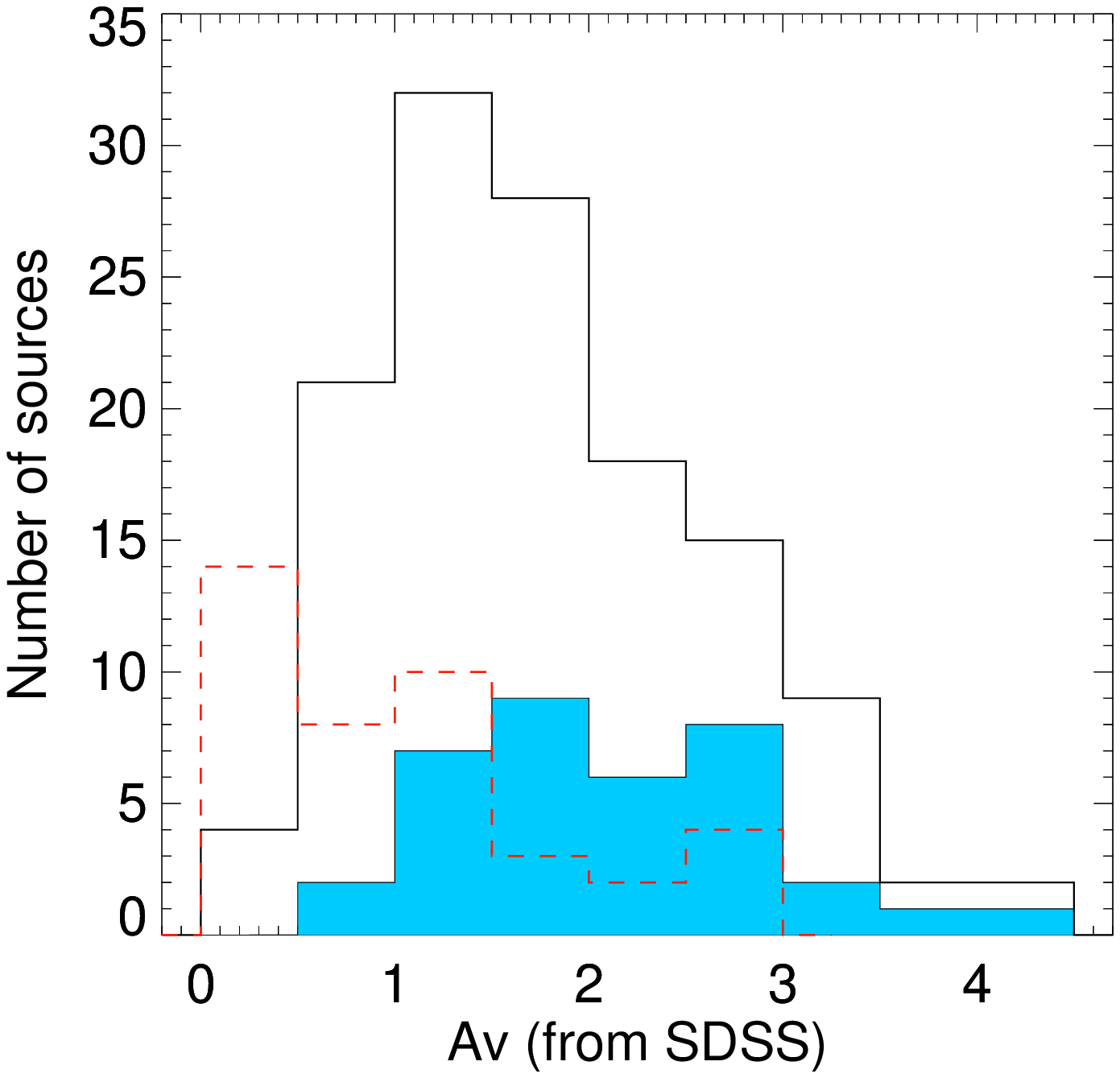}
\end{minipage}
\caption[ ]{\textit{Top:} Distribution of Av in magnitudes of MLLINERs measured from emission lines. \textit{Bottom:} Distributions of SDSS fibre Av in magnitudes measured through emission lines of: MLLINERs (filled blue histogram), the entire population of LLINERs (solid black lines), and \cite{ho97} sample of nearby LINERs (red dashed lines). }
\label{fig_extinction}
\end{figure}

\indent We found that our MLLINERs can be hosted by galaxies with a wide range of extinctions. When using the Av measurements based on the H$\alpha$ and H$\beta$ emission lines, we find that most of them reside in galaxies with high extinctions. The median Av is 1.65\,mag, covering the range 0.49\,-\,3.46\,mag (see Fig.~\ref{fig_extinction}, top plot). For comparison, the bottom plot in Fig.~\ref{fig_extinction} shows the Av distributions of LLINERs and MLLINERs when taking into account the SDSS MPA-JHU DR7 3\,arcsec fibre measurements, where we measured Av in the same way as explained in Section~\ref{sec_emission_measure}, through H$\alpha$ and H$\beta$ lines. Both samples cover similar range of extinctions, having the majority of galaxies with higher values of Av\,$>$\,1.0. These values are higher than the extinctions of the nearby and low-luminosity LINERs in \cite{ho97}, with median value of Av\,=\,0.97. 54\% and 78\% of all nearby LINERs in \cite{ho97} have Av parameter $<$\,1.0 and $<$\,1.5, respectively . These comparisons between the two samples of LINERs are consistent with the general finding that the typical extinction increases with SFR \citep[e.g.,][]{kauffmann03a}. 

\subsubsection{Morphology}

\begin{figure}
\centering
\begin{minipage}[c]{0.33\textwidth}
\includegraphics[width=6.0cm,angle=0]{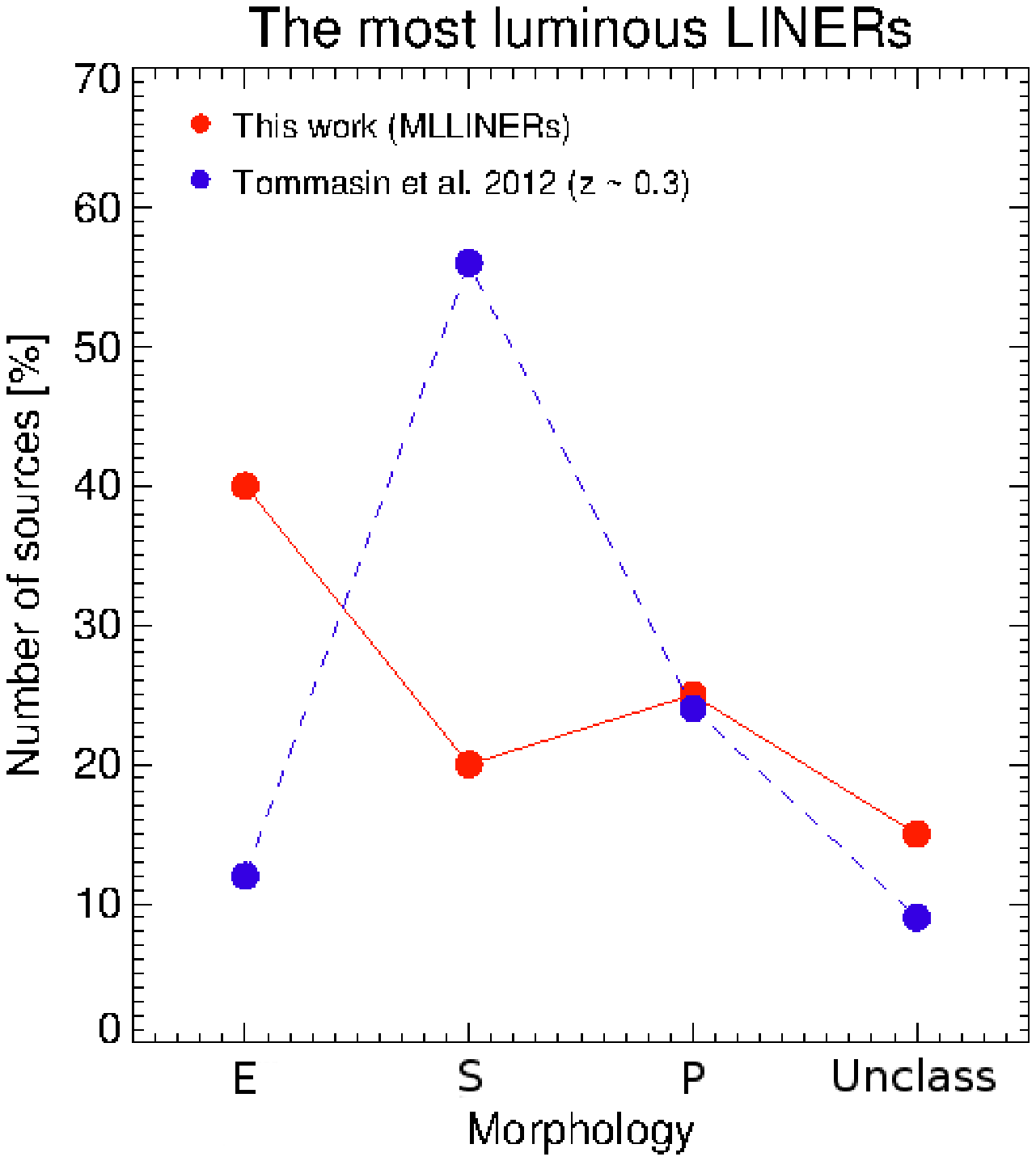}
\end{minipage}
\begin{minipage}[c]{0.33\textwidth}
\includegraphics[width=6.0cm,angle=0]{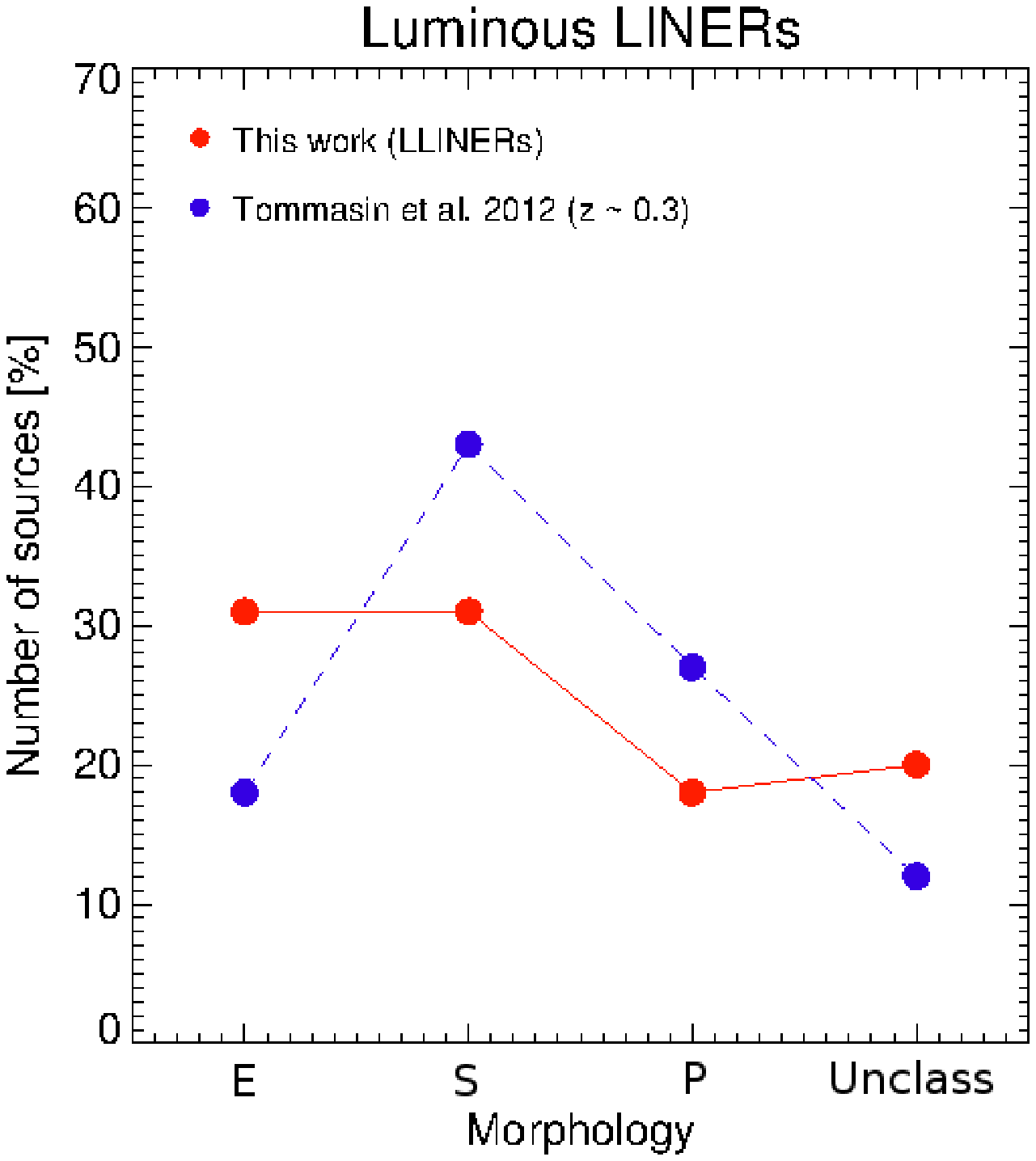}
\end{minipage}
\caption[ ]{\textit{Top:} Fraction of MLLINERs per morphological type (red filled circles)in our sample and in the \cite{tommasin12} FIR Herschel sample (blue filled circles). \textit{Bottom:} Entire LLINER sample and the entire \cite{tommasin12} optically selected sample (blue filled circles). E, S, P, and Unclass stand for Ell/S0, spiral, peculiar, and unclassified galaxies, respectively (see sec.~\ref{sec_opt_spec_data})}
\label{fig_comparison_morphology}
\end{figure}

\indent The MLLINERs studied in this work are hosted by galaxies with all morphologies, as shown in table~\ref{tab_obs} (see Section~\ref{sec_opt_spec_data} for classification details). Fig.~\ref{fig_comparison_morphology} shows comparisons between our MLLINERs (top plot) and LLINERs (bottom plot). While the differences per morphological type between MLLINERs and LLINERs are not significant ($\sim$\,10\% at most), by selecting MLLINERs we are selecting more E in comparison to S types.\\
\indent To compare our results with the sample by \cite{tommasin12} at z\,$\sim$\,0.3, we obtained the visual morphological classification in a completely consistent way as in our case, using the same classifiers and the same morphological types. We used HST/ACS images from the COSMOS\footnote{http://cosmos.astro.caltech.edu/} survey \citep{scoville07}, but we previously worsen their resolution to map the same physical size of $\sim$\,2kpc as in the case of SDSS images, and to have therefore comparable classifications. The fractions of E, S, P and unclassified galaxies can be seen in Fig.~\ref{fig_comparison_morphology} for FIR detected sample (top plot) and the entire optically-selected sample (bottom). When comparing z\,$\sim$\,0.3 and our samples, it seems that the fraction of galaxies classified as peculiar is similar at both redshifts and in both plots. On the other hand, we find higher fraction ($\sim$\,20\%) of early-type galaxies in our samples and of spiral galaxies in \cite{tommasin12}. To confirm if the observed differences are significant, we need better statistics. Incompleteness of the sample at z\,$\sim$\,0.3, plus the selection effects could be responsible for the observed differences. The most luminous \cite{tommasin12} galaxies were selected in the FIR using Herschel data, while our MLLINERs selection was carried out in optical. This could be the reason for the differences observed in the top plot of Fig.~\ref{fig_comparison_morphology}. If we check the morphological classification of our MLLINERs with the available FIR data (Table~\ref{tab_obs_fir}), we also observe that most sources are later types (68\%), classified either as S or peculiar. The sample is again too small for providing any reliable conclusions. On the other side, spectroscopic classification methods applied on the entire \cite{tommasin12} sample also differs from ours, and were based on NII-BPT and/or SII-BPT diagrams, while we used NII-BPT and OI-BPT diagrams (see sec.~\ref{sec_sample_selection}). Moreover, the apertures used in our and in \cite{tommasin12} samples cover different physical sizes of the observed galaxies. \\ 
\indent Different criteria were used in \cite{ho08} and this work to classify galaxies morphologically. While we are dealing with low-resolution data (and therefore only a rough classification in three morphology groups, E, S and P, is made) Ho's sample of nearby LINERs provides very detailed information on morphological structures. Therefore, since we are not dealing with samples classified in a consistent way, we are not able to provide any direct comparison with Ho's sample. In general, we would like to stress that our MLLINERs show higher fractions of later-types in comparison to nearby LINERs. Moreover, a significant fraction ($\sim$\,25\%) of MLLINERs are hosted by peculiar systems, showing unusual structures and clear signs of interactions, at both low- and higher-redshifts, which is again in contrast with the morphology of nearby LINERs.

\subsubsection{SFRs}

\begin{figure}
\centering
\begin{minipage}[c]{0.43\textwidth}
\includegraphics[width=6.0cm,angle=0]{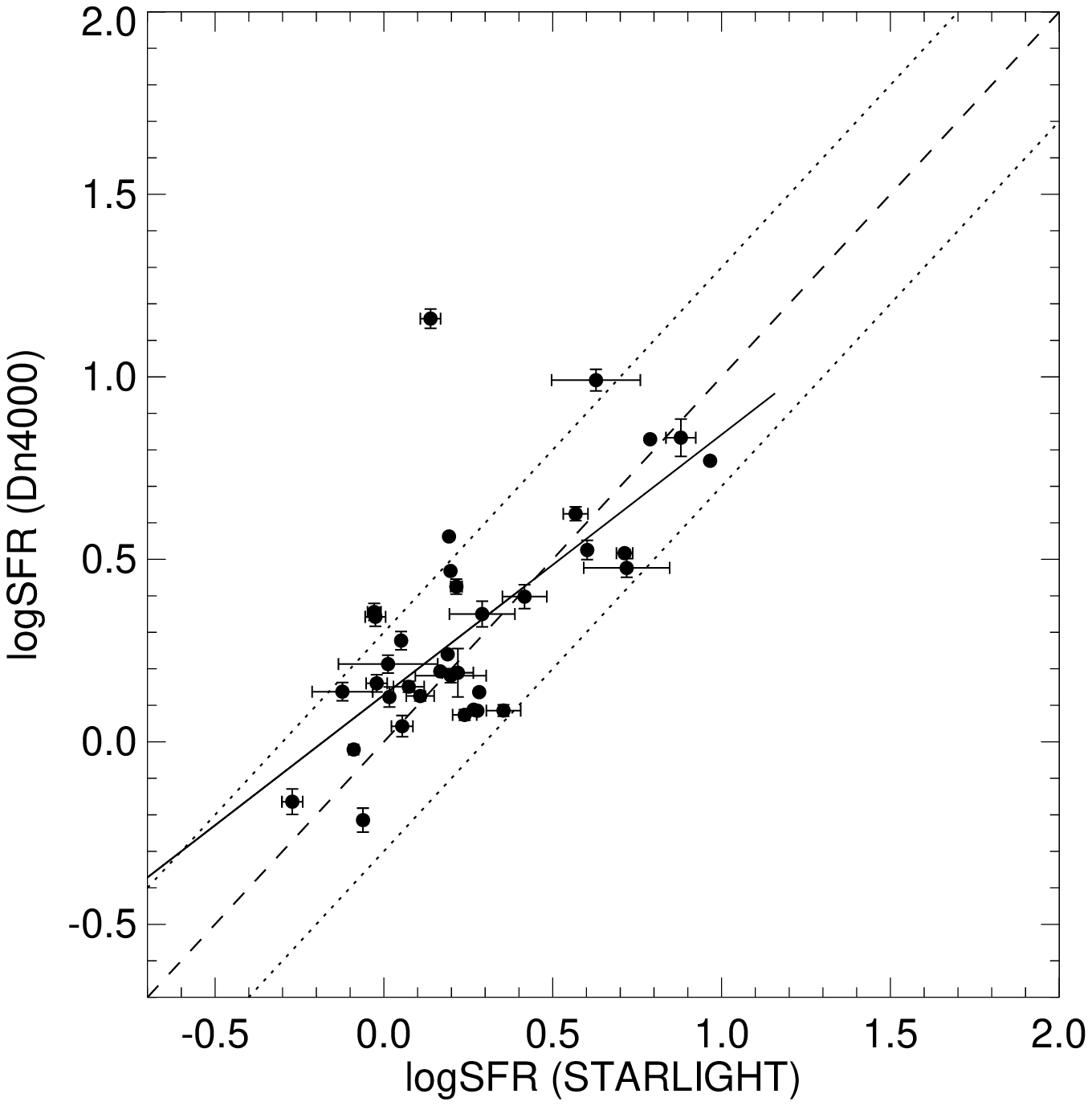}
\end{minipage}
\begin{minipage}[c]{0.43\textwidth}
\includegraphics[width=6.0cm,angle=0]{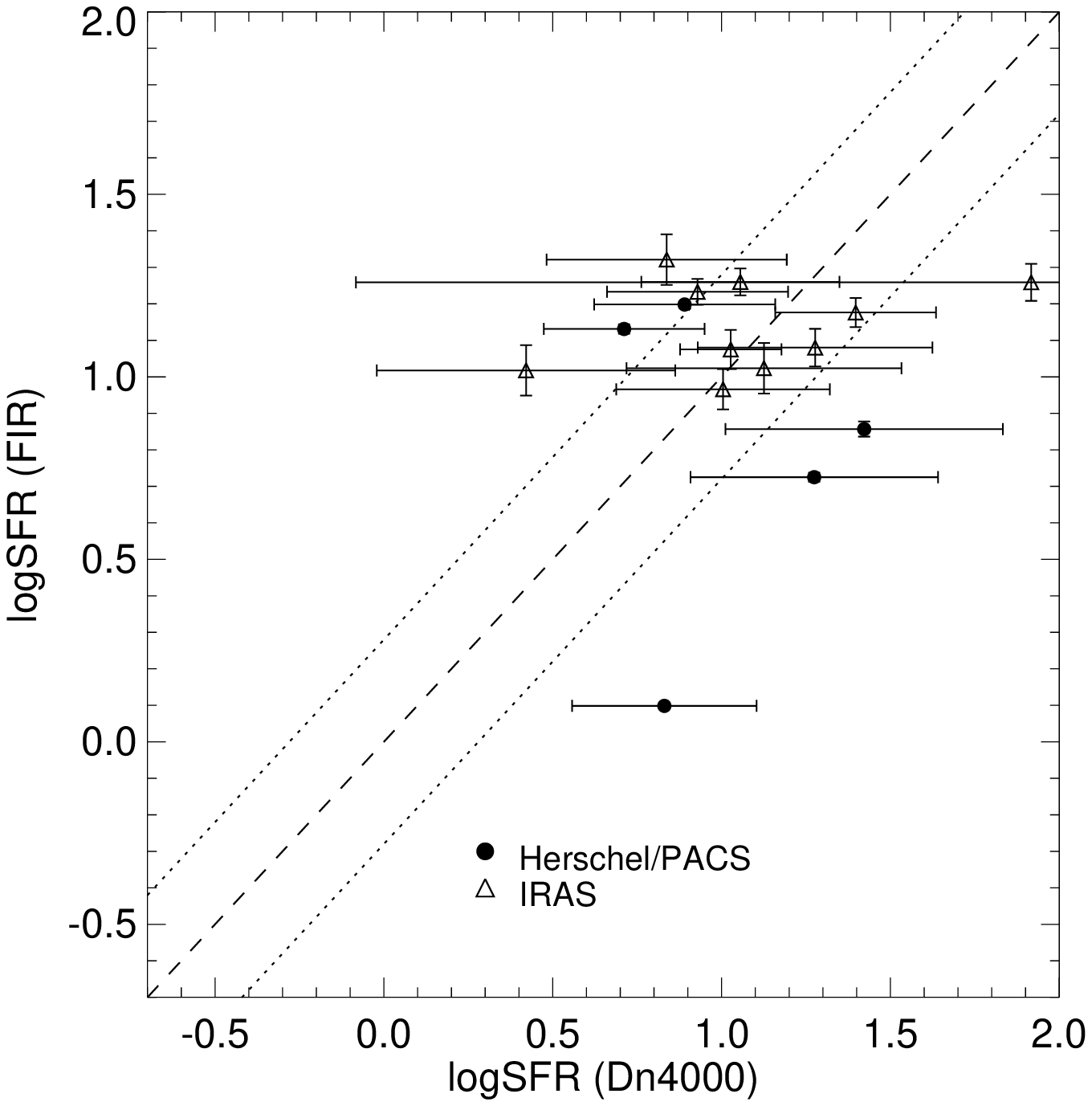}
\end{minipage}
\caption[ ]{\textit{(From top to bottom)} Comparison between SFRs measured with different methods: STARLIGHT and Dn4000 index for nuclear spectra, and Dn4000 (scaled) and FIR data. FIR data contain information from both Herschel\,-\,PACS (filled circles) and IRAS (open triangles). }
\label{fig_SFR_comparison}
\end{figure}

\indent In this work we use three different measurements of SFRs (see Section~\ref{sec_sfr}), two based on optical data (spectral fitting and Dn4000 index) and one on FIR (Herschel and IRAS). The average nuclear SFRs measured with STARLIGHT and Dn4000 index is $\sim$\,3\,[M$_{\odot}$/yr], which is significantly smaller than the SFR inferred from FIR observations with an average of $\sim$\,13\,[M$_{\odot}$/yr]. Most of the difference must be due to the fact that the nuclear region, in all sources, is considerably smaller than the size of the galaxy. If we scale the optical measurements of SFRs to the entire galaxy, assuming that the sSFR is constant (see Sec.~\ref{sec_sfr}), the difference between the optical and FIR methods becomes smaller: the average SFRs in this case are $\sim$\,9\,[M$_{\odot}$/yr] and $\sim$\,11\,[M$_{\odot}$/yr] when using STARLIGHT best-fit models and Dn4000 index, respectively. \\
\indent Fig.~\ref{fig_SFR_comparison} shows the comparison between SFRs obtained through different methods. With two different and independent methods based on optical data (spectral fitting and strength of 4000\,\AA\,Balmer break), we obtained consistent measurements of SFR, as can be seen on the top plot. As explained in Section~\ref{sec_sfr}, we used simulations from \cite{brinchmann04} to extract the mode sSFR for our nuclear measurements of Dn4000. This could be a source of several uncertainties. First, we are using just the mode values while for each Dn4000 the range of possibilities is much wider. In addition, Dn4000 measurements are based on nuclear spectra in this work while the authors used the information from SDSS aperture which is larger (see table~\ref{tab_obs} and Section~\ref{sec_observations}). Finally, in this work we are dealing with MLLINERs while the simulations were done for star-forming galaxies. Despite all this, we find a good agreement between STARLIGHT and Dn4000 SFR measurements, with $\sim$\,90\% of the sample being inside a difference of 1\,$\sigma$. \\
\indent When comparing the FIR estimations with the optical ones, but scaled to match the entire galaxy, the dispersion is larger, as shown in Fig.~\ref{fig_SFR_comparison} (bottom plot). We found $\sim$\,50\% of the sample with differences higher than 1\,$\sigma$, however we don't see any systematic trend. Several possibilities can explain the differences. First, as mentioned above we are dealing with different apertures, not only when comparing optical and FIR estimations, but for Herschel and IRAS. Secondly, the scaling assumed here, that the sSRF for the slit and the entire galaxy is the same, can lead to large uncertainties. There are other possibilities related to the geometry of the obscuring dust that affect the optically-based method much more than the FIR-based methods.\\
\indent Discrepancies based on optical and FIR SFR measurements were reported in previous works, usually finding smaller optical values in comparison to FIR \citep[e.g.][]{rigopoulou00,cardiel03,wuyts11,tommasin12}, but the scatter in most of these works is larger than in our case. In sample of the most luminous LINERs at z\,$\sim$\,0.3 by \cite{tommasin12}, the H$\alpha$ and UV measurements of SFRs are $\sim$\,30 times smaller than the FIR measurements. In contrast, the typical FIR SFRs in their sample are $\sim$\,10\,[M$_{\odot}$/yr], similar to our FIR estimations.    

\subsubsection{Stellar populations and star formation histories}

\indent As shown in Section~\ref{sec_starlight_fits} the nuclear regions of our MLLINERs are mainly characterised by intermediate (10$^8$\,$<$\,age\,[yr]\,$\le$\,10$^9$) and old (age\,[yr]\,$>$\,10$^9$) stellar populations. In $\sim$\,30\% of the sources the contribution of both intermediate and old stars is similar. In $\sim$\,20\% and 45\% of MLLINERs intermediate and old stellar populations are dominant, respectively. A young (age\,[yr]\,$\le$\,10$^8$) stars population is found in the nuclear regions of our sources in 43\% of MLLINERs, but for most of these galaxies the young stellar populations represent only $<$\,10\% of all stars. The median age of MLLINERs is logt\,=\,8.97\,[yr], covering the range logt\,=\,8.17\,-\,9.82\,[yr]. Our results are consistent with previous findings for low-luminous AGN (LINERs included) whose nuclear regions contain intermediate and old stellar populations \citep{cid04,gonzalez04}. Most of SF measured in FIR is possibly related to circumnuclear regions of MLLINERs, due to high stellar masses and/or young stars, since with our nuclear spectra in average we only cover $\sim$\,30\% of the total stellar mass. \\
\indent As mentioned in Section~\ref{sec_dn4000_hdelta}, the Dn4000 and H$\delta$ indices can be used as indicators of the SFH. The location of galaxies in the Dn4000 vs. H$\delta$ diagram, shown to be a powerful diagnostic of whether they have been forming stars continuously or in bursts over the past 1\,-\,2\,Gyr. Galaxies with continuous SFHs occupy a narrow strip in this plane (see Fig.~\ref{fig_Dn4000_Hdelta}). Following \cite{kauffmann03} models (Fig.~\ref{fig_Dn4000_Hdelta}), twelve MLLINERs (F12, F17, F21, F23, F24, B02, B15, B16, B17, B18, B19, and B23) might have experienced a burst of SF over more than 0.1\,Gyr ago (green circles). One source (B13) possibly experienced a burst of SF over less than 0.1\,Gyr ago (yellow circle). The other 17 sources (F02, F03, F06, F09, F14, F15, F20, F22, B01, B04, B05, B06, B08, B10, B11, and B14) could suffer both, burst and continuous SF (red circles). F04 has F$_{burst}$\,=\,0 (orange circle), and one can say with high confidence that this galaxy did not form a significant fraction of its stellar population in a burst over a past 2\,Gyr. Finally, five MLLINERs (F01, F16, F19, B07, and B12) lie outside the range covered by models (violet circles).

\subsection{AGN and SF luminosities of MLLINERs}
\label{sec_discussion_tommasinRelation}

\begin{figure}
\centering
\includegraphics[width=0.4\textwidth,angle=0]{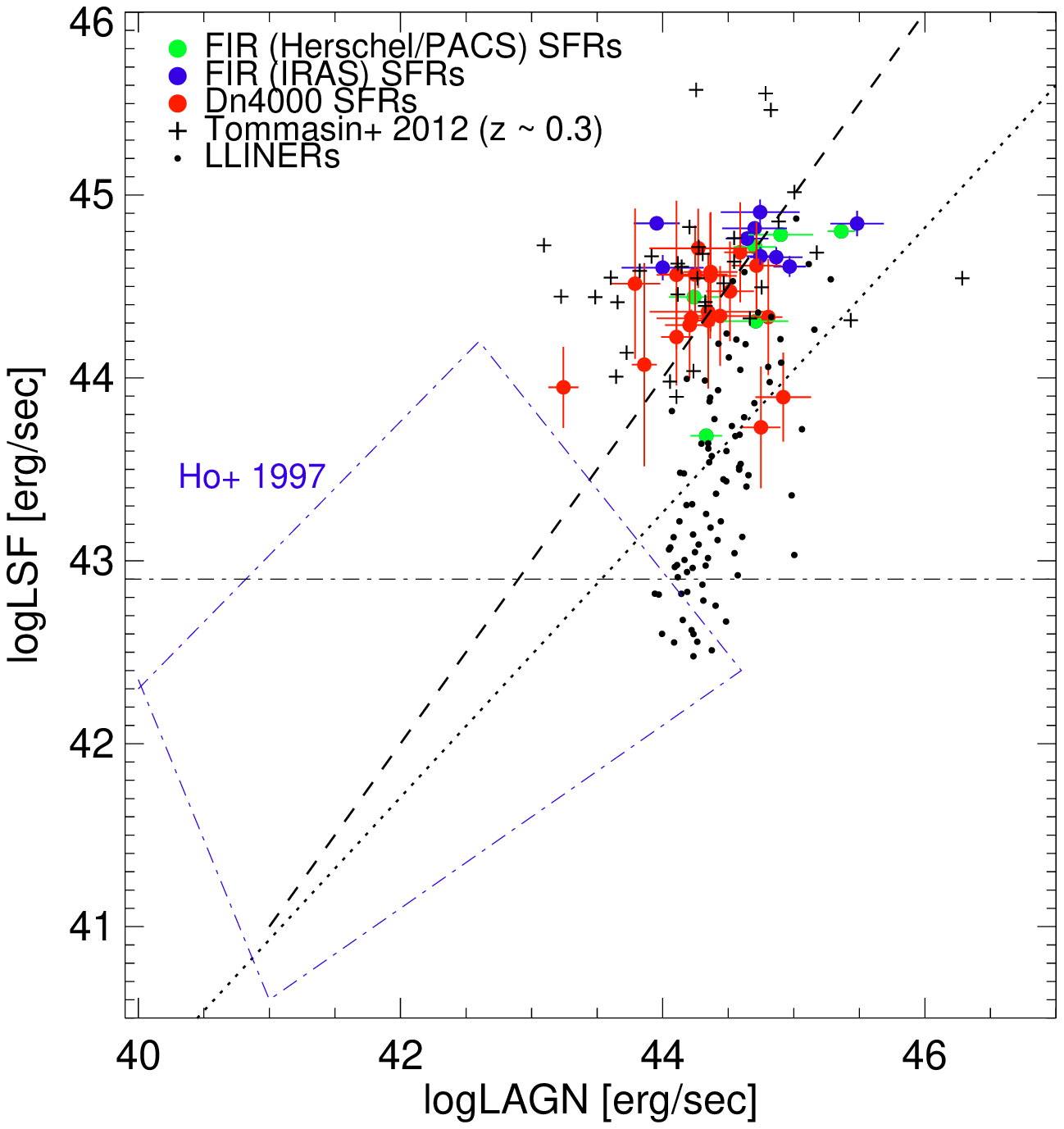}
\caption{The relationship between the AGN and SF luminosities of the most luminous local LINERs. LSF was measured in three different ways: with Herschel/PACS FIR data (big green filled circles), IRAS data (big dark blue filled circles), and through Dn4000 index (big red filled circles). For comparison, we plot the entire sample of LLINERs (small black dots), and Tommasin et al. (2012) sample of the most luminous LINERs at z\,$\sim$\,0.3 (black crosses). The blue dashed box shows the area where the nearby LINERs from Ho et al. (1997) are located. The dashed line shows the one-to-one LAGN-LSF relation, while the dotted line shows the empirical relationship for AGN-dominated sources from Netzer et al. (2009). The horizontal dashed-dot-dashed line shows the limit below which we do not trust LSF (at about 8\,$\times$\,10$^{42}$\,=\,0.2\,M$_{\odot}$/yr)
\label{fig_tommasin}}
\end{figure}

\indent The connection between LSF and LAGN was studied in many previous works, at different redshifts and for different samples of AGN, leading to somewhat inconsistent results \citep[e.g.,][and references therein]{netzer09,lutz10,rovilos12,page12,santini12,barger15,azadi15,netzer16}. Such relationships have been studied for AGN dominated sources (LAGN\,$>$\,LSF), SF dominated sources (LSF\,$>$\,LAGN) and the entire population. Some of the suggested correlations are clearly related to the sample selection (e.g., FIR or X-rays) and averaging (e.g., stacking) methods. In this section we study the relationship at low redshift for our samples of MLLINERs and LLINERs. Figure~\ref{fig_tommasin} shows LSF vs. LAGN for our two samples, where MLLINERs are represented with coloured filled circles and LLINERs with black dots. For MLLINERs, LAGN and LSF were measured as explained in previous sections. In the case of LLINERs, we used the SDSS/DR7 data and applied the H$\beta$ and OIII+OI methods to obtain LAGN, and the scaled Dn4000 method to obtain LSF. We note that in this case, some of the measured Dn4000 indices are very large (1.7 or larger) and hence cannot be used to obtained reliable SFRs \citep{kauffmann03}. We estimate this threshold to be equivalent to $\sim$\,logLSF\,=\,42.9 erg/sec (about 0.2\,M$_{\odot}$/yr).\\
\indent Figure~\ref{fig_tommasin} shows that MLLINERs tend to lie on the one-to-one LSF-LAGN relation (indicated on the diagram with a dashed line). About 90\% of all MLLINERs have values of LSF and LAGN in the range 10$^{44}$\,-\,10$^{45}$\,erg/sec. For comparison, we plotted also the line indicating the location of AGN-dominated galaxies from \cite{netzer09} (dotted line) which, by definition, are located below the 1:1 line. Our MLLINERs are located clearly above this line, and remain closer to the 1:1 relationship. On the other side, LLINERs are located below the one-to-one LSF-LAGN line, showing a wide range of LSF for the same LAGN. We suggest that this is again related to the stellar mass differences between MLLINER and LLINER samples discussed in sec.~\ref{sec_discussion_masses}. Although having the same LAGN, LLINERs with stellar masses higher than the critical one (of 6\,-\,7\,$\times$\,10$^{10}$\,M$_{\odot}$) seem to have already lower relative growth rates of stellar populations, and therefore lower LSF. As shown in \cite{perez13}, the differences in the growth rate can be even 50\%\,-\,100\%, which could explain significant differences in LSF between LLINERs and MLLINERs for the same LAGN.\\
\indent We compared our results with those for the most-luminous LINERs at z\,$\sim$\,0.3 using again the sample of \cite{tommasin12}. We used their measurements of LAGN and LSF, where LAGN were derived from the H$\beta$ and O[III]$\lambda$5007 methods and LSF from Herschel observations. In general, the location of MLLINERs at z\,$\sim$\,0.04\,-\,0.11 and at z\,$\sim$\,0.3 are very similar. \cite{tommasin12} compared their results with nearby LINERs from \cite{ho97}, finding that the later are characterised by considerably lower LAGN and LSF. In Fig.~\ref{fig_tommasin} we marked the region that corresponds to the location of nearby LINERs (blue dashed box). As can be seen, although both AGN and SF luminosities show lower values, in this case the dispersion from 1:1 relation is much larger. While some sources are distributed around 1:1 relation the others lie more around the AGN-dominated line. Note also that for low LAGN ($\sim$\,10$^{41}$\,erg/s) the difference between 1:1 and AGN-dominated relations becomes less significant \citep{netzer09}. \\
\indent In order to explain the differences between nearby and z\,$\sim$\,0.3 LINERs, \cite{tommasin12} pointed out several possibilities. First, the aperture difference, which is much smaller in the case of the Ho's sample, where only the very central regions of the galaxies are included. Second, the FIR selection of the z\,$\sim$\,0.3 sample in comparison to the Ho's LINERs, enforces higher values of LSF. Third, they argue that LINERs with such high LSF could be present in the local universe, but have not been studied yet systematically. Finally, \cite{tommasin12} suggested that there might be a real evolution in AGN and SF luminosities between z\,$\sim$\,0 and z\,$\sim$\,0.3. With our work we can provide more information about some of the questions made by \cite{tommasin12}. We can confirm the existence of LINERs in the local universe with the same SF and AGN properties as at z\,$\sim$\,0.3, discarding therefore the pure evolutionary scenario.\\

\begin{figure}
\centering
\includegraphics[width=0.4\textwidth,angle=0]{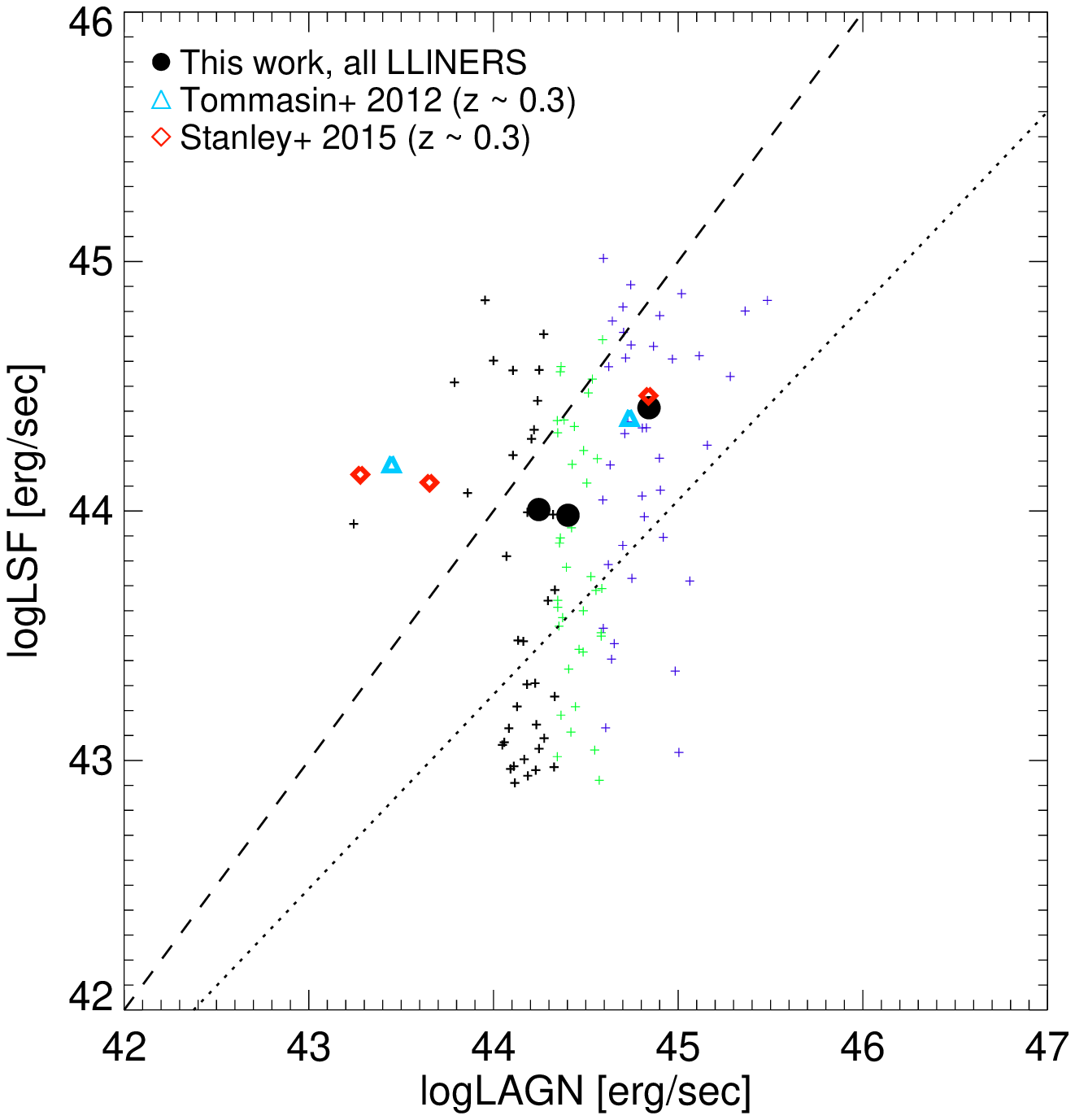}
\caption{The relationship between AGN and SF luminosities for all luminous LINERs, divided in three different LAGN bins (black, green, and blue crosses). The average values of LSF and LAGN per bin are represented with black filled circles. For comparison we show the average values of Stanley et al. (2015) for X-ray detected AGN in their first redshift bin of z\,$\sim$\,0.4 (red diamonds) and the average of the entire sample of Tommasin et al. (2012) at z\,$\sim$\,0.3 (blue triangles).
\label{fig_stanley}}
\end{figure}

\indent Recently, \cite{stanley15} studied the relationship between LSF and LAGN for a sample of $\sim$\,2000 X-ray detected (10$^{42}$\,$<$\,L$_{2-8keV}$\,$<$\,10$^{45.5}$\,erg/sec) AGN at redshifts z\,=\,0.2\,-\,2.5. They divided all galaxies in four redshift ranges, and for each redshift range they measured the average LSF and LAGN in bins of 40 galaxies. LSF was measured using FIR data, and was based mostly on Herschel upper limits (which is why they could only discuss mean LSF). LAGN is based on X-ray 2\,-8\,keV measurements. They found that the relationship between the average LSF and LAGN is mainly flat, independently of redshift and AGN luminosity. To test the flatness of the observed relationship, the authors tested their results with two empirical models \citep{aird13, hickox14} that predict $<$\,LSF\,$>$ as a function of LAGN. They suggested that the flat relationship is due to short-time scale variations in LAGN caused by changes in mass accretion rate onto the BH. These variations are shorter than those related to SF, and therefore for a given value of mean LSF, AGN luminosity can take different values and flatten the correlation. \\
\indent Here we are able to test, for the first time, \cite{stanley15} results for LINERs. We used the entire sample of LLINERs (MLLINERs included), dividing it in three LAGN bins (with 43 galaxies in the first bin and 44 in the other two bins) and measured mean LAGN and LSF in each bin. Figure~\ref{fig_stanley} shows all sources with crosses, while the mean LAGN and LSF values in the three LAGN bins are marked with filled black circles. For comparison, we plotted \cite{stanley15} averaged values for their first redshift bin at z\,$\sim$\,0.4 (red diamonds). We also show the results for the LINERs in \cite{tommasin12}. Only 34 out of the 97 objects in the Tommasin sample have measured (Herschel) SFRs. Since we are comparing averaged properties, we assume that all other LINERs in that sample have LSF\,=\,0. This would mean that the numbers we use are somewhat smaller than the actual mean LSF. \\
\indent Our results considering LLINERs are in general agreement with the \cite{stanley15} results. However, we do not have to rely on mean properties and can look at the entire LSF distribution in each bin of LAGN. The measured range in LSF is large, about 1.5 dex, similar to the overall range in LAGN. Obviously, using mean values will tend to emphasize the larger number of low SFR sources in each bin. However, the sources with the highest LSF in each LAGN bin certainly have different properties than the ones with the lowest LSF, as discussed in the following section.

\subsection{MLLINERs and the main sequence of SF galaxies}
\label{sec_discussion_ms}

\begin{figure}
\centering
\includegraphics[width=0.4\textwidth,angle=0]{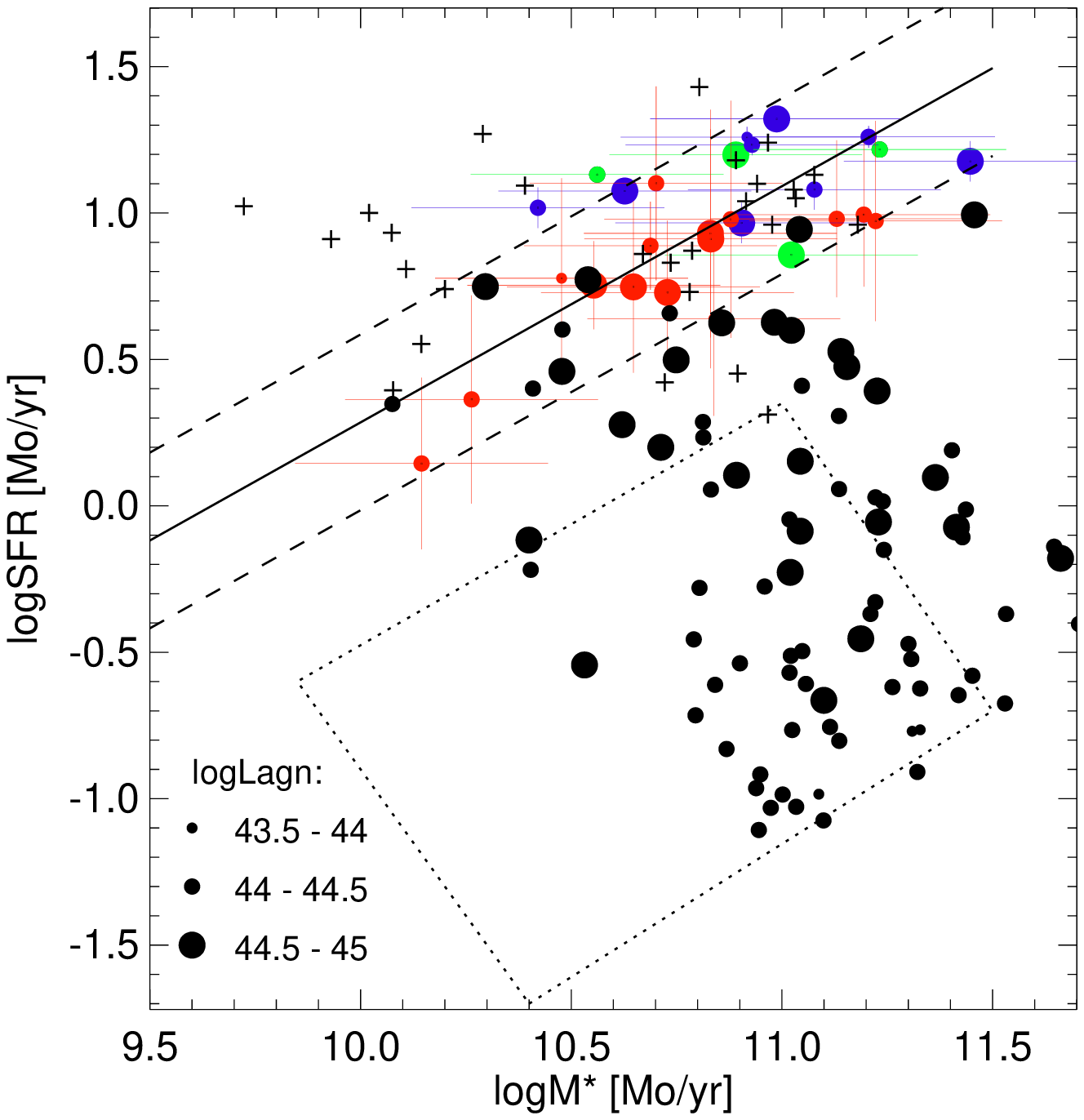}
\caption{The relationship between SFR and total stellar mass. SFRs were measured in three different ways: with Herschel/PACS FIR data (big green filled circles), IRAS data (big dark blue filled circles), and through Dn4000 index (big red filled circles). The solid black line shows the Whitaker et al. (2012) fit for the main sequence, and the dashed lines its typical width (see the text). The entire sample of luminous LINERs (small black dots), and Tommasin et al. (2012) sample of the most luminous LINERs at z\,$\sim$\,0.3 (black crosses) are shown for comparison. The dotted area is reproduced from Leslie et al (2016) and represents the typical location of 60\% of all LINERs at low redshifts. Depending on their AGN luminosity, MLLINERs and LLINERs are represented with symbols of different sizes (see sec. \ref{sec_discussion_ms_LAGN}). 
\label{fig_ms}}
\end{figure}

\indent SF galaxies show a tight and well-defined relationship called the 'main sequence' (MS) between SFR and stellar mass. This relationship depends on redshift and has been studied at different cosmic time \citep[e.g.][and references therein]{brinchmann04,noeske07,elbaz07,daddi07,gonzalez10,whitaker12,guo13,leslie16}. Figure~\ref{fig_ms} shows all the objects studied in this work on the SFR\,-\,M$_*$ diagram. For the SFRs we used exactly the same data as in Fig.~\ref{fig_tommasin}. For the stellar mass we used the mass of the entire galaxy, recovered from the MPA-JHU DR7 catalogue. For the MS, we used the fit obtained by \cite{whitaker12}, whose SFRs are also based on Kroupa IMF. We plotted the MS (solid line) for z\,=\,0.07, which is the average value in our sample. For the width of the MS we used $\pm$\,0.3\,dex (dashed lines), found in many previous works to be the typical 1\,$\sigma$ boundaries \citep[e.g,][]{elbaz07,rodighiero10,whitaker12,whitaker14,shimizu15}. More than 90\% of our MLLINERs lie along the main sequence of SF galaxies (within the dashed lines). 

\indent Once again our MLLINERs at z\,=\,0.04\,-\,0.11 show the same properties as the most luminous LINERs at z\,$\sim$\,0.3 (black crosses in Fig.~\ref{fig_ms}) in \cite{tommasin12}. At both redshifts, the most luminous LINERs represent $\sim$\,1/3 of all LLINER. Most remaining 2/3 of LLINERs lie below the MS (black dots), having lower SFRs for masses typical of MLLINERs or even higher. Considering morphological types, we found that the different types are located on the MS. This sample seems to be different from the general galaxy population where later-types are mainly located on the MS, while earlier-types lie below it \citep[e.g.][and references therein]{gonzalezdelgado16}. \\
\indent Recently, \cite{leslie16} studied the SFR\,-\,stellar mass plane for different types of low-redshift galaxies from the SDSS survey. They classified all galaxies into star-forming, composite, Sy2, LINERs\footnote{Note that this work does not take into account the separation of LINERs into systems excited by AGN and by pAGB stars.}, and ambiguous, using the emission line ratios from MPA-JHU DR7 catalogues. 6.5\% of of the sources studied in this work are LINERs. We assumed this sample (of 13,176 galaxies) to be representative of LINERs at low redshifts and plot in Fig.~\ref{fig_ms} a dotted box representing $>$\,60\% of the sources in \cite{leslie16}. The average stellar masses and SFRs they found are $<$\,log(M$_*$)\,$>$\,=\,10.74 and $<$\,log(SFR)\,$>$\,=\,-0.79, respectively. These values are smaller than for our MLLINERs, $<$\,log(M$_*$)\,$>$\,=\,10.82 and $<$\,log(SFR)\,$>$\,=\,0.86, respectively. This is not surprising given that our MLLINERs were selected according to both LAGN and LSF. 

\subsubsection{\textbf{Relation between the fraction of SF galaxies and AGN luminosity}}
\label{sec_discussion_ms_LAGN}
The more important issue of the location of LINERs in the SFR vs. M$_*$ plane as a function of LAGN, as found here, was not considered by \cite{leslie16}. To illustrate this we consider the properties of all SDSS/DR7 LINERs in the redshift range 0.04\,-\,0.11. We measured LAGN as described above and used the scaled Dn4000 method to estimate LSF. We then estimated their fraction on the MS using different bins of LAGN, where the MS is defined exactly as in Fig.~\ref{fig_ms}. The fraction of z\,=\,0.04\,-\,1.11 LINERs located on the MS is 2\%, 3\%, 11\%, and 37\% in the bins of logLAGN\,=\,43\,-\,43.5, 43.5\,-\,44, 44\,-\,44.5, 44.5\,-\,45, respectively. Thus we can safely conclude that the fraction of SF galaxies among low redshift LINERs is LAGN-dependent. While studies like those of Leslie et al. (2016) are not available at higher redshifts, it seems that for the most luminous LINERs, this difference from the rest of the population extends at least to z\,=\,0.3.\\

\section*{Summary and conclusions}
\label{sec_summary}

\indent In this work we analyse the properties of the 42 most-luminous LINERs (in terms of AGN and star-formation luminosities) at z\,=\,0.04\,-\,0.11 from the entire SDSS DR4 survey. We obtained long-slit spectroscopy of the nuclear regions for all sources, and FIR data (Herschel and IRAS) for 30\% of the sample. We carried out spectral fitting using the STARLIGHT code and templates from \cite{bruzual03}, testing 25 ages and solar metallicity. From the best-fit models we obtained the emission spectra, stellar masses, SFRs, stellar populations, and ages. We used the spectra to measure the emission lines, extinction, and extinction corrected luminosities. We also measured the Dn4000 and H$\delta$ indices. The AGN luminosities were measured through extinction-corrected emission lines, and SFRs using different indicators (both optical and FIR).\\
\indent Previous works characterised the population of local LINERs as: hosted by old and massive early-type galaxies, with low extinctions, massive black holes, old stellar populations and little star-formation \citep{ho97,ho08,heckman14}. In contrast, our most-luminous LINERs are hosted by both early- and late-types. Moreover, $\sim$\,25\% of sources are peculiar systems, with clear signs of sub-structures and interactions or mergers. We found higher values of extinction than typical for most low-redshift LINERs. The nuclear regions mainly consist of intermediate (10$^8$\,$<$\,age\,[yr]\,$\le$\,10$^9$) and old (age\,[yr]\,$>$\,10$^9$) stellar populations, while young stars are present only in 43\% of sources, similar to what has been found for nearby LINERs \citep{cid04}. The median SFRs are $\sim$\,10\,[M$_{\odot}$/yr], much higher than those for most local LINERs. However, it is interesting that they do not have the highest stellar masses, and in general show lower masses than other luminous LINERs. We found that the median stellar mass of our most-luminous LINERs corresponds to the mass of 6\,-\,7\,$\times$\,10$^{10}$\,M$_{\odot}$ measured in different works to be critical for the peak of relative growth rates of stellar populations (highest SFRs and LSF). Other LINERs although showing the same AGN luminosities, show lower SF luminosities. \\
\indent LINERs with these kind of properties were previously studied only at z\,$\sim$\,0.3 \citep{tommasin12}. With our work we confirmed the existence of such LINERs also at low-redshifts (z\,$\sim$\,0.07). They show the same properties in terms of stellar mass, SFRs, and AGN luminosity at both redshifts. Our most luminous LINERs tend to lie along the LAGN\,=\,LSF line hinting for co-evolution of the two properties. In addition, most of them are found on the MS of SF galaxies, with stellar masses $\gtrsim$\,10$^{10}$\,M$_{\odot}$. Finally, using the entire DR7 sample, we present evidence that the fraction of LINERs on the MS depends on their AGN luminosity.

\section*{Acknowledgments}

\indent We thank the anonymous referee for accepting to review this paper, giving us constructive comments that improved our paper. This research was supported by the Junta de Andaluc\'ia through project TIC114, and the Spanish Ministry of Economy and Competitiveness (MINECO) through project AYA2013-42227-P. The work was also supported by the Israel Science Foundation grant 284/13. MP acknowledge financial support from JAE-Doc program of the Spanish National Research Council (CSIC), co-funded by the European Social Fund. We used the observations collected at the Centro Astron\'omico Hispano Alem\'an (CAHA) at Calar Alto, operated jointly by the Max-Planck Institut fur Astronomie and the Instituto de Astrof\'isica de Andaluc\'ia (CSIC). The work is also based on observations made with the Nordic Optical Telescope, operated by the Nordic Optical Telescope Scientific Association at the Observatorio del Roque de los Muchachos, La Palma, Spain, of the Instituto de Astrofisica de Canarias. All data reduction and part of analysis were done using IRAF, the Image Analysis and Reduction Facility made available to the astronomical community by the National Optical Astronomy Observatories, which are operated by the Association of Universities for Research in Astronomy (AURA), Inc., under contract with the US National Science Foundation. We made use of Virtual Observatory Tool for OPerations on Catalogues And Tables (TOPCAT). We also made use of the sipl spectral analysis code made by Jaime Perea, and we thank the author for making the code available. We also thank to STARLIGHT team for making their code public.

\appendix

\section[]{STARLIGHT fits and emission spectra}

\indent \indent In this section we show the example of flux calibrated nuclear spectra (blue lines), STARLIGHT fits (red lines), and the emission spectra (black lines) of MLLINERs. The emission spectra were obtained after subtracting the best model found by STARLIGHT from the flux calibrated spectra.   

\begin{figure*}
\centering
\begin{minipage}[c]{0.49\textwidth}
\includegraphics[width=9.0cm,angle=0]{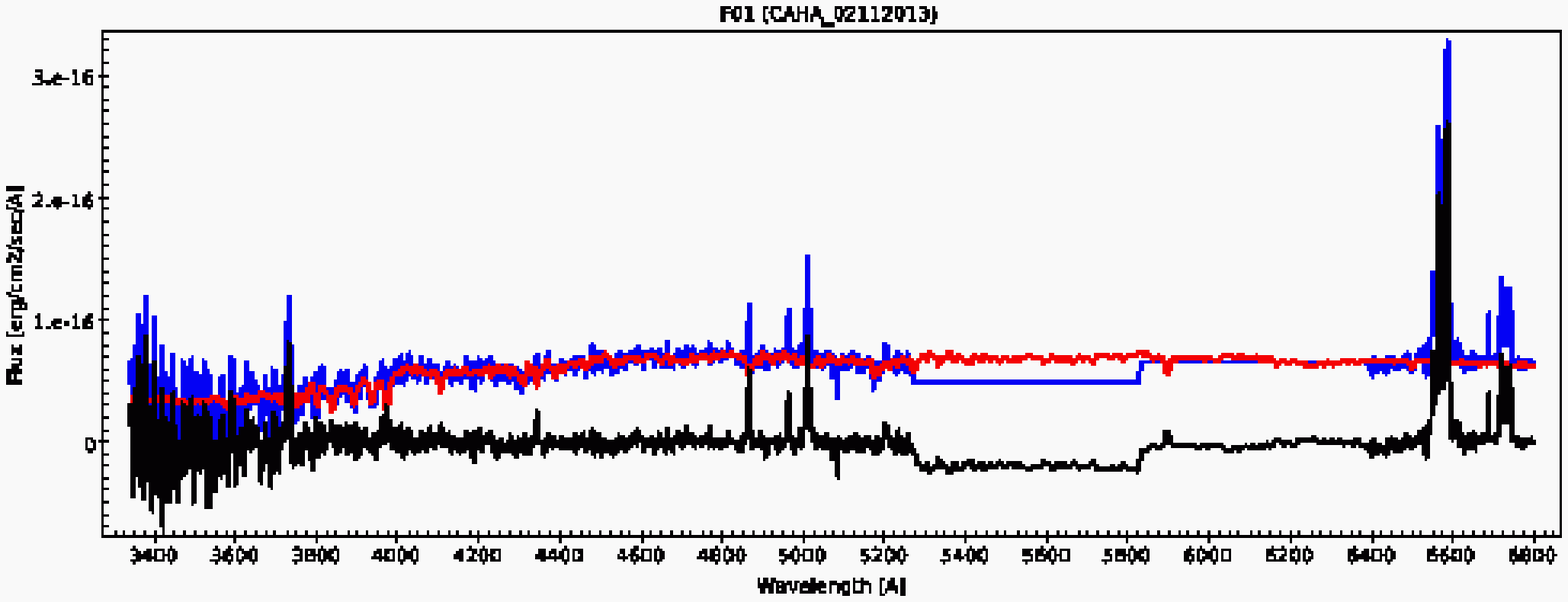}
\end{minipage}
\begin{minipage}[c]{0.49\textwidth}
\includegraphics[width=9.0cm,angle=0]{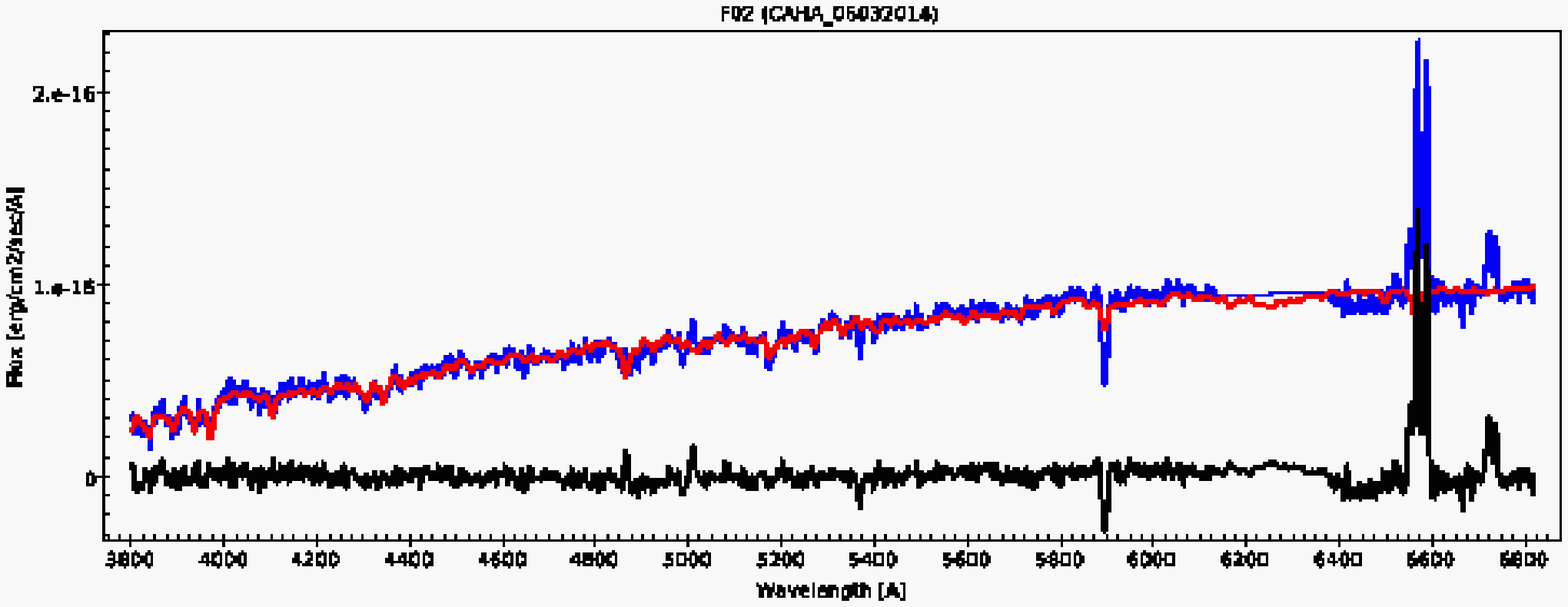}
\end{minipage}
\begin{minipage}[c]{0.49\textwidth}
\includegraphics[width=9.0cm,angle=0]{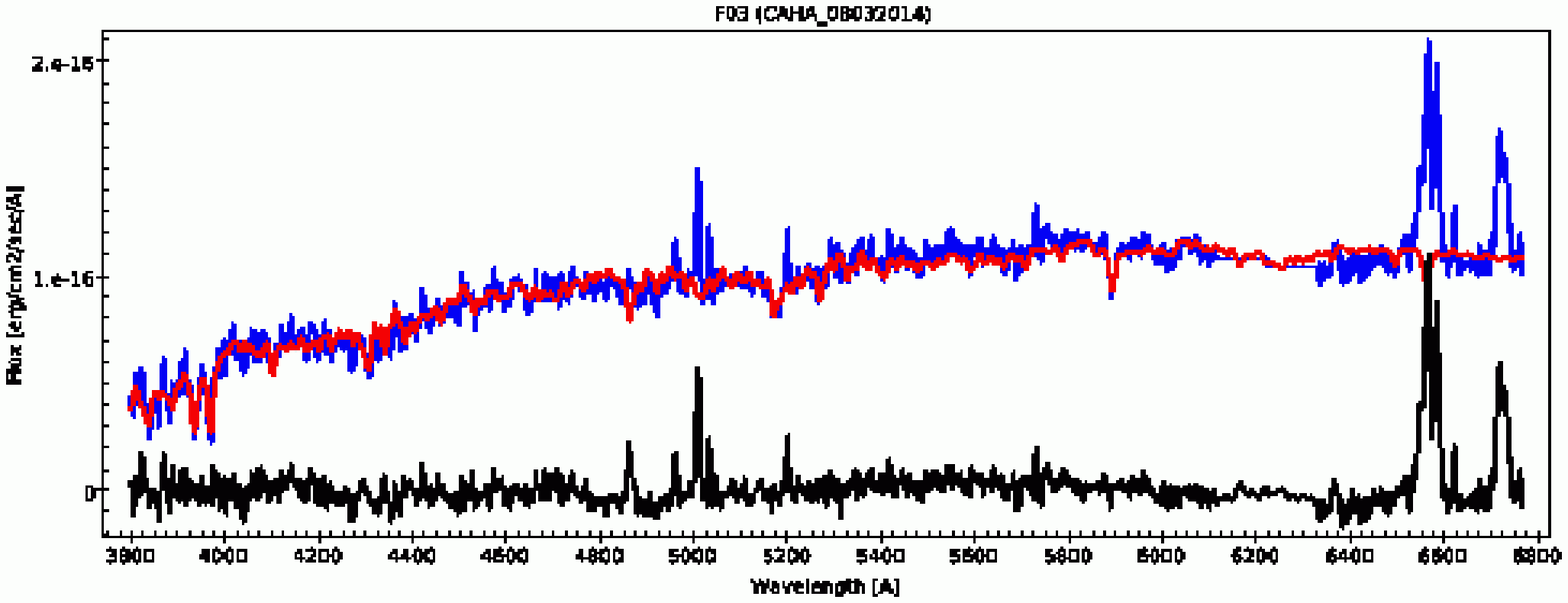}
\end{minipage}
\begin{minipage}[c]{0.49\textwidth}
\includegraphics[width=9.0cm,angle=0]{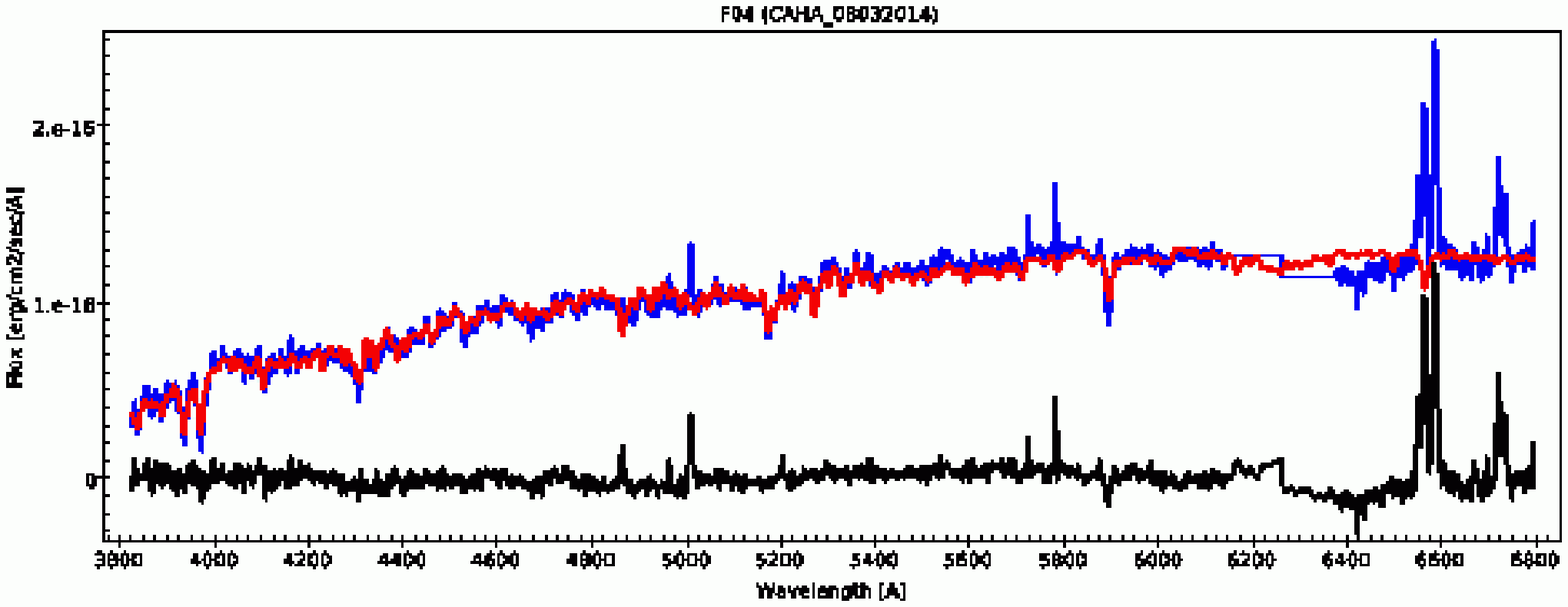}
\end{minipage}
\begin{minipage}[c]{0.49\textwidth}
\includegraphics[width=9.0cm,angle=0]{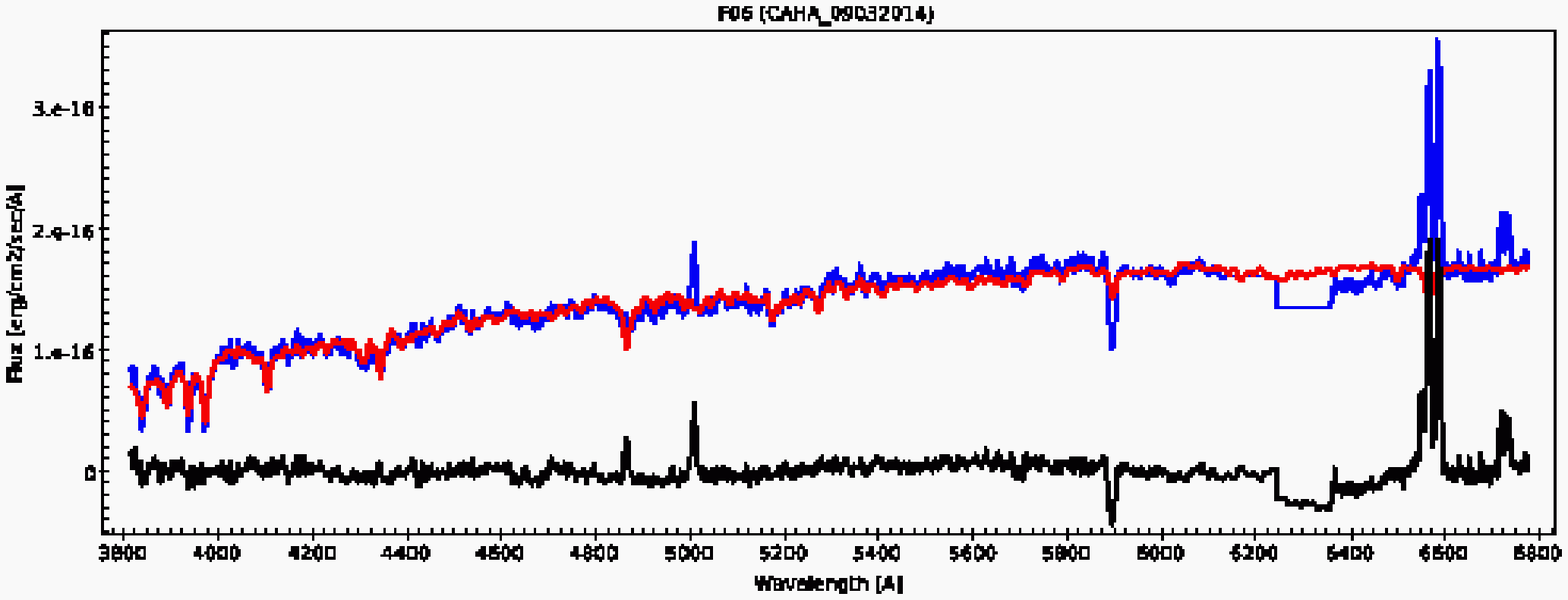}
\end{minipage}
\begin{minipage}[c]{0.49\textwidth}
\includegraphics[width=9.0cm,angle=0]{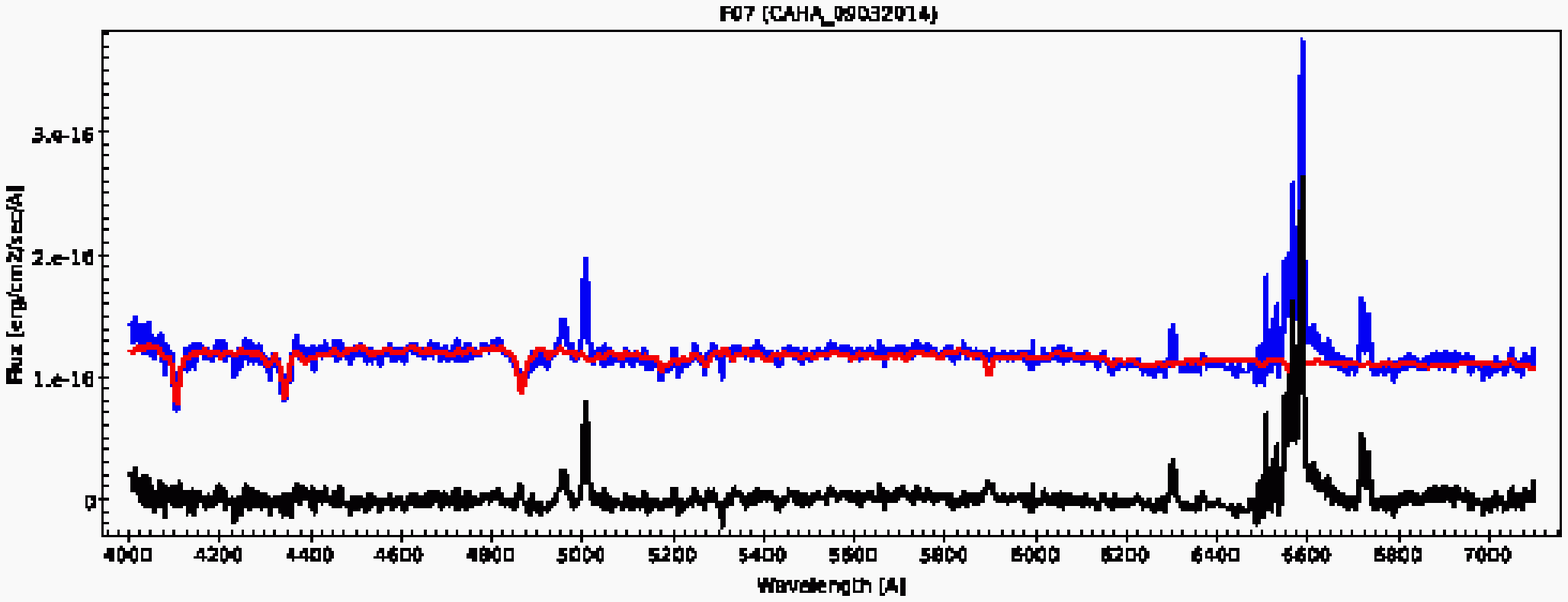}
\end{minipage}
\begin{minipage}[c]{0.49\textwidth}
\includegraphics[width=9.0cm,angle=0]{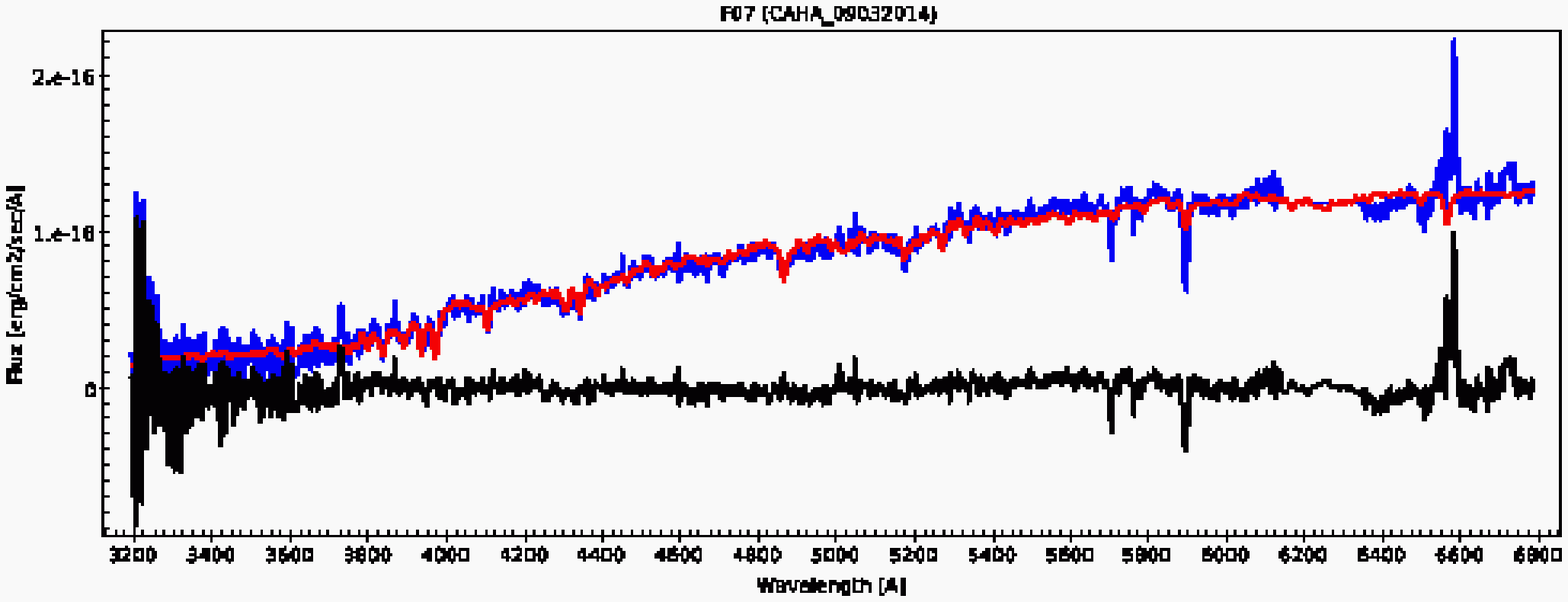}
\end{minipage}
\begin{minipage}[c]{0.49\textwidth}
\includegraphics[width=9.0cm,angle=0]{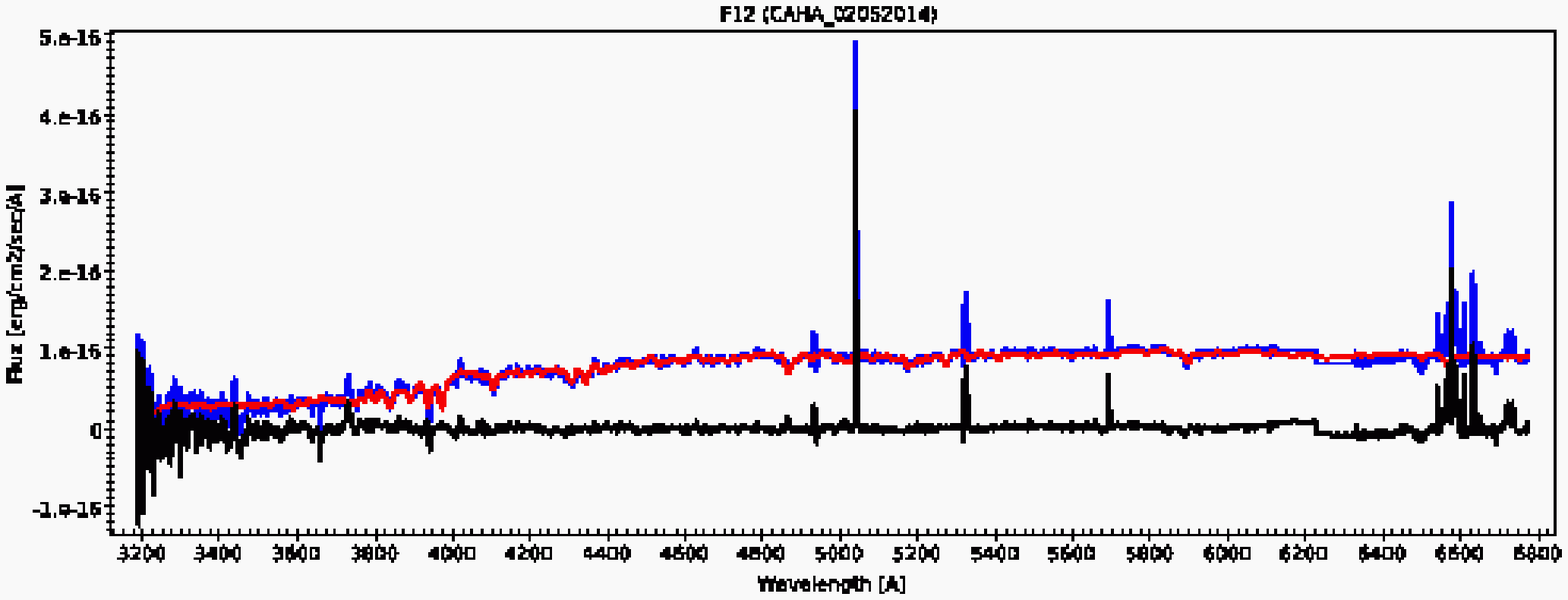}
\end{minipage}
\begin{minipage}[c]{0.49\textwidth}
\includegraphics[width=9.0cm,angle=0]{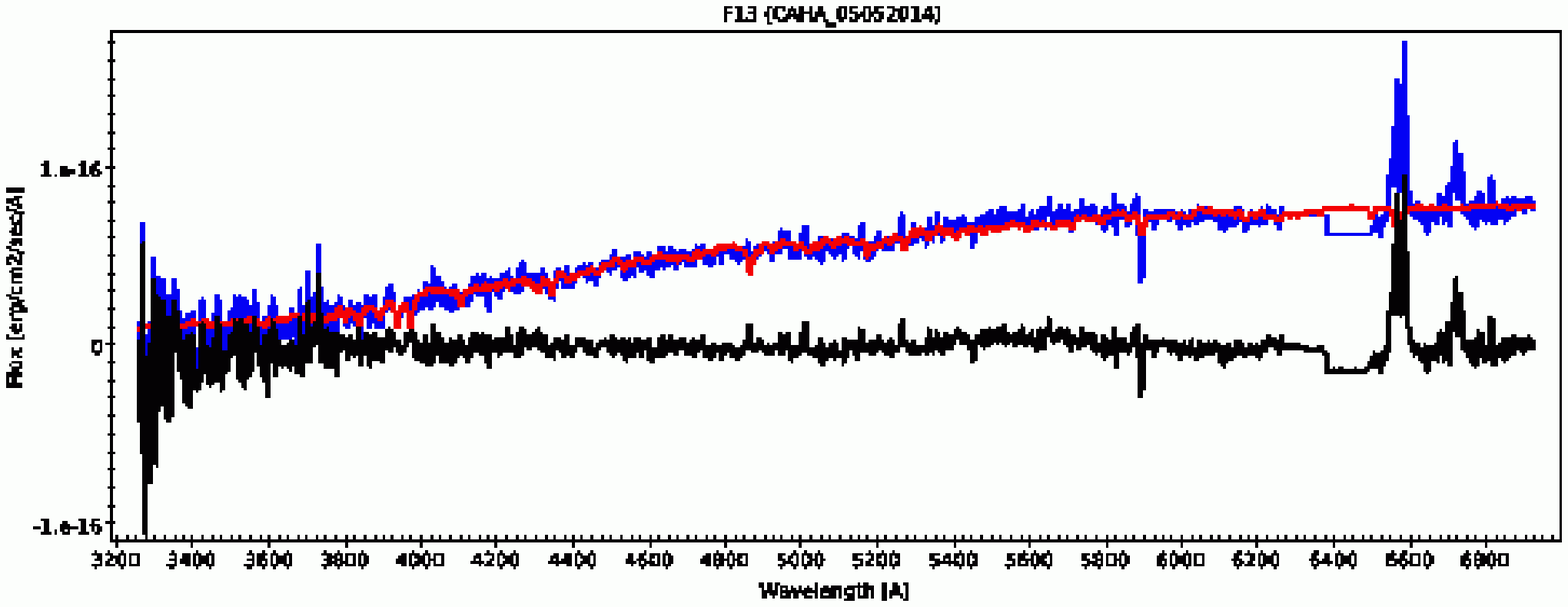}
\end{minipage}
\begin{minipage}[c]{0.49\textwidth}
\includegraphics[width=9.0cm,angle=0]{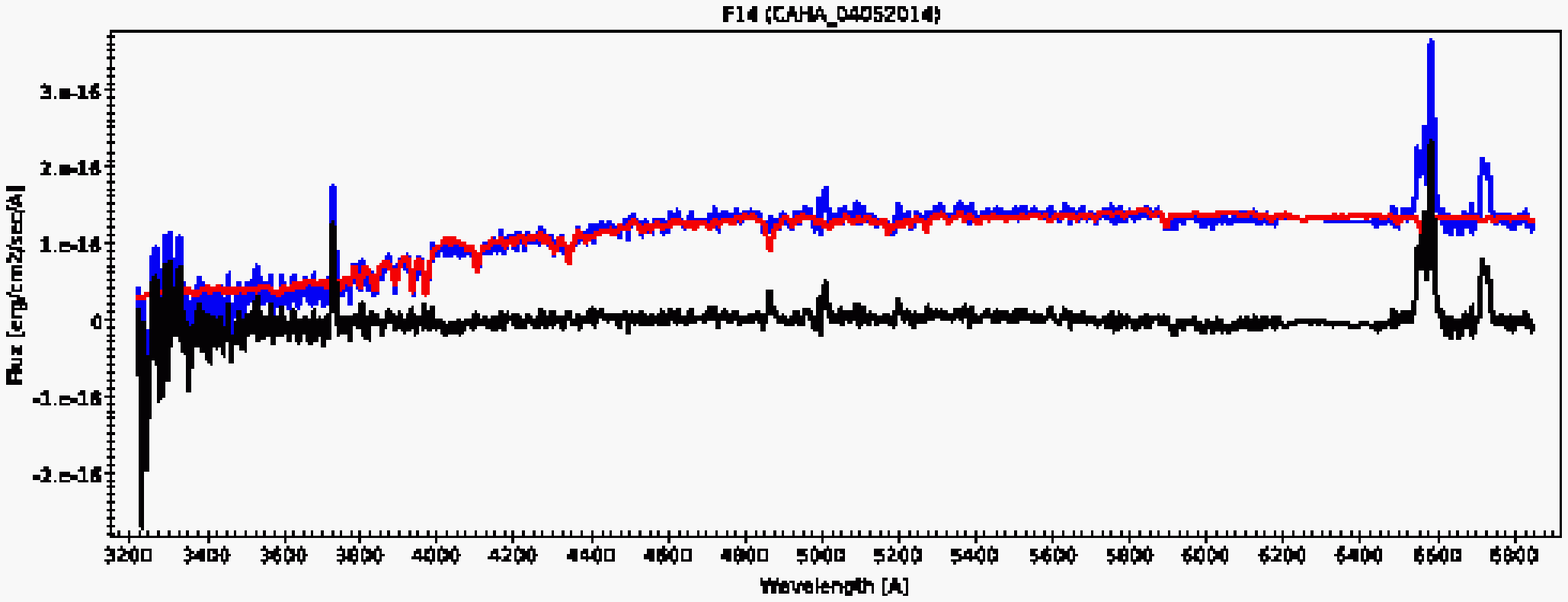}
\end{minipage}
\caption[ ]{Original (blue), best-model fit (red), and emission (black) spectra of (from top to bottom, and from left to right): F01, F02, F03, F04, F06, F07, F09, F12, F13, and F14 LINERs. }
\label{fig_spectra_1}
\end{figure*} 

\renewcommand{\thefigure}{\arabic{figure}}

\label{lastpage}

\end{document}